\documentclass[iop, twocolumn]{aastex63}
\hypersetup{citecolor=blue,urlcolor=red}

\usepackage{multirow}
\usepackage{soul}
\usepackage{xspace}
\usepackage{subfiles}
\usepackage{ulem}
\usepackage{amssymb}
\usepackage{amsmath}
\usepackage{bm}

\usepackage{natbib}
\usepackage{booktabs}
\usepackage{amsmath}
\usepackage[figuresright]{rotating}
\usepackage{url}
\usepackage{booktabs}

\usepackage[para,online,flushleft]{threeparttable}


\newcommand{\Msun}{\ensuremath{M_{\odot}} }

\newcommand{\nustar}{{\it NuSTAR}}

\newcommand{\swift}{{\it Swift}}
\newcommand{\fermi}{{\it Fermi}}
\newcommand{\planck}{{\it Planck}}
\newcommand{\spizer}{{\it Spizer}}
\newcommand{\herschel}{{\it Herschel}}

\shorttitle{The origin of multiwavelength emissions of PKS\,1510$-$089}
\shortauthors{Lei et al.}

\begin{document}
\title{Insight into the origin of multiwavelength emissions of PKS\,1510$-$089 through modeling 12 SEDs from 2008 to 2015}

\correspondingauthor{Maichang Lei}
\email{maichanglei83@163.com}

\author[0000-0002-2224-6664]{Maichang Lei}
\affiliation{College of Physics and Engineering Technology, Minzu Normal University of Xingyi, Xingyi, Guizhou 562400, People’s Republic of China}

\author[0000-0002-2224-6664]{Yuan Zheng}
\affiliation{College of Physics and Engineering Technology, Minzu Normal University of Xingyi, Xingyi, Guizhou 562400, People’s Republic of China}

\author[0000-0002-2224-6664]{Jianfu Zhang}
\affiliation{Department of Physics, Xiangtan University, Xiangtan, Hunan 411105, People’s Republic of China}

\author[0000-0002-2224-6664]{Yuhai Yuan}
\affiliation{Center for Astrophysics, Guangzhou University, Guangzhou, Guangdong, 510006, People’s Republic of China}

\author[0000-0002-2224-6664]{Jiancheng Wang}
\affiliation{Yunnan Observatories, Chinese Academy of Sciences, Kunming 650216, People’s Republic of China}

\begin{abstract}\label{sec:abstra}
PKS\,1510$-$089 is one of the most peculiar sources among the FSRQs, exhibiting a notable big blue bump (BBB). This provides an unique opportunity to explore the coupling between the activity of the central engine and the relativistic jet, offering further insight into the origin of the multiwavelength emissions. To this end, we collected multiwavelength data spanning four periods from 2008 to 2015 and performed the spectral energy distribution (SED) modeling using a one-zone homogeneous leptonic model. In the model, a multichromatic accretion disk (AD) is used to fit the optical/UV data sets, while the external radiation fields from the broad-line region (BLR) and dusty torus (DT) are properly considered to produce the high-energy $\gamma$-ray emissions. Our best fit to 12 SEDs yields the following results: (i) The innermost stable orbit ($R_{\rm ISO}$) of the AD is not stable but varies between $3\,R_{\rm S}$ and $18\,R_{\rm S}$ during these observations. (ii) The high-energy hump of the SED is well dominated by Compton scattering of the BLR photons, while the X-ray flux may be comprised of multiple radiation components. (iii) The $\gamma$-ray emitting regions are generally matter-dominated, with low magnetization, and are located beyond the BLR but within the DT. At such distance, the multiwavelength emissions are likely to originate from shock accelerations; (iv) For the energization of the relativistic jet, our study supports the Blandford$-$Znajek (BZ) mechanism, instead of the Blandford$-$Payne (BP) mechanism, as the latter fails to power the jet.
\end{abstract}


\section{Introduction}\label{sec:part-1}
Blazars are radio-loud active galactic nuclei (AGNs) characterized by highly relativistic and collimated jets that are closely aligned with our line of sight \citep{1995PASP..107..803U}. To date, it is commonly believed that these jets are powered by the central supermassive black hole (BH, $10^{6}$\,\Msun$\sim 10^{10}$\,$M_{\odot}$, where \Msun is the solar mass) through Blandford $\&$ Znajek (BZ) mechanism \citep{1977MNRAS.179..433B}, with a mass loading efficiency $\eta_{\rm jet}\equiv \dot{M}_{\rm jet}/\dot{M}_{\rm acc}<1$ \citep{2016ApJ...828...13I}, or by the inner regions of the accretion disk via Blandford $\&$ Payne (BP) mechanism \citep{1982MNRAS.199..883B}, with $\eta_{\rm jet}\sim 0.1-1$ \citep[][]{2008MNRAS.385..283C,2011MNRAS.414.2674G}, where $\dot{M}_{\rm jet}$ and $\dot{M}_{\rm acc}$ are the outflow and mass accretion rates, respectively. In either case, the strong large-scale magnetic fields that thread the highly spinning BH or the rotating accretion disk will play a vital role.

The spectral energy distribution (SED) of blazars typically exhibits a double-bumped morphology. The low-energy component, spanning from radio to ultraviolet (UV) or X-rays, is generally attributed to synchrotron emissions from relativistic electron populations moving down the jet. The high-energy component, which covers a wider energy range from X-rays to GeV or even up to multi-TeV energies, has a more debated origin. Within the leptonic framework, the $\gamma$-ray emissions are thought to originate from inverse Compton scattering of the synchrotron photons (synchrotron Self-Compton, SSC) and/or of several external radiation fields (external radiation Compton, ERC). These fields may be provided by the accretion disk (AD), the broad-line region (BLR), and the dusty torus (DT). On the other hand, some hadronic scenarios, e.g., the synchrotron emission of the high-energy protons, as well as the decays of pions and muons, are also proposed to interpret the high-energy $\gamma$-ray emissions \citep[][]{2000NewA....5..377A,2001APh....15..121M,2013ApJ...768...54B,2015ApJ...802..133D,2015MNRAS.448..910C,2019Galax...7...20B,2020Galax...8...72C}.

The blazar class has two members, i.e., flat spectrum radio quasars (FSRQs) and BL Lac objects (BL Lacs), both of which exhibit dramatically different optical properties. FSRQs have noticeable emission lines with equivalent widths usually larger than 5~\AA\,, whereas BL Lacs are characterized by the absence or weakness of broad emission lines \citep{1995PASP..107..803U}. This discrepancy may be related to difference in accretion rate. Previous studies have suggested that FSRQs have higher accretion rate compared to BL Lacs \citep[][]{2011MNRAS.414.2674G,2012MNRAS.421.1764S,2014ARA&A..52..589H,2014ARA&A..52..529Y}, with $L_{\rm d}/L_{\rm Edd}\geq 10^{-2}$ \citep[][]{2011AIPC.1381..180G,2014Natur.515..376G}, where $L_{\rm d}$ is the AD luminosity, and the Eddington luminosity is given by $L_{\rm Edd}\simeq1.26\times 10^{46}$\,$M_{\rm BH,8}$\,erg~$\rm s^{-1}$, with $M_{\rm BH,8}$ representing the central BH mass in units of $10^{8}$\,$M_{\odot}$. Based on synchrotron peak frequency $\nu_{\rm p,s}$, blazars can be further subdivided into low- ($\nu_{\rm p,s}<10^{14}$\,Hz, LSPs), intermediate- ($10^{14}$\,Hz$\leq \nu_{\rm p,s}\leq 10^{15}$\,Hz, ISPs) and high-synchrotron-peaked sources ($\nu_{\rm p,s}>10^{15}$\,Hz, HSPs) \citep{2010{\natexlab{a}}ApJ...716...30A}. In contrast, FSRQs typically show a lower $\nu_{\rm p,s}$ and are thus classified as LSPs. The lower $\nu_{\rm p,s}$ also corresponds to a lower inverse Compton (IC) peak. This behavior of FSRQs is presumably related to efficient cooling of the relativistic electrons due to intense diffuse radiation fields from the BLR and DT \citep{1998MNRAS.301..451G}, which are roughly located at $\sim 10^{3}-10^{4}$\,$R_{\rm S}$ and at $1\sim 10$\,pc, respectively (where $R_{\rm S}=2GM_{\rm BH}/c^{2}$ is the Schwarzschild radius, G is the gravitational constant).

PKS\,1510$-$089 is one of the very high energy ($E_{\rm \gamma}\gtrsim 100$\,GeV, VHE) $\gamma$-ray FSRQs, located at $z=0.361$ \citep[][]{1990PASP..102.1235T,2013A&A...554A.107H}. The radio images show a large misalignment in the arcsecond and milliarcsecond scales of its jet, with a superluminal velocity of $v_{\rm app}\simeq 20$\,c \citep{2002ApJ...580..742H}. Interestingly, this source has the most fastest jet ever recorded, with apparent speeds of $\sim 46$\,c \citep{2005AJ....130.1418J}, indicating an extreme ability to accelerate particles to higher energy. Like other blazars, PKS\,1510$-$089 exhibits strong flux variability throughout entire electromagnetic spectrum, from radio to $\gamma$-rays, with the observed variability timescales ranging from months down to 20\,minutes \citep[][]{2013A&A...555A.138F,2013ApJ...766L..11S,2017ApJ...844...62P}. Therefore, it has been the target of many multiwavelength campaigns. Since the operation of the \fermi-Large Area Telescope (\fermi-LAT), the first multiwavelength observation was conducted in 2008 September and 2009 June. During this period, the isotropic $\gamma$-ray luminosity reached up to $\sim 10^{48}$\,erg~$\rm s^{-1}$, and a BH mass of $\simeq 5.4\times 10^{8}$\,$M_{\odot}$ was obtained by using UV photometry, corresponding to an accretion rate $\simeq 0.5$\,$M_{\odot}$\,$\rm yr^{-1}$. The results of the discrete cross-correlation function (DCCF) analysis suggest a weak correlation between the $\gamma$-ray and the UV fluxes, but no correlation with the X-rays \citep{2010{\natexlab{b}}ApJ...721.1425A}. In 2011, the source was observed in a quiet state, and the broadband investigations indicated the corona likely contributed to the X-ray flux \citep{2012ApJ...760...69N}. In 2015, this source came into a high-activity state, exhibiting significant VHE $\gamma$-ray variability for the first time. This flaring state was accompanied by a rotation of the optical polarization angle. During this period, a new emission component was detected in radio observations \citep{2017A&A...603A..29A}. Meanwhile, another mutifrequency observation found a strong radio flare following the $\gamma$-ray one, where two laterally separated knots of emission originated from radio core, and an edge-brightened line polarization was observed \citep{2015ApJ...804..111M}. This suggests that the $\gamma$-ray flare may have originated from the compression of the knots by a standing shock, indicating that the jet has a multiple complex layer system. At radio wavelengths, very long baseline interferometry (VLBI) observations showed that the highly polarized and variable cores are correlated with $\gamma$-ray flares in variability \citep[][]{2010ApJ...710L.126M,2014{\natexlab{a}}A&A...569A..46A}. These radio cores are usually identified with recollimation shocks that are arised as a result of the collisions between the jet and confining medium \citep[][]{1983ApJ...266...73S,1988ApJ...334..539D,1997MNRAS.288..833K,2015ApJ...809...38M,2016A&A...588A.101F,2018ApJ...860..112P}, occurring at distances $\gtrsim 1$\,pc from the central engine \citep[][]{2009MNRAS.400...26O,2012A&A...545A.113P}. \cite{2013MNRAS.428.2418O} found a $\gamma$-ray flare followed by a strong radio flare and constrained the $\gamma$-ray emitting site to be $\sim 10$\,pc downstream of the jet. At this distance, the $\gamma$-ray flares may arise from internal shocks occurring when the inhomogeneities of the jet collide \citep{2001MNRAS.325.1559S}. In fact, the ejection of double knots during the strong $\gamma$-ray flares have been observed \citep{2019ApJ...877..106P}. Moreover, the SED modeling imposes severe constraint on the location of the $\gamma$-ray emitting region. \cite{2015ApJ...809..171S} suggested that the $\gamma$-ray flare observed in 2011 occurred at 0.3\,pc or 3\,pc, depending on whether a free-expanding jet or a collimated outflow is considered, with the seed photons mainly provided by the DT. Similar distances are also favored by several investigations \citep[][]{2010{\natexlab{b}}ApJ...721.1425A,2012ApJ...760...69N,2017A&A...603A..29A,2018A&A...619A.159M,2019ApJ...883..137P,2021A&A...648A..23H}. An exception is the flare occurred during 2009 January and 2010 January, and the modeling to the resulting SEDs requires the $\gamma$-ray emissions are produced roughly at $10^{3}$\,$R_{\rm S}$ ($\sim 3\times 10^{-2}$\,pc) for a BH mass of $3\times 10^{8}$\,$M_{\odot}$, which is well within the BLR \citep{2017A&A...601A..30C}.

Thanks to persistent efforts in spectroscopic monitoring, significant progress has been made in estimating the central BH mass of PKS\,1510$-$089. \cite{2001MNRAS.327.1111G} estimated a BH mass of $\sim 1.3\times 10^{9}$\,$M_{\odot}$. \cite{2002ApJ...576...81O} used the $H_{\beta}$ FWHM and luminosity to derive a relatively smaller BH mass of $3.86\times 10^{8}$\,$M_{\odot}$. We also note that other mass estimates from the literatures are roughly consistent with these values, e.g., \cite{2005AJ....130.2506X} provided two values for the BH mass, $1.1\times 10^{8}$\,$M_{\odot}$ and $1.6\times 10^{8}$\,$M_{\odot}$, using the minimum variability timescale and single-epoch spectrum, respectively. Meanwhile, the reverberation mapping (RM) technique has been used to measure the size of the BLR and the BH mass. \cite{2020A&A...642A..59R} used 8.5\,years of the spectroscopic monitoring data to obtain $R_{\rm BLR}\simeq 61.1$\,light-days and inferred a BH mass of $5.71\times 10^{7}$\,$M_{\odot}$ using line dispersion $\sigma_{\rm line}\simeq 1262$\,km~$\rm s^{-1}$. More recently, \cite{2025ApJ...979..227A} analyzed nearly a decade of the optical spectroscopic data within the disk dominance regime, yielding a BH mass of $\sim 2.85\times 10^{8}$\,$M_{\odot}$. As an important quantity, the BH mass is closely related to the size of the BLR, the Eddington luminosity, and the production mechanism of the relativistic jets. Unfortunately, the BH mass of PKS\,1510$-$089 remains highly uncertain.

Given the notable optical/UV excess and multiple SED observations, these will help constrain the central BH mass and provide important insights into the origin of the multiwavelength emissions. To achieve this, we collected 12 SEDs from four periods spanning 2008 to 2015 and performed SED modelings detailed. This paper is organized as follows. Section~\ref{sec:part-2} provides details on the model setup. Section~\ref{sec:part-3} presents the BZ and BP mechanisms. Section~\ref{sec:part-4} provides the details on SED modeling, and Section~\ref{sec:part-5} highlights some results from spectropolarimetric observations, while the discussion and conclusions are given in Section~\ref{sec:part-6} and Section~\ref{sec:part-7}. In this paper, we adopt the following cosmological parameters: $H_{\rm 0}=67.4$\,km~$\rm s^{-1}$~$\rm Mpc^{-1}$, $\Omega_{\rm m}=0.315$, and $\Omega_{\Lambda}=0.685$ \citep{2020A&A...641A...6P}.

\section{The Model Setup}\label{sec:part-2}
In the model, the multiwavelength emissions are assumed to originate from an isotropic and homogeneous spherical blob, moving down the relativistic jet with dimensionless bulk speed $\beta_{\rm j}=v_{\rm j}/c$ and the viewing angle $\theta_{\rm v}$. The corresponding bulk Lorentz factor and Doppler factor are given by $\Gamma_{\rm j}=(1-\beta_{\rm j}^{2})^{-1/2}$ and $\delta_{\rm D}=[\Gamma_{\rm j}(1-\beta_{\rm j}\cos\theta_{\rm v})]^{-2}$, respectively. Provided that $\Gamma_{\rm j}\sim \delta_{\rm D}$ holds, that is, $\theta_{\rm v}\simeq 1/\Gamma_{\rm j}$, this is reasonable to blazars because their jets are nearly aligned with our line of sight, $\theta_{\rm v}\lesssim 5^{\circ}$ \citep{1995PASP..107..803U}. The emitting region is permeated with tangled magnetic field $B^{'}$, and the relativistic electron population with broken power-law distribution as follows,
\begin{equation} \label{eq:1}
n_{\rm e}^{'}\left(\gamma \right) = \left\{ \begin{array}{cc}
n_{\rm 0} \gamma^{\rm -s_{\rm 1}} & \hspace{5mm}  \gamma_{\rm min}\leq \gamma \leq \gamma_{\rm br}, \\
n_{\rm 0} \gamma^{\rm -s_{\rm 2}}\gamma_{\rm br}^{(\rm s_{\rm 2}-s_{\rm 1})}  & \hspace{5mm} \gamma_{\rm br}<\gamma \leq \gamma_{\rm max}, \\
\end{array} \right.
\end{equation}
where $n_{\rm 0}$ is the normalization (cm$^{-3}$), $\gamma$ is the dimensionless energy of electron in the comoving frame, $s_{\rm 1}$ and $s_{\rm 2}$ are the spectral indices below and above the broken energy $\gamma_{\rm br}$, respectively.

The electron population is energized instantaneously via some acceleration mechanisms and injected into an extended emitting region, the so-called $\gamma$-ray emitting region, whose radius $R_{\rm b}^{'}$ can be constrained by minimum variability timescale $t_{\rm v,min}$, i.e., $R_{\rm b}^{'}\simeq c t_{\rm v,min}\delta_{\rm D}/(1+z)$, whereas its distance from the central engine can be determined by model parameter $R_{\rm H}$. After injecting, electron population inevitably suffers from energy losses through synchrotron, SSC and ERC processes. In calculations, we use approximate Compton scattering cross section as follows \citep[][]{1968PhRv..167.1159J,1985ApJ...298..128B},
\begin{equation} \label{eq:2}
\frac{d\sigma(\gamma,\nu^{'})}{d\nu_{\rm s}^{'}} = \left\{\begin{array}{cc}
\frac{3\sigma_{\rm T}}{16\gamma^{4}\nu^{'}}\Big(\frac{4\gamma^{2}\nu_{\rm s}^{'}}{\nu^{'}}-1\Big)&\frac{1}{4\gamma^{2}} \leq \frac{\nu_{\rm s}^{'}}{\nu^{'}}\leq 1 \\
\frac{3\sigma_{\rm T}}{4\gamma^{2}\nu^{'}}F_{\rm C}(q,\Gamma_{\rm e}) & 1\leq \frac{\nu_{\rm s}^{'}}{\nu^{'}}\leq \frac{4\gamma^{2}}{1+\Gamma_{\rm e}}  \\
\end{array} \right.
\end{equation}
where $\sigma_{\rm T}$ and $m_{\rm e}$ are the Thomson cross section and rest mass of electron, respectively. $\nu^{'}$ is the frequency of the incident photon, and $\nu_{\rm s}^{'}$ is the corresponding scattered frequency. Whereas $F_{\rm C}$ is the Compton scattering kernel for the isotropic electron and photon distributions, given by
\begin{eqnarray}
F_{\rm C}(q,\Gamma_{\rm e})&=& 2q\ln(q)+(1+2q)(1-q)
\nonumber\\
&+&\frac{1}{2}(1-q)\frac{(\Gamma_{\rm e}q)^{2}}{1+\Gamma_{\rm e}q}
\end{eqnarray}
where
\begin{eqnarray}
q = \frac{\nu_{\rm s}^{'}}{\Gamma_{\rm e}(\gamma m_{\rm e}c^{2}/h-\nu_{\rm s}^{'})},~~~~ \Gamma_{\rm e}=\frac{4\gamma h\nu^{'}}{m_{\rm e}c^{2}},
\end{eqnarray}
the limits to q are $0\leq q \leq 1$, h is the Planck constant.

The AD is thought to be Shakura$-$Sunyaev type \citep{1973A&A....24..337S}, which extends from $R_{\rm in}=f_{\rm dic}(3R_{\rm S})$ to $R_{\rm out}=500$\,$R_{\rm S}$ and produces a luminosity $L_{\rm d}=\eta_{\rm f}\dot{M}_{\rm acc}c^{2}$, where the accretion efficiency $\eta_{\rm f}=1/12$ is taken. Here a correction factor, $f_{\rm dic}$, is multiplied to $R_{\rm in}$ for adjusting the peak location of the AD spectrum by changing the innermost stable orbit ($R_{\rm ISO}$). The temperature profile, being a function of the radius $R$, is given by
\begin{eqnarray}
T_{\rm d}^{\rm 4}(R) &=& \frac{R_{\rm in}L_{\rm d}}{16\pi \eta_{\rm f}\sigma_{\rm SB}R^{3}}\Bigg(1-\sqrt{\frac{R_{\rm in}}{R}}\Bigg),
\end{eqnarray}
where $\sigma_{\rm SB}$ is the Stefan-Boltzmann constant. The emissions from each annulus are represented by a blackbody spectrum, and the specific intensity seen in the blob frame is
\begin{eqnarray}
I_{\rm d}^{'}(\nu^{'}) &=& 2\pi\int_{\mu_{\rm d}}^{\rm 1}\frac{I_{\rm d}^{'}(\nu^{'})d\mu}{[\Gamma_{\rm j}(1-\beta_{\rm j}\mu)]^{2}},
\end{eqnarray}
where $\mu=\cos\vartheta$ and $\vartheta$ is the angle of the incoming photons from each annulus of the AD with respect to the jet axis. The $\mu_{\rm d}$ denotes the lower limit of integration on $\mu$, and can be determined by $\mu_{\rm d}=[1+R_{\rm out}^{2}/R_{\rm z}^{2}]^{1/2}$, where $R_{\rm z}$ represents the radial distance of a point on the jet.

The radiation field of the BLR is treated by applying a spherical geometry proposed by \cite{2003APh....18..377D}, with the inner radius is determinated in terms of $L_{\rm d}$ as follows \citep{2010MNRAS.402..497G},
\begin{eqnarray} \label{eq:R-BLR-in}
R_{\rm BLR,in} &\simeq& 1.0\times 10^{17} \sqrt{\frac{L_{\rm d}}{10^{45}\,\rm erg~s^{-1}}}\,\rm ~~cm.
\end{eqnarray}
The BLR is assumed to reflect the AD emissions only into the continuum, with the specific emissivity as
\begin{eqnarray}
j(\nu;r) &=& \frac{L_{\rm d}(\nu)\xi_{\rm BLR}}{16\pi^{2}r^{2}},
\end{eqnarray}
where the specific luminosity $L_{\rm d}(\nu)$ is related to the $L_{\rm d}$ via $L_{\rm d}=\int L_{\rm d}(\nu)d\nu$, $\xi_{\rm BLR}$ is the covering factor. The specific radiative intensity at $\theta$ to the jet axis can be given by \citep{2003APh....18..377D}
\begin{eqnarray}
I(\nu;R_{\rm z},\theta) &=& \int_{\rm 0}^{\zeta_{\rm min,1}}j(\nu;r)d\zeta+\int_{\zeta_{\rm min,2}}^{\zeta_{\rm max}}j(\nu;r)d\zeta,
\end{eqnarray}
where the relation $r^{2}=R_{\rm z}^{2}-2R_{\rm z}\zeta\cos\theta+\zeta^{2}$ is satisfied between $R_{\rm z}$ and $r$. The GeV $\gamma$-rays from the emitting region will interact with BLR photons via photon-annihilation processes, $\gamma\gamma\to e^{\pm}$, and will be attenuated. The corresponding optical depth is given as
\begin{eqnarray}\label{eq:tau_att}
\tau_{\rm BLR}[E_{\gamma}(\rm GeV)] &=& \int\int\int \sigma(x)n_{\rm BLR}(R_{\rm z},\nu,\Omega)
\nonumber\\
&\times& (1-\cos\theta)d\nu d\Omega dR_{\rm z},
\end{eqnarray}
where $n_{\rm BLR}(R_{\rm z},\nu,\Omega)$ is the number density of diffuse photon per unit frequency per unit solid angle at position $R_{\rm z}$, the cross section is \citep{1967PhRv..155.1404G}
\begin{eqnarray}
\sigma(x) &=& \frac{3\sigma_{\rm T}}{16}(1-x^{2})\biggr[(3-x^{4})\ln\frac{1+x}{1-x}
\nonumber\\
&-&2x(2-x^{2})\biggr],
\end{eqnarray}
where $x$ is given by
\begin{eqnarray}
x \equiv \sqrt{1-\frac{2(m_{\rm e}c^{2})^{2}}{h^{2}\nu_{\gamma}\nu(1-\cos\theta)}},
\end{eqnarray}
and the frequency of $\gamma$-ray photon $\nu_{\gamma}$ is related to its energy $E_{\gamma}$ via $E_{\gamma}\simeq 4.14\times 10^{-24}\nu_{\gamma}$/(1+z).

In addition, the DT structure is treated simply as a spherically thin shell. Inside the DT, the comoving specific intensity can be estimated as
\begin{eqnarray}
I_{\rm DT}'(\nu') &=& \frac{\xi_{\rm DT}L_{\rm d}h \Gamma_{\rm j}^{2}}{4R_{\rm DT}^{2}c^{2}\sigma_{\rm SB}} \frac{\nu'^{3}}{{T_{\rm DT}^{'4}}}\frac{1}{\exp(h\nu'/k_{\rm B}T_{\rm DT}^{'})-1}
\nonumber\\
&\times& \Big(\frac{2}{3}\beta_{\rm j}^{2}-2\beta_{\rm j}+2\Big),
\label{I_DT_prime}
\end{eqnarray}
where $\xi_{\rm DT}$ is the covering factor, $k_{\rm B}$ is the Boltzmann constant. The size of the DT is determined by \citep{2011ApJ...732..116M}
\begin{eqnarray}
R_{\rm DT} &\simeq& 3.5\times 10^{18} \sqrt{\frac{L_{\rm d}}{10^{45}\,\rm erg~s^{-1}}}\Bigg(\frac{T_{\rm sub}}{1000\,\rm K}\Bigg)^{-2.6}\,\rm cm,
\label{DT_radius}
\end{eqnarray}
where $T_{\rm sub}$ is the sublimation temperature of dust grains and is fixed at 1000\,K \citep[][]{2008ApJ...685..160N,2011A&A...525L...8C,2017A&A...603A..29A}, while $T_{\rm DT}'$ is the DT temperature in the comoving frame,
\begin{eqnarray}
T_{\rm DT}^{'} &=& 100\Gamma_{\rm j}(1-\beta_{\rm j}\cos\alpha)\,\rm K\sim 100\Gamma_{\rm j}\,\rm K.
\end{eqnarray}
Here a pre-factor of 100 is used, which significantly differs from 370 given in \cite{2009MNRAS.397..985G}, instead it is consistent with the one adopted by \cite{2012ApJ...760...69N}. Whereas $\alpha$ is the angle between the incoming photon and jet axis. In the model, we also consider the contribution from the corona to X-rays, and assume that the emitting photon frequencies extend from $5.0\times 10^{15}$\,Hz to $1.0\times 10^{21}$\,Hz, with the total luminosity $L_{\rm Corona}=\xi_{\rm Corona}L_{\rm d}$, whose spectrum is assumed to follow a cutoff power law, i.e., $L_{\rm X}(\nu)\sim \nu^{-\alpha_{\rm x}}\exp(\nu/\nu_{\rm c})$, where $\alpha_{\rm x}=1.0$ and the cutoff energy $E_{\rm c}=h\nu_{\rm c}\sim 150$\,keV are adopted \citep{2009MNRAS.397..985G}.

\section{Two Energized Mechanisms To The Jet}\label{sec:part-3}
The accretion of material onto a highly spinning BH is an ubiquitous phenomenon in blazar, a large fraction of accretion power will be extracted through strongly twisting magnetic field to accelerate the plasma in the bipolar jets at relativistic speed. Up to date, two kinds of launching mechanism have been proposed to feed the jet, i.e., BZ and BP mechanisms \citep[][]{1977MNRAS.179..433B,1982MNRAS.199..883B}. In the following, we present in detail the main frameworks used to calculate the maximal jet power to PKS\,1510$-$089.

\subsection{The BZ Mechanism}
\cite{1977MNRAS.179..433B} has shown that if a highly spinning BH is threaded by large-scale magnetic field, the spin energy of the BH will be efficiently extracted and the obtained maximal power of the jet can be given by \citep[][]{2010ApJ...711...50T,2011MNRAS.418L..79T,2016ApJ...828...13I}
\begin{eqnarray}\label{eq:BZ}
P_{\rm BZ} &=& 4.0\times 10^{-3}\frac{1}{c}\Omega_{\rm H}^{2}\Phi_{\rm BH}^{2}f(\Omega_{\rm H}),
\end{eqnarray}
where $\Omega_{\rm H}=ac/2r_{\rm H}$ is the angular frequency of the BH horizon, $r_{\rm H}=R_{\rm g}(1+\sqrt{1-a^{2}})$ is the horizon radius, where $R_{\rm g}=GM_{\rm BH}/c^{2}$ is the gravitational radius. Moreover, $\Phi_{\rm BH}\simeq \phi_{\rm BH}\sqrt{\dot{M}_{\rm acc}R_{\rm g}^{2}c}\approx \phi_{\rm BH}\sqrt{L_{\rm d}R_{\rm g}^{2}/\eta_{\rm f}c}$ is the net magnetic flux in the central region, and the dimensionless magnetic flux $\phi_{\rm BH}$ takes typical value of 50 \citep{2012MNRAS.423.3083M}. By defining $x_{\alpha}\equiv R_{\rm g}\Omega_{\rm H}/c$, then $f(x_{\alpha})=1+1.38x_{\alpha}^{2}-9.2x_{\alpha}^{4}$.

\subsection{The BP Mechanism}
The maximal jet power that can be extracted by large-scale magnetic fields threading the AD is \citep{2003ApJ...599..147C}
\begin{eqnarray}\label{eq:BP}
P_{\rm BP} &=& 4\pi \int\frac{B_{\rm pd}^{2}(R)}{4\pi}R^{2}\Omega(R)dR,
\end{eqnarray}
where the toroidal and poloidal components of magnetic field have been assumed approximately to be equal, $B_{\rm \phi d}\approx B_{\rm pd}$. The poloidal magnetic fields are related to the one produced by dynamo processes in the following form \citep[][]{1989MNRAS.238..897L,1999ApJ...512..100L,2002MNRAS.332..999C}
\begin{eqnarray}
B_{\rm pd}(R) &\simeq& \frac{H}{R}B_{\rm dynamo}(R)
\nonumber\\
&\simeq& 26.345\dot{m}_{\rm acc}r^{-1}\mathcal{C}_{2}(R)B_{\rm dynamo}(R),
\end{eqnarray}
where $H/R$ stands for scale-height of the AD, $\dot{m}_{\rm acc}\equiv \dot{M}_{\rm acc}/\dot{M}_{\rm Edd}\simeq L_{\rm d}/L_{\rm Edd}$ is the Eddington ratio, $r=R/R_{\rm g}$ is the dimensionless radius, $\mathcal{C}_{2}$ is the relativistic correction factor \citep{1973blho.conf..343N}. For a relativistic AD, the magnetic field $B_{\rm dynamo}$ has given as \citep{2002MNRAS.332..999C}
\begin{eqnarray}
B_{\rm dynamo}(R) &=& 3.56\times 10^{8}r^{-3/4}m_{\rm BH}^{-1/2}\mathcal{A}^{-1}\mathcal{B}\mathcal{E}^{1/2},
\end{eqnarray}
where $m_{\rm BH}=M_{\rm BH}/M_{\odot}$ is the dimensionless mass of the BH. Defining the varible $x\equiv \sqrt{r}$, the relativistic correction factors $\mathcal{A}$, $\mathcal{B}$, $\mathcal{E}$ can be writtten as \citep[][]{1973blho.conf..343N,1974ApJ...191..499P}
\begin{eqnarray}
\mathcal{A} &=& 1+a^{2}x^{-4}+2a^{2}x^{-6},
\nonumber\\
\mathcal{B} &=& 1+ax^{-3},
\nonumber\\
\mathcal{E} &=& 1+4a^{2}x^{-4}-4a^{2}x^{-6}+3a^{4}x^{-8}.
\end{eqnarray}

Considering a relativistic AD, the Keplerian angular velocity is \citep{2008bhad.book.....K}
\begin{eqnarray}
\Omega(R) &=& \sqrt{\frac{GM_{\rm BH}}{R^{3}}}\Bigg[1+\frac{a}{r^{3/2}}\Bigg]^{-1},
\end{eqnarray}
where $a$ is the spin of the BH.

\begin{table*}
	\centering
	\tablenum{1}
	\caption{ Details of the energy coverage to the studied 12 SEDs, the
last column gives the reference related to the SED data sets: (1) \cite{2010{\natexlab{b}}ApJ...721.1425A}; (2) \cite{2012ApJ...760...69N}; (3) \cite{2017A&A...603A..29A}; (4) \cite{2019ApJ...883..137P}.}
	\label{tab:Table-Detail-seds}
	\begin{tabular}{lccccccccc}
		\hline
		 SED-ID & T$_{\rm start}$ & T$_{\rm stop}$ & Energy range covered & Refs \\
				     & Y/M/D (MJD) & Y/M/D (MJD)  \\
		\hline
		Ab-10(A)  & 2009/01/04 (54835) & 2009/01/27 (54858) & $5-230$\,GHz, $(0.47-1.48)\times 10^{15}$\,Hz, $0.75-54.1$\,keV, $0.23-13$\,GeV & 1\\
		Ab-10(B)  & 2009/03/10 (54900) & 2009/04/09 (54930) & $5-43$\,GHz, $(0.18-1.48)\times 10^{15}$\,Hz, $0.51-54.1$\,keV, $0.23-13$\,GeV  & 1\\
		Ab-10(C)  & 2009/04/15 (54936) & 2009/05/14 (54965) & $5-43$\,GHz, $(0.47-1.48)\times 10^{15}$\,Hz, $0.51-54.1$\,keV, $0.23-13$\,GeV  & 1\\
		Na-12(A)  & 2011/07/24 (55766) & 2011/07/29 (55769) & $37-4280$\,GHz, $(0.14-1.48)\times 10^{15}$\,Hz, $0.4-10$\,keV, $0.14-171.1$\,GeV  & 2\\		
		Na-12(B)  & 2011/08/17 (55790) & 2011/08/18 (55791) & $22.5-4280$\,GHz, $(0.68-1.48)\times 10^{15}$\,Hz, $0.3-10$\,keV, $0.18-17.8$\,GeV & 2\\
		Pr-19(A)  & 2015/03/19 (57100) & 2015/05/08 (57150) & $(0.55-1.55)\times 10^{15}$\,Hz, $1.04-6.03$\,keV, $0.14-37$\,GeV  & 4\\
		Pr-19(B)  & 2015/05/08 (57150) & 2015/06/07 (57180) & $(0.55-1.55)\times 10^{15}$\,Hz, $1.0-7.5$\,keV, $0.14-37$\,GeV & 4\\
		Ah-17(A)  & 2015/05/18 (57160) & 2015/05/19 (57161) & $37$\,GHz, $(0.47-1.46)\times 10^{15}$\,Hz, $0.48-4.2$\,keV, $0.14-130.2$\,GeV  & 3\\
		Ah-17(B)  & 2015/05/22 (57164) & 2015/05/24 (57166) & $37$\,GHz, $(0.14-1.45)\times 10^{15}$\,Hz, $0.43-3.93\times10^{3}$\,keV, $0.14-123.1$\,GeV  & 3\\		
		Pr-19(Q2)  & 2015/06/07 (57180) & 2015/07/05 (57208) & $(0.55-1.55)\times 10^{15}$\,Hz, $1.02-6.12$\,keV, $0.14-3.32$\,GeV  & 4\\			
		Pr-19(C)  & 2015/07/05 (57208) & 2015/08/01 (57235) & $(0.54-1.55)\times 10^{15}$\,Hz, $1.0-7.02$\,keV, $0.14-16.5$\,GeV   & 4\\
		Pr-19(D)  & 2015/08/01 (57235) & 2015/08/26 (57260) & $(0.54-1.55)\times 10^{15}$\,Hz, $1.03-5.99$\,keV, $0.14-16.8$\,GeV  & 4\\		
		\hline
	\end{tabular}
\end{table*}

\begin{figure*}
\hbox{
\includegraphics[width=0.5\textwidth]{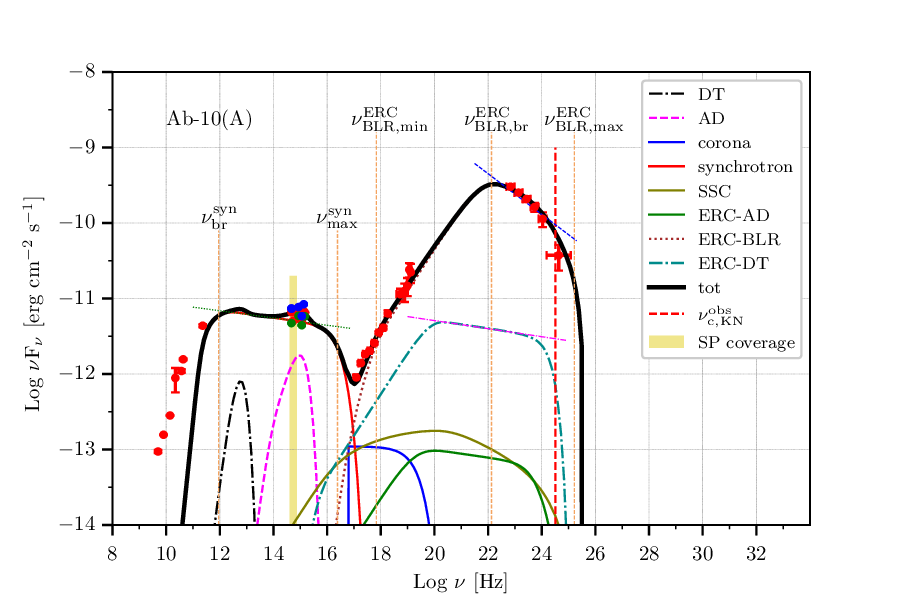}
\includegraphics[width=0.5\textwidth]{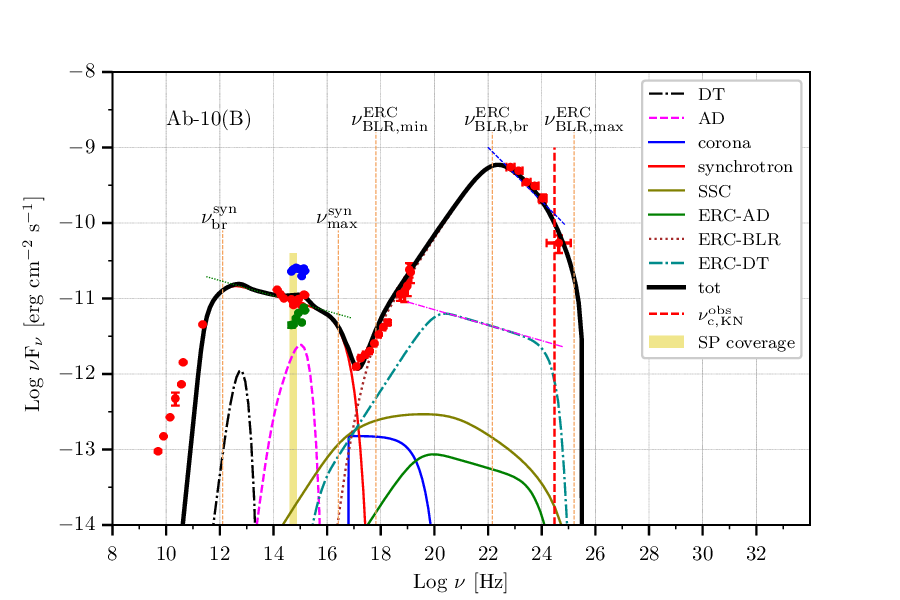}
}
\hbox{
\includegraphics[width=0.5\textwidth]{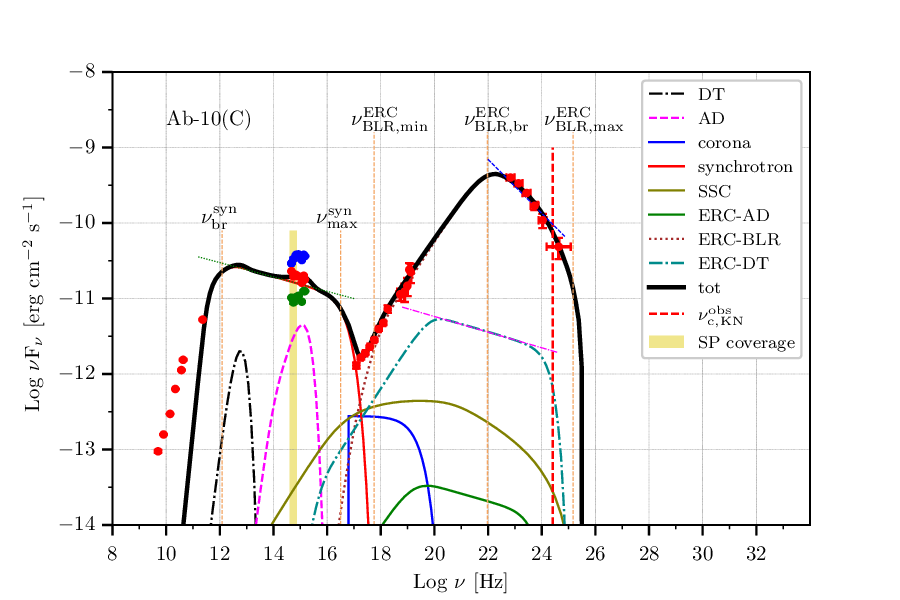}
}
\caption{SED modelings of the quasi-simultaneous data sets for the low-, high- and mediate-active states, labeled respectively as Ab-10(A), Ab-10(B) and Ab-10(C). These states occurred during the high-activity period from 2008 September to 2009 June. The black dot-dashed, magenta dashed, darkcyan dot-dashed, brown dotted lines indicate emissions from the DT, AD, ERC-DT, ERC-BLR, respectively. The red, olive, green, blue solid lines correspond to synchrotron, SSC, ERC-AD, and the corona emissions, whereas the thick solid line represents the sum of all emission components. The khaki strip around $10^{15}$\,Hz designates the frequency range covered by the optical spectropolarimetric (SP) observation; moreover, the vertically sandybrown lines mark the position of five characteristic frequencies, and the vertically red line indicates the critical frequency $\nu_{\rm c,KN}^{\rm obs}$ between the Thomson and KN regimes in the ERC-BLR process. The green dotted, magenta dashdot and blue dashed lines represent the linear fittings to the synchrotron, ERC-DT and ERC-BLR components, respectively.}
\label{fig:SED-1}
\end{figure*}

\begin{figure*}
\hbox{
\includegraphics[width=0.5\textwidth]{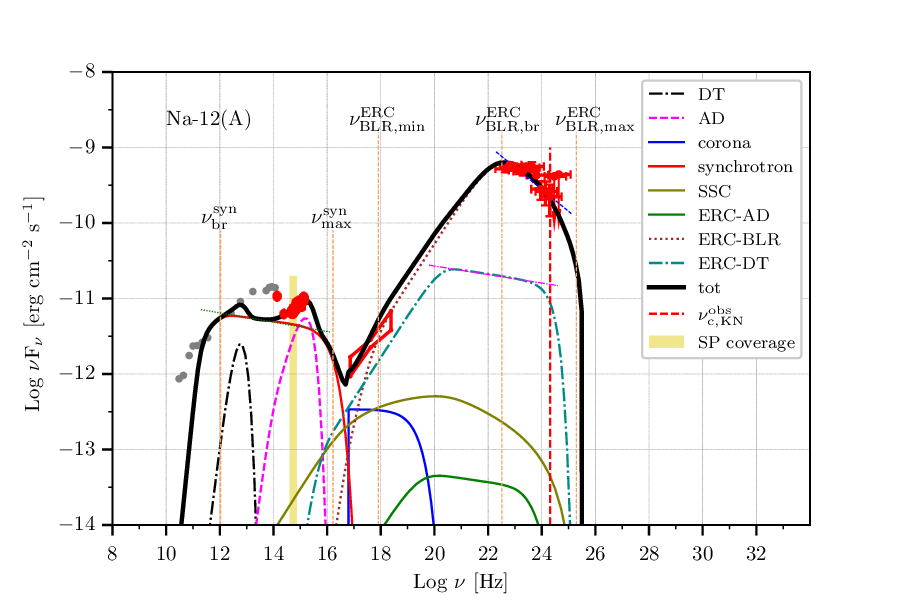}
\includegraphics[width=0.5\textwidth]{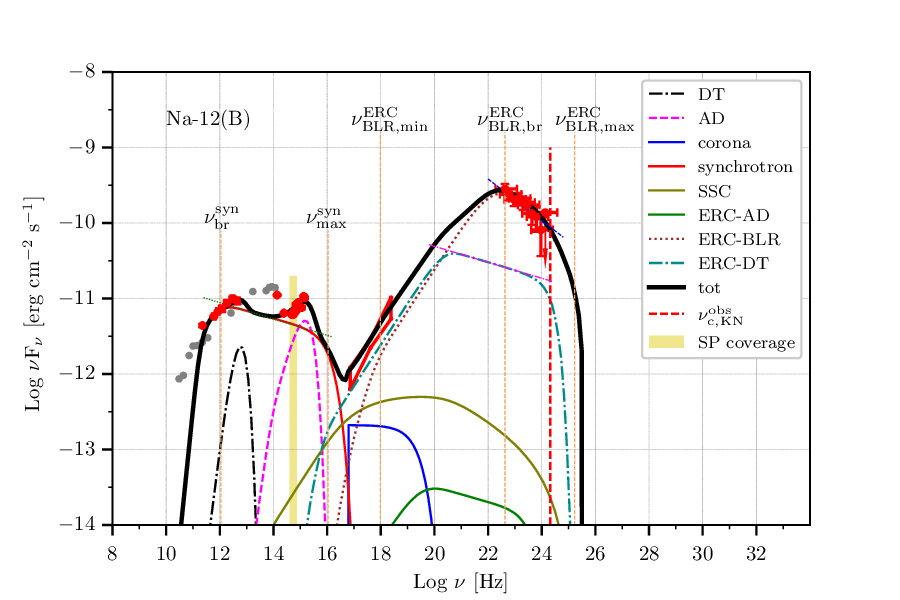}
}
\caption{SED modelings of the quasi-simultaneous data sets for high- and low-active states, labelled respectively as Na-12(A) and Na-12(B), the SED data are collected during a $\gamma$-ray flare and the \herschel\ observation, respectively. At radio/optical band in both panels, the \planck\ and \spizer\ data are added and indicated in gray as a reference to the current flux level \citep{2011A&A...536A..15P,2011ApJ...732..116M}. The black dot-dashed, magenta dashed, darkcyan dot-dashed, brown dotted lines indicate DT, AD, ERC-DT, ERC-BLR emissions, respectively. The red, olive, green, blue solid lines correspond to synchrotron, SSC, ERC-AD, and the corona emissions, while the thick solid line represents the sum of all the emission components. Moreover, the khaki strip around $10^{15}$\,Hz designates the frequency range covered by the optical spectropolarimetric (SP) observation, and the vertically sandybrown lines mark the position of five characteristic frequencies, while the vertically red line indicates the critical frequency $\nu_{\rm c,KN}^{\rm obs}$ between the Thomson and KN regimes in the ERC-BLR process. The green dotted, magenta dashdot and blue dashed lines are the linear fittings to the synchrotron, ERC-DT and ERC-BLR components, respectively.}
\label{fig:SED-2}
\end{figure*}

\begin{figure*}
\hbox{
\includegraphics[width=0.5\textwidth]{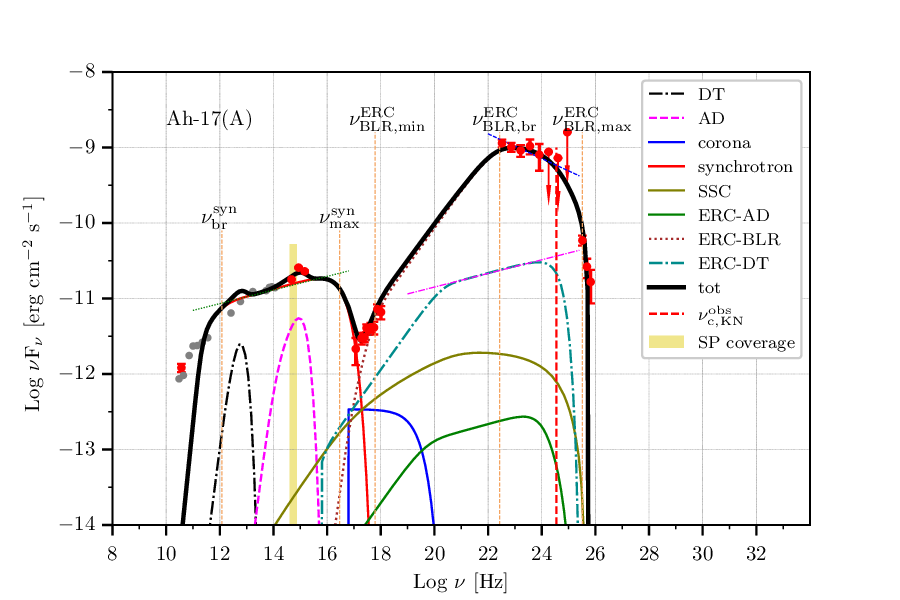}
\includegraphics[width=0.5\textwidth]{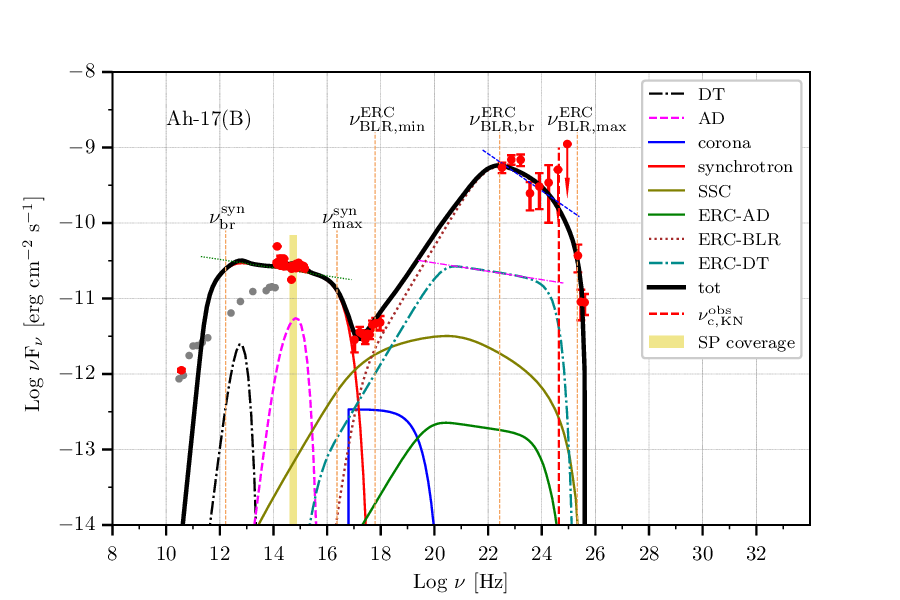}
}
\caption{SED modelings of the quasi-simultaneous data sets for high- and low-active states, labelled respectively as Ah-17(A) and Ah-17(B), which occurred during a long, high $\gamma$-ray state in May 2015. At radio/optical band in both panels, the \planck\ and \spizer\ data are added and indicated in gray as a reference to the current flux level \citep{2011A&A...536A..15P,2011ApJ...732..116M}. The black dot-dashed, magenta dashed, darkcyan dot-dashed, brown dotted lines indicate DT, AD, ERC-DT, ERC-BLR emissions, respectively. The red, olive, green, blue solid lines correspond to synchrotron, SSC, ERC-AD, and the corona emissions, whereas the thick solid line represents the sum of all the emission components. The khaki strip around $10^{15}$\,Hz designates the frequency range covered by the optical spectropolarimetric (SP) observation; moreover, the vertically sandybrown lines mark the position of five characteristic frequencies, and the vertically red line indicates the critical frequency $\nu_{\rm c,KN}^{\rm obs}$ between the Thomson and KN regimes in the ERC-BLR process. The green dotted, magenta dashdot, and blue dashed lines are the linear fittings to the synchrotron, ERC-DT and ERC-BLR components, respectively.}
\label{fig:SED-3}
\end{figure*}

\begin{figure*}
\hbox{
\includegraphics[width=0.5\textwidth]{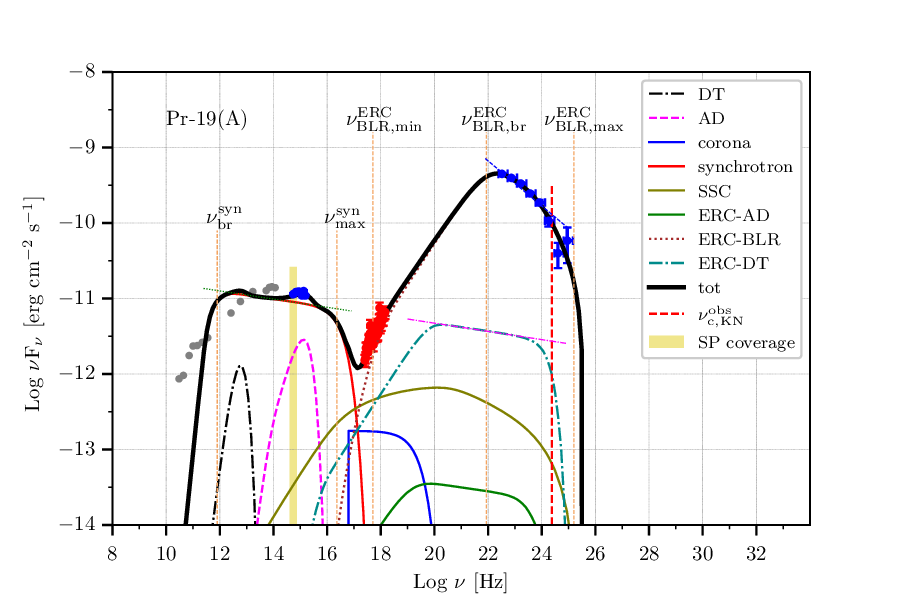}
\includegraphics[width=0.5\textwidth]{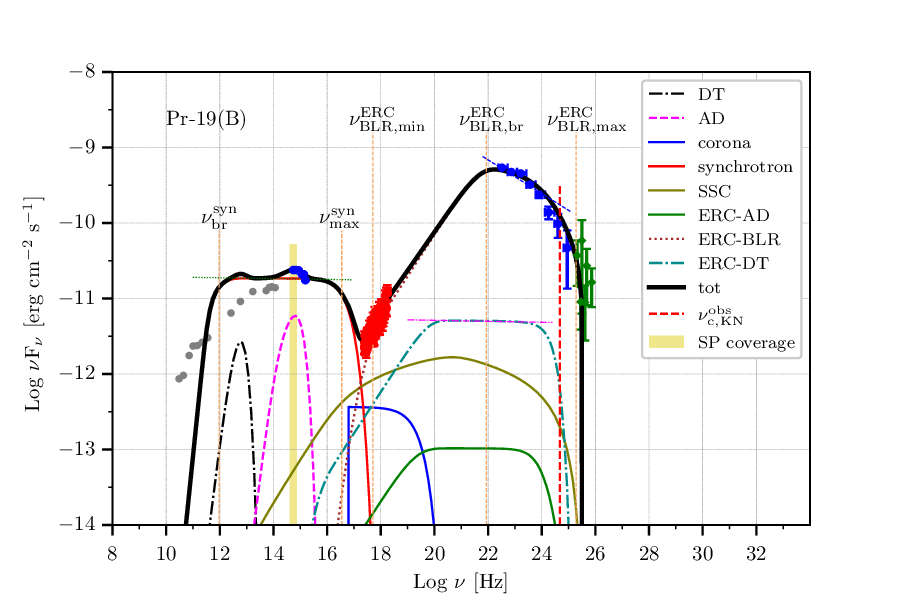}
}
\hbox{
\includegraphics[width=0.5\textwidth]{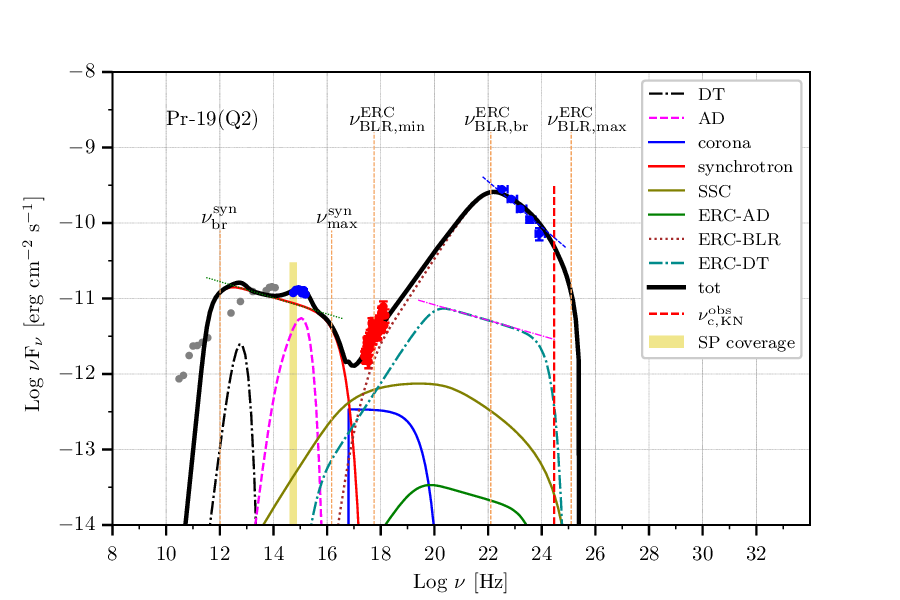}
\includegraphics[width=0.5\textwidth]{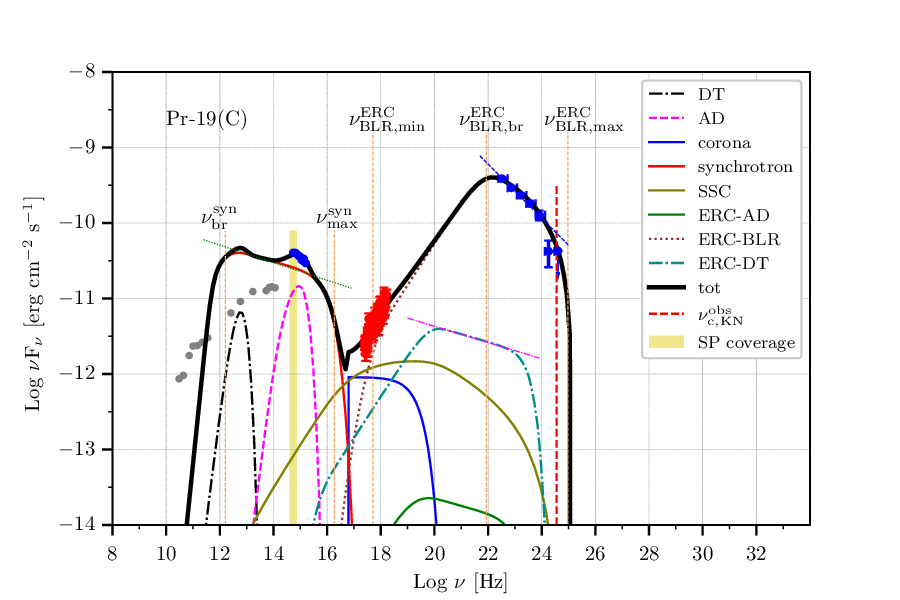}
}
\hbox{
\includegraphics[width=0.5\textwidth]{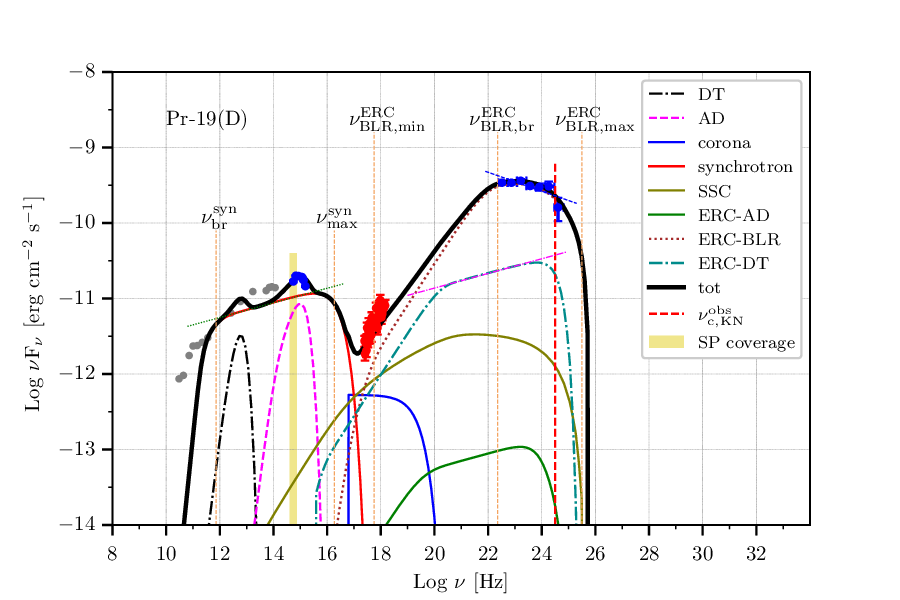}
}
\caption{SED modelings of the quasi-simultaneous data for four flaring episodes and a quiescent state in 2015, labeled respectively as Pr-19(A), Pr-19(B), Pr-19(Q2), Pr-19(C) and Pr-19(D). In Pr-19(B), the contemporaneous MAGIC observations, presented in \cite{2017A&A...603A..29A}, are shown in green solid diamonds at the high-energy tail. At radio/optical band in all the panels, the \planck\ and \spizer\ data are added and indicated in gray as a reference to the current flux level \citep{2011A&A...536A..15P,2011ApJ...732..116M}. The black dot-dashed, magenta dashed, darkcyan dot-dashed, brown dotted lines indicate DT, AD, ERC-DT, ERC-BLR emissions, respectively. The red, olive, green, blue solid lines correspond to synchrotron, SSC, ERC-AD, and the corona emissions. The thick solid line represents the sum of all emission components. The khaki strip around $10^{15}$\,Hz designates the frequency range covered by the optical spectropolarimetric (SP) observation; moreover, the vertically sandybrown lines mark the position of five characteristic frequencies, and the vertically red line indicates the critical frequency $\nu_{\rm c,KN}^{\rm obs}$ between the Thomson and KN regimes in the ERC-BLR process. The green dotted, magenta dashdot, and blue dashed lines are the linear fittings to the synchrotron, ERC-DT and ERC-BLR components, respectively.}
\label{fig:SED-4}
\end{figure*}

\section{Multiwavelength Modeling} \label{sec:part-4}
We performed SED modeling on 12 multiwavelength data sets, including: three SEDs labeled as Ab-10(A), Ab-10(B), Ab-10(C), taken from \cite{2010{\natexlab{b}}ApJ...721.1425A}; two SEDs labeled as Na-12(A), Na-12(B), taken from \cite{2012ApJ...760...69N}; two SEDs labeled as Ah-17(A), Ah-17(B), taken from \cite{2017A&A...603A..29A}. The remaining five SEDs are taken from \cite{2019ApJ...883..137P} and labeled as Pr-19(A), Pr-19(B), Pr-19(Q2), Pr-19(C) and Pr-19(D), respectively. The details regarding the energy coverage of these SEDs are provided in Table \ref{tab:Table-Detail-seds}, which lists the SED identity (SED-ID), the start and stop times of the observations ($T_{\rm start}$, $T_{\rm stop}$), and the corresponding references.

In the numerical implementations, we utilized a portion of C codes encapsulated in `JetSeT', an open-source C/Python framework designed to reproduce radiative and accelerative processes acting in relativistic jets of the extragalactic and galactic objects, allowing to fit the numerical models to observed data sets, this code is publicly available on GitHub\footnote{\url{https://jetset.readthedocs.io/en/latest/}}. Based on these codes, we incorporated some new considerations mentioned in Sections \ref{sec:part-2} and \ref{sec:part-3}, primarily related to the following aspects: first, we treat $R_{\rm in}$ as an adjustable quantity (via changing the parameter $f_{\rm dic}$) to well pin down the peak location of the big blue bump (BBB) spectrum for different SEDs. Secondly, $R_{\rm BLR,in}$ is related to $L_{\rm d}$ by Equation (\ref{eq:R-BLR-in}). Thirdly, the DT structure is modeled as a thin spherical shell, with its radiative strength and geometrical size determined by Equations (\ref{I_DT_prime}) and (\ref{DT_radius}), respectively. In addition, we also included the calculations on jet power provided by the BZ and BP processes, as well as the $\gamma$-ray optical depth, Equation (\ref{eq:tau_att}), along the path toward the observer.

The intensely external radiation fields cause high-energy electrons to cool efficiently within a timescale shorter than the light-crossing time ($\sim R_{\rm b}^{'}/c$). Under such conditions, the overall electron distribution can be approximated by Equation (\ref{eq:1}). Consequently, the mutiwavelength SEDs are thought to be a series of snapshots contributed by these electrons. The low-energy hump of the SED originates from synchrotron emissions, while the high-energy emissions are dominated by the Compton scattering of external photon fields from the BLR (ERC-BLR) and DT (ERC-DT). These external radiation fields significantly complicate the model description compared to a one-zone homogeneous SSC model, increasing the number of model parameters to eighteen. Among these parameters, $\Gamma_{\rm j}$ or $\delta_{\rm D}$, $B^{'}$, $R_{\rm b}^{'}$ and $R_{\rm H}$ are used to constrain the environment of the $\gamma$-ray emitting region. Six parameters, $\gamma_{\rm min}$, $\gamma_{\rm max}$, $\gamma_{\rm br}$, $n_{\rm 0}$, $s_{1}$ and $s_{2}$, specify the electron distribution. As for another eight parameters, three are covering factors for the BLR, DT and corona ($\xi_{\rm BLR}$, $\xi_{\rm DT}$ and $\xi_{\rm Corona}$), one is the outer radius of the BLR ($R_{\rm BLR,out}$), two are related to the AD ($f_{\rm dic}$ and $L_{\rm d}$), and the remaining two are the BH mass ($M_{\rm BH}$) as well as spin parameter $a$. 

To reduce the free parameters, some of them are properly fixed. Specifically, $\xi_{\rm BLR}$ and $\xi_{\rm DT}$, as commonly used, are fixed at 0.1 and 0.6, respectively \citep{2009MNRAS.392L..40T,2009ApJ...704...38S,2014{\natexlab{b}}A&A...567A..41A,2015ApJ...815L..23A,2017A&A...603A..29A,2018A&A...619A.159M,2022A&A...660A..18N}. Similar to the choices made by \cite{2016ApJ...821..102B} and \cite{2022MNRAS.513.4645S}, $R_{\rm BLR,out}$ is taken as $2R_{\rm BLR,in}$. The minimum Lorentz factor of the electrons, $\gamma_{\rm min}$, is not well constrained, but some values close to 1 are often adopted \citep{2010{\natexlab{b}}ApJ...721.1425A,2012ApJ...760...69N,2017A&A...603A..29A,2022MNRAS.515.1655B,2023ApJ...952L..38A}; therefore, we fix it at 1. This choice extends the Compton spectrum to lower energy and will contribute to a better fit to the X-ray data sets. On the other hand, setting $\gamma_{\rm min}$ to 1 significantly increases the resulting jet power, as shown by Equations (\ref{eq:Pe}) and (\ref{eq:Ne}) in Subsection \ref{sec:jet-power}. In this sense, our model provides a maximal estimate to the conversion efficiency from accreted matter to the jet power. Regarding the BH spin, past studies have shown that radio-loud quasars generally have higher radiative efﬁciencies, indicating that these sources have higher BH spins in contrast with radio-quiet populations \citep{2012JPhCS.355a2016M,2017ApJ...849....4S}. As far as PKS\,1510$-$089 is considered, the presence of a highly relativistic jet suggests that there is a rapidly spinning BH at the galaxy's center. Under this condition, the BH spin parameter $a$ is fixed at a high value of 0.95 \citep{2001Sci...291...84M,2002MNRAS.332..999C,2003ApJ...599..147C}. At last, we use the same value to $R_{\rm b}^{'}$ as the paper that originally presented the multiwavelength SEDs. With these constraints, only twelve parameters remain adjustable in the model.

\begin{deluxetable*}{cccccccccccccccc}
\tablenum{2}
\tablecaption{The best-fit parameters used to reproduce the multiwavelength SEDs\label{tab:mod-parameters}}
\tablewidth{0pt}
\tabletypesize{\footnotesize}
\tablehead{
 \colhead{Parameters} &  \colhead{Ab-10} &  \colhead{Ab-10} &  \colhead{Ab-10} &  \colhead{Na-12} &  \colhead{Na-12} & \colhead{Ah-17}  & \colhead{Ah-17} & \colhead{Pr-19}& \colhead{Pr-19} & \colhead{Pr-19} & \colhead{Pr-19} & \colhead{Pr-19} \\
\colhead{}  & \colhead{(A)}  & \colhead{(B)} &   \colhead{(C)} &    \colhead{(A)}  & \colhead{(B)} & \colhead{(A)} &  \colhead{(B)} & \colhead{(A)} &   \colhead{(B)} &   \colhead{(Q2)} &   \colhead{(C)} &   \colhead{(D)}
}
\startdata
$R_{\rm b}^{'}$($\times 10^{16}\,\rm cm$)	&  3.2   &  3.2  &   3.2    &  3.0   &  3.0   &  2.8  &  2.8   &  2.1   &  2.1 & 2.1 &  2.1  &  2.1 \\
$\delta_{\rm D}$    &  21   &  22  &   22    &  23   &  25   &  23  &  23   &  21   &  21 & 22 &  21  &  22 \\
$B^{'}$(G)    &  0.80   &  1.1  &  1.6     &  0.37   &  0.36   &  0.57  &  0.79   &  1.1   &  1.3 & 1.0 & 2.2 &  0.40 \\
$\gamma_{\rm max}(\times 10^{4})$    &  2.3   &  2.3  &   2.1    &   2.3  &  2.1   &  3.3  &  2.5   &  2.2   & 2.2   & 1.8  &  1.4  & 3.2 \\
$\gamma_{\rm br}$    &  140   &  160  &   130    &  200   &  210   &  210  &  210   &  130   & 130  & 150 &  130  &  200 \\
$n_{\rm 0}(\times 10^{3}\, \rm cm^{-3})$    &  1.8   &  1.6  &  1.7     &  4.3   &  4.2   &  4.4  &  3.6   &  6.4   & 6.8   &  7.1  & 5.9   & 11.0 \\
$s_{\rm 1}$    &  1.9   &  1.9  &   1.9    &  1.9   &  1.9   &  2.0  &  1.8   &  1.9   &  1.9   & 1.9  & 1.9   & 1.9 \\
$s_{\rm 2}$    &  3.1   &  3.2  &   3.2    &  3.1   &  3.2   &  2.8  &  3.1   &  3.1   & 3.0  & 3.2  &  3.2  & 2.8  \\
$\xi_{\rm Corona}$    &  0.06   & 0.06  &  0.06     &  0.06   &  0.06  &  0.06  &  0.06   &  0.06   &  0.06 & 0.06  & 0.06   &  0.06 \\
$f_{\rm dic}$ & 1.5 & 1.5 & 1.5 & 1.0 & 1.0 & 3.0 & 5.0 & 1.0 & 6.0 & 2.0 & 5.0 & 3.0 \\
$L_{\rm d}(\times 10^{45}\,\rm erg~s^{-1})$    & 0.87    &  1.2  &  2.2     &  2.7   &  2.5   &  2.7 & 2.7    &   1.4 & 2.9  & 2.7  &  7.2  & 4.2 \\
$R_{\rm H}(\times 10^{17}\,\rm cm)$    &  1.92   & 2.2   &  3.0     &  3.7   &  4.3   & 3.4   &  3.8   &  2.37   &   3.41 & 3.6  &  5.37  &  4.9 \\
\enddata
\end{deluxetable*} 

\begin{figure*}
\hbox{
\includegraphics[width=0.5\textwidth]{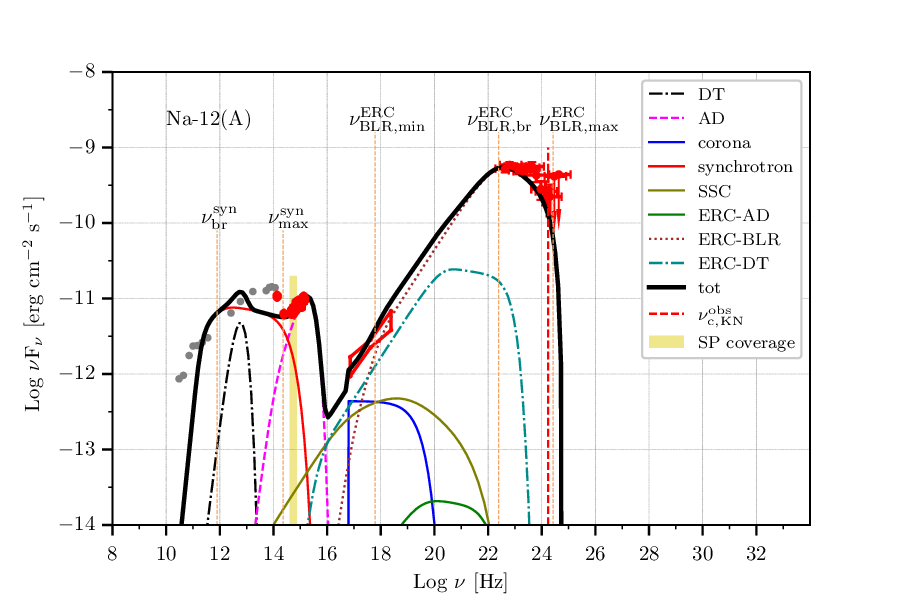}
\includegraphics[width=0.5\textwidth]{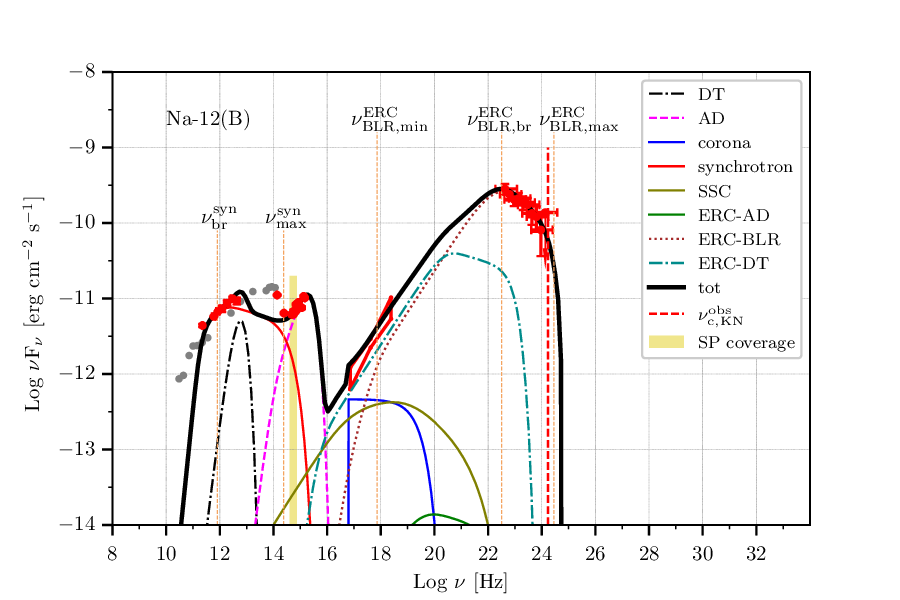}
}
\caption{Similar to Figure \ref{fig:SED-2}, but we re-perform SED modeling for Na-12(A) and Na-12(B) by changing the part of model parameters. Specifically, Na-12(A): $\gamma_{\rm br}=2.0\times 10^{2}$, $\gamma_{\rm max}=3.4\times 10^{3}$, $\tau_{\rm Corona}=0.04$, $L_{\rm d}=5.2\times 10^{45}$\,erg s$^{-1}$, $R_{\rm H}=5.2\times 10^{17}$\,cm. Whereas for Na-12(B): $\gamma_{\rm br}=2.1\times 10^{2}$, $\gamma_{\rm max}=3.6\times 10^{3}$, $\xi_{\rm Corona}=0.04$, $L_{\rm d}=5.5\times 10^{45}$\,erg s$^{-1}$, $R_{\rm H}=6.3\times 10^{17}$\,cm.}
\label{fig:SED-55}
\end{figure*}

The notable flux excess at BBB is directly related to $M_{\rm BH}$ and $L_{\rm d}$, and subsequently to $R_{\rm BLR,in}$. We began with our SED modeling efforts by fitting the optical/UV data using a multi-temperature blackbody spectrum with the peak frequency at $\nu_{\rm max}\simeq 1.39\times 5.88\times 10^{10}T_{\rm in}$\,(Hz), where $T_{\rm in}$ is the temperature of the AD at $R_{\rm ISO}$. The factor 1.39 is included because $\nu_{\rm max}$ represents the peak in SED distribution. By fitting the data set around the BBB, both the amplitude ($L_{\rm d}$) and the peak location ($\nu_{\rm max}$ or $R_{\rm in}$) of the BBB component can be determined. Then, we perform SED modeling to other multiwavelength data. It is here emphasis that in the model the radio spectrum is not reproduced, since the radio emissions $\lesssim 10^{12}$\,Hz are strongly self-absorbed, and generally thought to be produced by some more extended jet components. At last, through the visual inspection, further fine tuning the model parameters, including $L_{\rm d}$, to achieve a better agreement between the model curve and data points. The resulting SED modelings are shown in Figures \ref{fig:SED-1} $-$ \ref{fig:SED-4}, and the best-fitting model parameters are listed in Table \ref{tab:mod-parameters}. We note that there is an infrared data point that significantly deviates the fitting curve in the SEDs of Na-12(A) and Na-12(B). This data point is obtained by the photometry using the K filter by the \swift-UVOT telescope. As shown by \cite{2012ApJ...760...69N}, the K band shows a high flux level and relatively stable variability throughout whole observational period. This suggests that the flux in K band may have some contributions from other emitting components that beyond the $\gamma$-ray emitting region. In the SED modeling, in order to ensure that the resulting fitting curve passes through the data points at the low-energy end of the X-ray spectrum, we use a value of $\gamma_{\rm max}$ as large as possible. This extends the $\gamma$-ray spectrum to higher energy. Nonetheless, it has been shown by Pr-19(B) that the model still cannot provide a good fit to the contemporaneous MAGIC data, shown by green solid diamonds at the high-energy tail. This implies that the MAGIC flux may originate from some emitting regions compatible with Ah-17(A) and Ah-17(B), but significantly different from the one required for Pr-19(B). (Please note that the MAGIC data points presented in Pr-19(B) are in fact the superposition of the SEDs of Ah-17(A) and Ah-17(B), respectively, as presented in \cite{2019ApJ...883..137P}, and taken from \cite{2017A&A...603A..29A}). 

\begin{table*}
	\centering
	\tablenum{3}
	\caption{Two characteristic times as well as three important quantities used to analyze the polarimetric properties at optical band to PKS\,1510$-$089.}
	\label{tab:table-uqQ}
	\begin{tabular}{lccccccccc}
		\hline
		SED-ID & $T_{\rm \gamma-ray}^{\rm peak}$ & $T_{\rm optical}^{\rm obs}$ & $q_{\rm m}$ & $u_{\rm m}$ & $Q(\%)$  \\
		  (1) & (2) & (3) & (4) &  (5)   & (6)   \\
			
		\hline 
		Ab-10(A)&$54846.76~(2009/01/15)$  &$54846.76~(2009/01/15)$&   N &   N  &   N \\
		Ab-10(B)&$54916.58~(2009/03/26)$  &$54917.42~(2009/03/27)$&   -0.1151$\pm$0.0145&    0.0446$\pm$0.0136 &  12.34$\pm$0.06 \\
		Ab-10(C)&$54946.50~(2009/04/25)$  &$54946.37~(2009/04/25)$&   0.0036$\pm$0.0121 &    0.0749$\pm$0.0132 &   7.50$\pm$0.06 \\
		Na-12(A)&$55767.51~(2011/07/25)$  &$55769.22~(2011/07/27)$&  -0.0656$\pm$0.0195 &   -0.0115$\pm$0.0169 &   6.66$\pm$0.09 \\		
		Na-12(B)&$55790.39~(2011/08/17)$  &$55790.39~(2011/08/17)$&      N   &     N     &    N   \\
		Pr-19(A)&$57125.30~(2015/04/13)$  &$57125.39~(2015/04/13)$&  0.0468$\pm$0.0194 &   0.0230$\pm$0.0184  &  5.22$\pm$0.09  \\		
		Ah-17(A)&$57160.19~(2015/05/18)$  &$57160.30~(2015/05/18)$&  -0.0038$\pm$0.0152 &  -0.0445$\pm$0.0153  &   4.46$\pm$0.07 \\
		Ah-17(B)&$57165.46~(2015/05/23)$  &$57165.30~(2015/05/23)$&   0.1486$\pm$0.0102 &  -0.1011$\pm$0.0092  &  17.98$\pm$0.05 \\
		Pr-19(B)&$57167.73~(2015/05/25)$  &$57167.73~(2015/05/25)$&   0.0993$\pm$0.0088 &  -0.1627$\pm$0.0090  &  19.06$\pm$0.04 \\
		Pr-19(Q2)&$57192.64~(2015/06/19)$ &$57192.30~(2015/06/19)$&   -0.0146$\pm$0.0339 &   0.0467$\pm$0.0378  &   4.89$\pm$0.17 \\	  		
		Pr-19(C)&$57220.76~(2015/07/17)$  &$57219.27~(2015/07/16)$&  -0.0012$\pm$0.0084 &  -0.0157$\pm$0.0078  &   1.57$\pm$0.04 \\
	    Pr-19(D)&$57245.46~(2015/08/11)$  &$57245.46~(2015/08/11)$&     N    &     N     &     N  \\		
		\hline
	\end{tabular}
	\tablecomments{
(1) the identity of the SED. (2) the day with the highest $\gamma$-ray flux during the SED observation. (3) the nearest day with the spectropolarimetric observations to the day in column (2). (4) and (5) are the median values of Stokes parameters u, q distributions, with uncertainties given based on the photon statistics. ``N" indicates no spectropolarimetric observations during the SED observation. (6) the polarization degree, with uncertainty determined through the uncertainties of u and q.
}
\end{table*}

In principle, the SSC+ERC model can be constrained by using simultaneous multi-instrument observations, with sufficient energy coverage to low- and high-energy bumps, as well as the well-determinated $t_{\rm v,min}$. Under this scenario, the emission region radius $R_{\rm b}^{'}$ can be constrained by $t_{\rm v,min}$ as $R_{\rm b}^{'}\simeq c t_{\rm v,min}\delta_{\rm D}/(1+z)$. Based on the rising part of the ERC hump and the flat region of the synchrotron hump, the spectral indices $\alpha_{\rm 1}$ and $\alpha_{\rm 2}$ can be determined. These indices are directly related to the spectral indices of electron distribution, i.e., $\alpha_{\rm 1}=(s_{\rm 1}-1)/2$ and $\alpha_{\rm 2}=(s_{\rm 2}-1)/2$. Also, the peak frequencies and luminosities (or fluxes) of the synchrotron and ERC bumps, along with $t_{\rm v,min}$, are linked to the Doppler factor $\delta_{\rm D}$, the magnetic field $B^{'}$, the break Lorentz factor $\gamma_{\rm br}$ and the electron density $n_{\rm 0}$. Here, we provide two relationships between the model parameters and observables by combining Equations (2), (4) and (9) from \cite{1998ApJ...509..608T}, that is,
\begin{eqnarray}\label{eq:B_delta_D_prim}
B^{'} & = & 4.84\times 10^{21} (1+z)\frac{\nu_{\rm p,s}^{3}}{\nu_{\rm p,ERC}^{3/2}}
\nonumber\\
&\times &\Bigg\{\frac{t_{\rm v,min}^{2}[\nu_{\rm p,ERC}L_{\rm ERC}(\nu_{\rm p,ERC})]}{[\nu_{\rm p,s}L_{\rm s}(\nu_{\rm p,s})]f(\alpha_{\rm 1},\alpha_{\rm 2})}\Bigg\}^{1/4},
\end{eqnarray}

\begin{eqnarray}
\delta_{\rm D} & = & 7.39\times 10^{-29} 
\nonumber\\
&\times&\Bigg\{\frac{[\nu_{\rm p,s}L_{\rm s}(\nu_{\rm p,s})]^{2}f(\alpha_{\rm 1},\alpha_{\rm 2})}{t_{\rm v,min}^{2}[\nu_{\rm p,ERC}L_{\rm ERC}(\nu_{\rm p,ERC})]}\Bigg\}^{1/4},
\end{eqnarray}
where $\nu_{\rm p,ERC}$ is the observed peak frequency of ERC emission, $[\nu_{\rm p,s}L_{\rm s}(\nu_{\rm p,s})]$ and $[\nu_{\rm p,ERC}L_{\rm ERC}(\nu_{\rm p,ERC})]$ are the observed peak luminosities of the synchrotron and ERC spectra, respectively. Here, the factor $f(\alpha_{\rm 1},\alpha_{\rm 2})=1/(1-\alpha_{\rm 1})+1/(\alpha_{\rm 2}-1)$. Moreover, by applying the $\delta$-approximation to synchrotron emission \citep{2002ApJ...575..667D}, we obtain
$\nu_{\rm p,s}L_{\rm s}(\nu_{\rm p,s})\simeq (2/9)c\sigma_{\rm T}R_{\rm b}^{'3}B^{'2}\delta_{\rm D}^{4}\gamma_{\rm br}^{3}n_{\rm e}^{'}(\gamma_{\rm br})$,
where $\gamma_{\rm br}=(0.75~\nu_{\rm p,ERC}/\nu_{\rm p,s})^{1/2}$. Then we obtain $n_{\rm 0}$ via plugging $\gamma_{\rm br}$ into Equation (\ref{eq:B_delta_D_prim}) as
\begin{eqnarray}\label{eq:n0_delta}
n_{\rm 0} &=& 2.97\times 10^{136}~t_{\rm v,min}^{-1/2}~ \gamma_{\rm br}^{s_{\rm 1}-3}~\frac{\nu_{\rm p,ERC}^{3}}{\nu_{\rm p,s}^{6}}~ \frac{1+z}{\big[\nu_{\rm p,s}L_{\rm s}(\nu_{\rm p,s})\big]^{2}}    
\nonumber\\
&\times & \big[\nu_{\rm p,ERC}L_{\rm ERC}(\nu_{\rm p,ERC})\big]^{5/4}~\big[f(\alpha_{\rm 1},\alpha_{\rm 2})\big]^{-5/4}.
\end{eqnarray}

Unfortunately, the collected multiwavelength data sets of 12 SEDs have only a narrow energy coverage, which does not allow us to determine these observables with high precision. Consequently, determining the model parameters becomes challenging, and leading to many parameter degeneracies. Other combinations of the model parameters can also reproduce these SEDs. As an illustration, a new fitting is shown in Figure \ref{fig:SED-55} for Na-12(A) and Na-12(B) by significantly decreasing $\gamma_{\rm br}$ and $\gamma_{\max}$, which results in the synchrotron spectrum cutting off at optical frequencies. To fit the optical/UV data, we need to increase $L_{\rm d}$ and $R_{\rm H}$ to match the \fermi\ data. Overall, this new set of parameters provides a comparatively good fit to the SED data. However, as shown in Subsection \ref{sec:part-5}, a higher polarization degree of approximately $6.66$\,\% observed during Na-12(A) observation, it is not compatible with the new SED fitting. In the latter case, the polarization degree of the synchrotron component would be largely dilluted by the higher flux level of the nonpolarized, thermal contribution from the AD. Therefore, we here cannot claim that the model parameters presented in Table \ref{tab:mod-parameters} are exclusive, but only claim that they are relatively reasonable and represent the ``best" among our SED modeling exercises.

\begin{figure}
\hbox{
\includegraphics[width=0.5\textwidth]{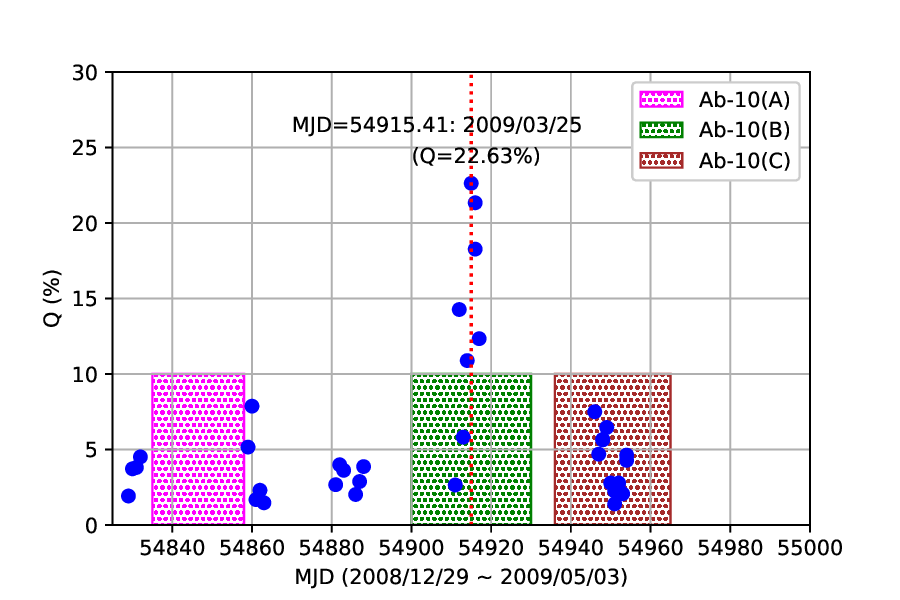}
}
\hbox{
\includegraphics[width=0.5\textwidth]{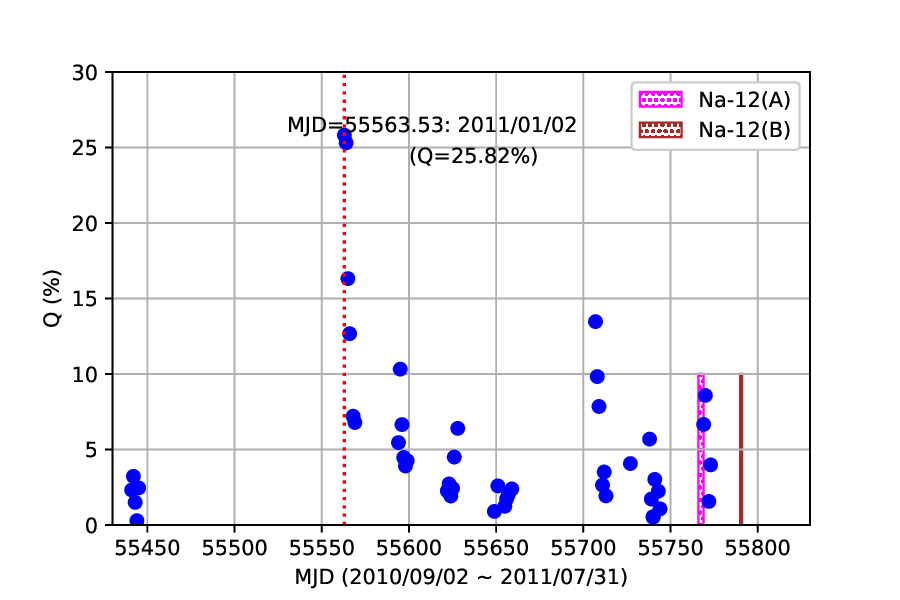}
}
\hbox{
\includegraphics[width=0.5\textwidth]{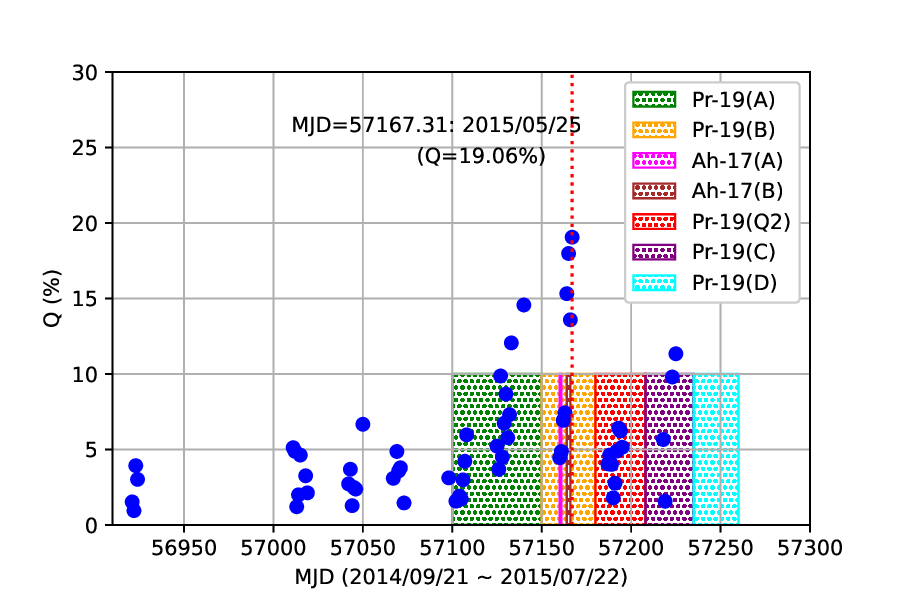}
}
\caption{The polarization degree $Q$ varies with the observation time during three periods, which well encompass 12 SED observations. The vertical dotted line corresponds to the flaring time with the highest polarization degree, and the corresponding time and polarization degree have been labeled in the plot. The shaded regions show the temporal coverage of the studied multiwavelength SEDs.}
\label{fig:Q-Time}
\end{figure}

\section{Spectropolarimetric Observations}\label{sec:part-5}
Long-term optical spectropolarimetric observations of PKS\,1510$-$089 have been regularly carried out at Steward Observatory since the launch of the \fermi-satellite \citep{2009arXiv0912.3621S}. The collected data are summarized in table and Gzipped Tar file, and are accessible publicly\footnote{\url{http://james.as.arizona.edu/~psmith/Fermi/}}. These files provide the flux $F_{\lambda}$, as well as the normalized Stokes parameters q and u. For each set of q and u versus wavelength $\lambda$ distributions, the median values $q_{\rm m}$ and $u_{\rm m}$ can be obtained and used to calculate the polarization degree via $Q=100\sqrt{q_{\rm m}^{2}+u_{\rm m}^{2}}$. This gives the ratio of polarized flux to the total flux detected between 5000-7000\,\AA\/, which includes any emission lines within the considered spectral region. In each observation cycle, the full-resolution flux and polarization spectra can be obtained, which span 4000 $-$ 7550\,\AA\, with a dispersion 4\,\AA\//pixel, and this simply gives the ``size'' of a pixel in wavelength space. All the spectra in the database are not binned and retain their full spectral resolution. Typically, the resolution is between 16$-$24\,\AA\, depending on the width of slit used for the observation.

\begin{figure}
\hbox{
\includegraphics[width=0.5\textwidth]{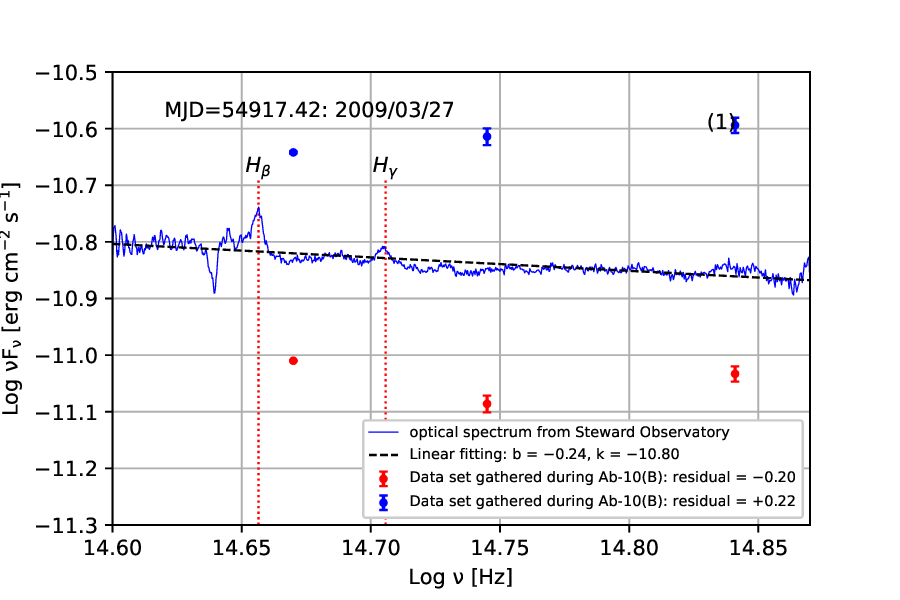}
}
\hbox{
\includegraphics[width=0.5\textwidth]{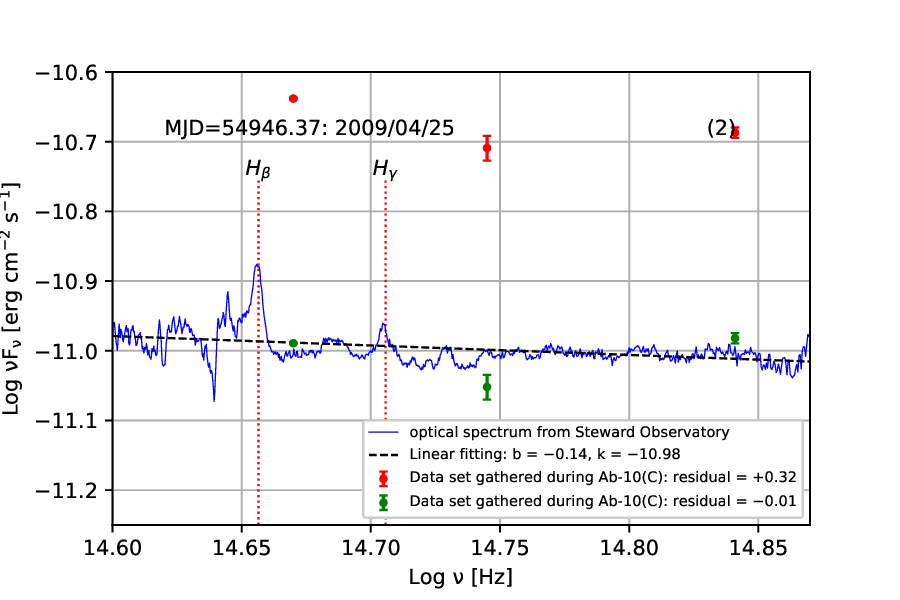}
}
\hbox{
\includegraphics[width=0.5\textwidth]{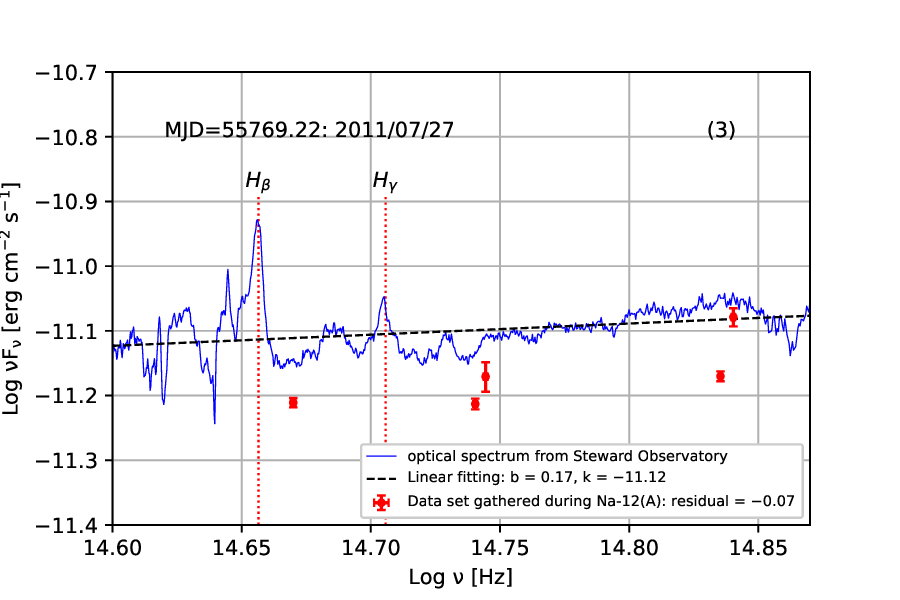}
}
\caption{Distribution of full-resolution optical flux with the frequency. Two notable spikes are marked by two red-dotted lines, which are presumably the emission lines $H_{\beta}$ and $H_{\gamma}$. The black dashed-line represents the linear fitting using a linear function, $f(x)$=b$x$+k, where the obtained values of slope b and intercept k are presented. The multiwavelength data sets are also shown in the same color as in the SED distribution. The residual of the data sets deviating from the linear function is also given, where ``+" represents the data sets located above the linear function systematically, whereas the opposite situation is designated by ``-".}
\label{fig:flux-nu-1}
\end{figure}

Simultaneous spectropolarimetric observations provide critical constraints on the contribution of thermal versus non-thermal radiation in optical band. To properly determine the spectropolarimetric observation during the SED observation, we note that each SED is related to a $\gamma$-ray flare characterized by high $\gamma$-ray flux. By identifying the time at which the highest $\gamma$-ray flux is observed, we can determine the corresponding time for the spectropolarimetric observations. The days with the highest $\gamma$-ray flux are listed in the second column of Table \ref{tab:table-uqQ}. 

Figure \ref{fig:Q-Time} illustrates three periods, i.e., 2008 December 29 to 2009 May 3, 2010 September 2 to 2011 July 31 and 2014 September 21 to 2015 July 22, which cover all SED observations. From these polts, it is clear that three SEDs, i.e., Ab-10(A), Na-12(A) and Pr-19(D), lack the corresponding spectropolarimetric coverage, whereas for the other SEDs, even with both the spectropolarimetric and SED observations, but most of them have no strict one-to-one correspondence in terms of the peak $\gamma$-ray flux time and the spectropolarimetric observation. Therefore, we use the nearest day with the spectropolarimetric observations as a counterpart to the $\gamma$-ray flaring activity. These days are listed in third column of Table \ref{tab:table-uqQ}, where ``N" indicates no spectropolarimetric observations during the entire SED observation period. Besides, the median values $q_{\rm m}$,  $u_{\rm m}$ as well as the polarization degree $Q$ are also provided.

\begin{figure*}
\hbox{
\includegraphics[width=0.5\textwidth]{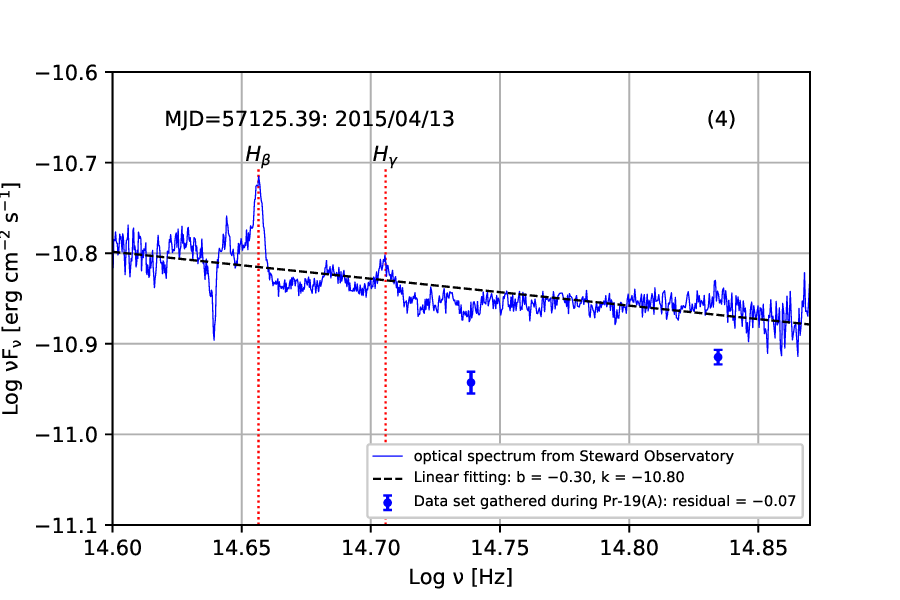}
\includegraphics[width=0.5\textwidth]{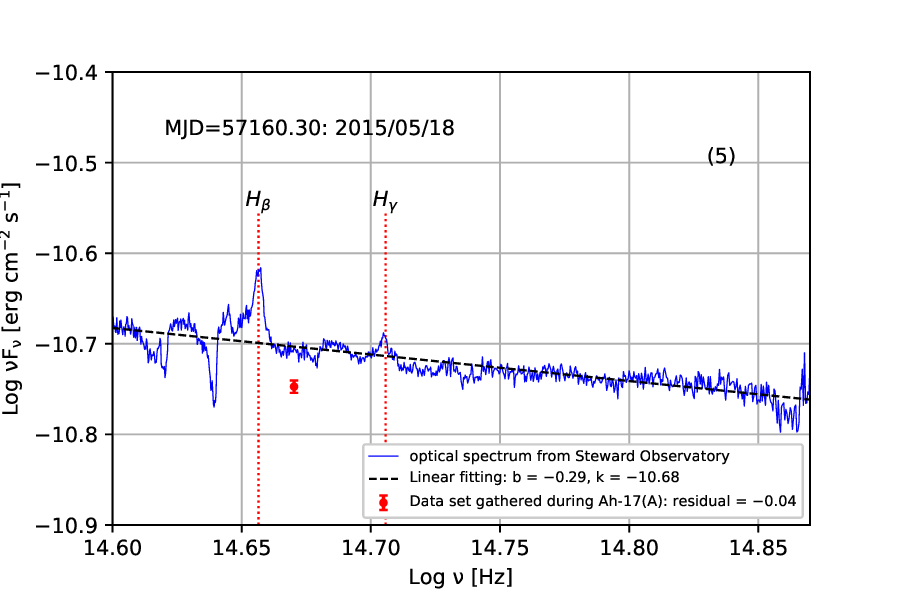}
}
\hbox{
\includegraphics[width=0.5\textwidth]{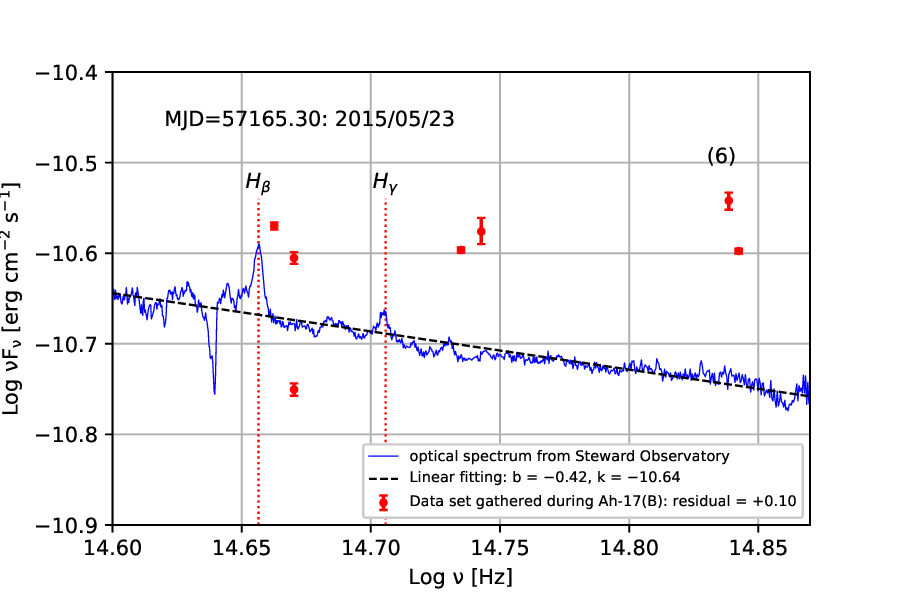}
\includegraphics[width=0.5\textwidth]{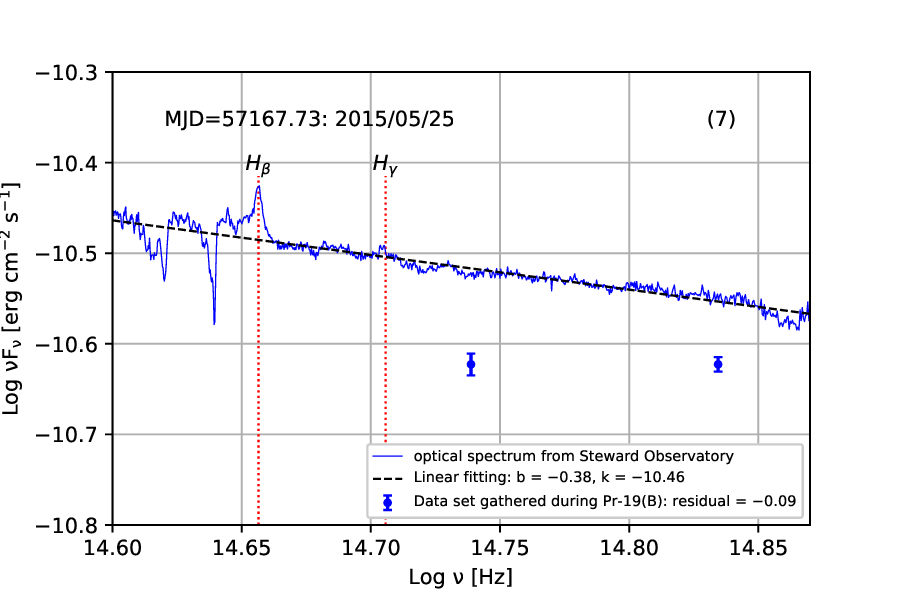}
}
\hbox{
\includegraphics[width=0.5\textwidth]{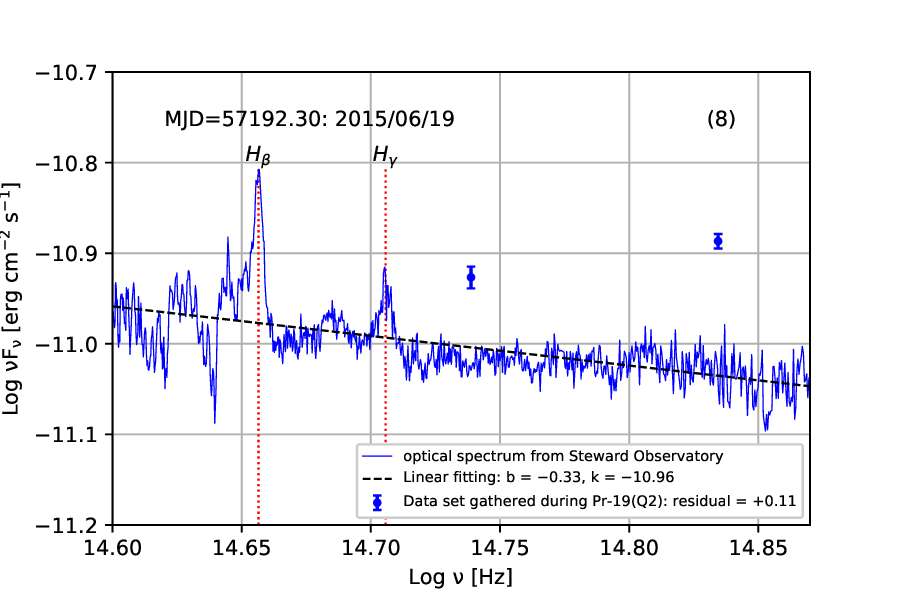}
\includegraphics[width=0.5\textwidth]{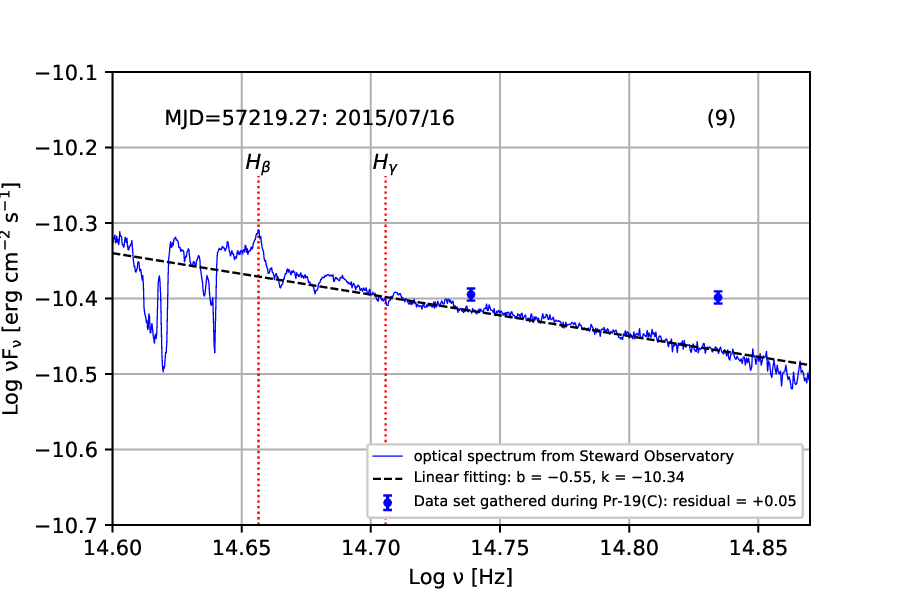}
}
\caption{Distribution of full-resolution optical flux with the frequency. Two notable spikes are marked by two red-dotted lines, which are presumably the emission lines $H_{\beta}$ and $H_{\gamma}$. The black dashed-line represents the linear fitting using a linear function, $f(x)$=b$x$+k, where the obtained values of slope b and intercept k are presented. The multiwavelength data sets are also shown in the same color as in the SED distribution. The residual of the data sets deviating from the linear function is also given, where ``+" represents the data sets located above the linear function systematically, whereas the opposite situation is designated by ``-".}
\label{fig:flux-nu-2}
\end{figure*}

The vertical dotted line in Figure \ref{fig:Q-Time} indicates the observation time with the highest polarization, where the corresponding time and the polarization degree are labeled. In top panel, it is evident that the observation of Ab-10(B) has captured an intense polarization state, likely related to the origin of the multiwavelength emission in Ab-10(B). The middle panel shows an exceptionally high polarization degree, reaching up to 25.82\,\%, which is the maximum ever recorded by Steward Observatory. After this peak, polarization degree continuously decreases to 2.39\,\% on 2011 April 8. Prior to this maximum, the polarization state is unclear due to a 3.8-month gap in spectropolarimetric coverage. Moreover, we note that the polarization degree of Na-12(A) was close to 6.66\,\% on 2011 July 27, whereas Na-12(B) has not corresponding spectropolarimetric observations. In the bottom panel, the polarization degree progressively increased, reaching a peak of 19.06\,\% on 2015 May 25. Given the lack of the spectropolarimetric monitoring in subsequent 20 days, the trend of polarization development cannot be conclusively determined. However, after 20 days, the polarization degree $Q$ drops to around 4\,\%. Subsequently, $Q$ further decreases to 1.8\,\% before bouncing back to 6.4\,\%. The high polarization state likely lasts for a month or more, spanning Pr-19(A), Pr-19(B), Ah-17(A), Ah-17(B), Pr-19(Q2) and Pr-19(C). This suggests that during this period, the particle population within the relativistic jet were undergoing a strong shock acceleration process, possibly consisting of multiple sub-shock processes that are associated with different flares, and are responsible for six SEDs. These sub-shocks propagated down the jet and persistently compressed the turbulent magnetic fields, thereby displaying a higher polarization degree \citep{1985ApJ...298..114M,1985ApJ...298..301H,2008ApJ...672...40H}. 

To more directly compare spectropolarimetric flux with the SEDs, the conversion from $F_{\lambda}-\lambda$ to $\nu F_{\nu}-\nu$ distributions has been performed and is shown in Figures \ref{fig:flux-nu-1} and \ref{fig:flux-nu-2}. The linear fitting was conducted using a linear function $f(x)$=b$x$+k, where b and k represent the slope and the intercept. The resulting b and k values, along with the SED-ID, are shown in the legend of each panel. Meanwhile, the corresponding multiwavelength data have been incorporated into the plots, using the same color as in the SED distributions. To assess the deviation of the SED data from the linear function, a residual is calculated using $\sum_{i=1}^n\big[y(x_{\rm i})-f(x_{\rm i})\big]/n$, where n is the number of data points, and $x_{\rm i}$ and $y(x_{\rm i})$ are the logarithm of the frequency ($\nu$) and flux ($\nu F_{\nu}$), respectively. The results show that only panel (3) has a positive slope, while the remaining spectra exhibit a negative slope, indicating that these spectra decline with increasing frequency. Comparing the spectropolarimetric and broadband SED fluxes, only several spectra, shown in panels (2) (green data), (3), (4), (5), (7) and (9), have very close level, with the residual less than 0.1 in magnitude. This may indicate that the spectropolarimetric and SED observations roughly correspond to the same flaring state. Particularly in panel (2)(green data), where both fluxes perfectly match each other, the residual is $- 0.01$, with the ``-" sign indicating that the SED data sets lie below the linear function. Overall, for panels (1) (blue data), (2)(red data), (6), (8) and (9), the SED flux is higher than the spectropolarimetric observations, resulting in positive residual values. The opposite situation occurs for the red data in panels (1) (red data), (3), (4), (5) and (7), which demonstrate a negitive residual. Here, it is worth mentioning that spectropolarimetric observations were unavailable for Na-12(A) during the peak $\gamma$-ray flux on 2011 July 25. The closest spectropolarimetric data were obtained on 2011 July 27. However, as shown in Figure \ref{fig:flux-nu-1}, the optical/UV flux levels measured by the \swift-UVOT telescope at MJD 55766.64 and 55768.43 for Na-12(A) closely match those from the spectropolarimetric measurements on 2011 July 27. These optical/UV observations are nearly simultaneous with the \fermi\ observations that spanned from MJD 55766.49 (2011/07/24) to MJD 55769.47 (2011/07/27). Given the proximity in both flux and observational time, the polarization degree of $(6.66 \pm 0.09)$\,\%, derived from the spectropolarimetric observation on 2011 July 27, may serve as a reasonable approximation to the polarization properties of the optical spectrum during Na-12(A). 

\begin{figure}
\hbox{
\includegraphics[width=0.5\textwidth]{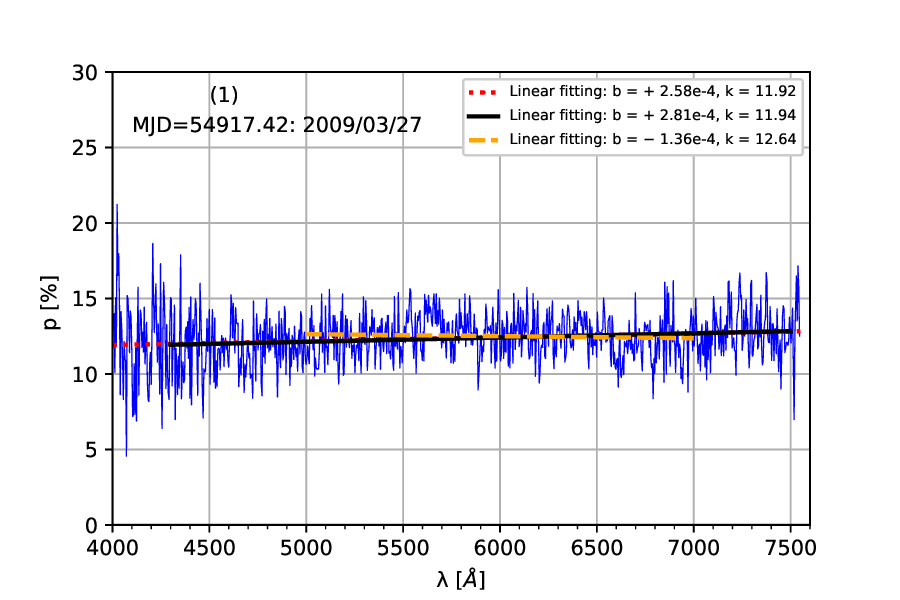}
}
\hbox{
\includegraphics[width=0.5\textwidth]{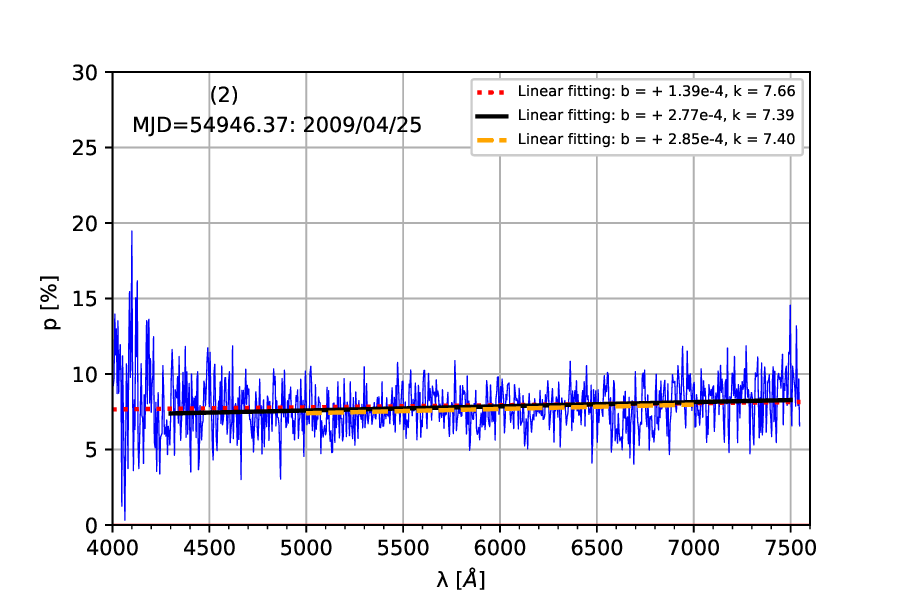}
}
\hbox{
\includegraphics[width=0.5\textwidth]{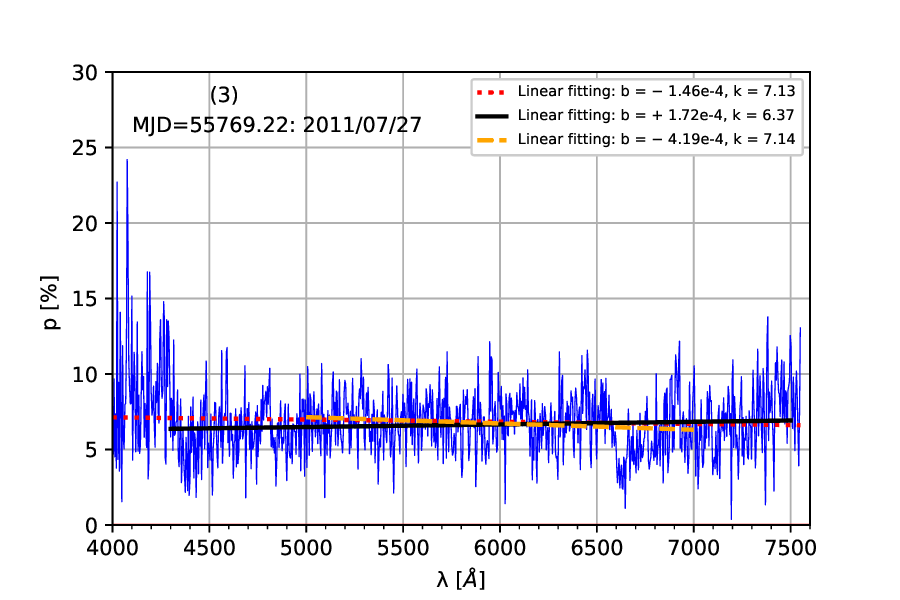}
}
\hbox{

}
\caption{Dependence of polarization degree on the wavelengths covered by optical spectropolarimetric observations. A linear fitting of form $p(\lambda)=b\lambda+k$ is applied to three wavelength ranges, [4000$-$7550]~\AA\,, [4300$-$7500]~\AA\, and [5000$-$7000]~\AA\,, shown as the red dotted, black solid as well as the orange dashed lines, respectively.}
\label{fig:p-lambda-1}
\end{figure}

\begin{figure*}
\hbox{
\includegraphics[width=0.5\textwidth]{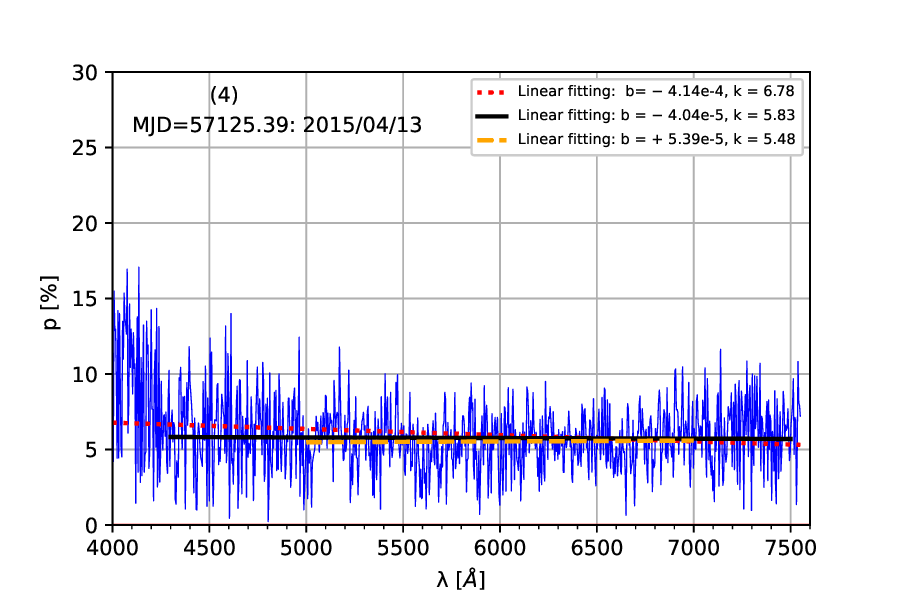}
\includegraphics[width=0.5\textwidth]{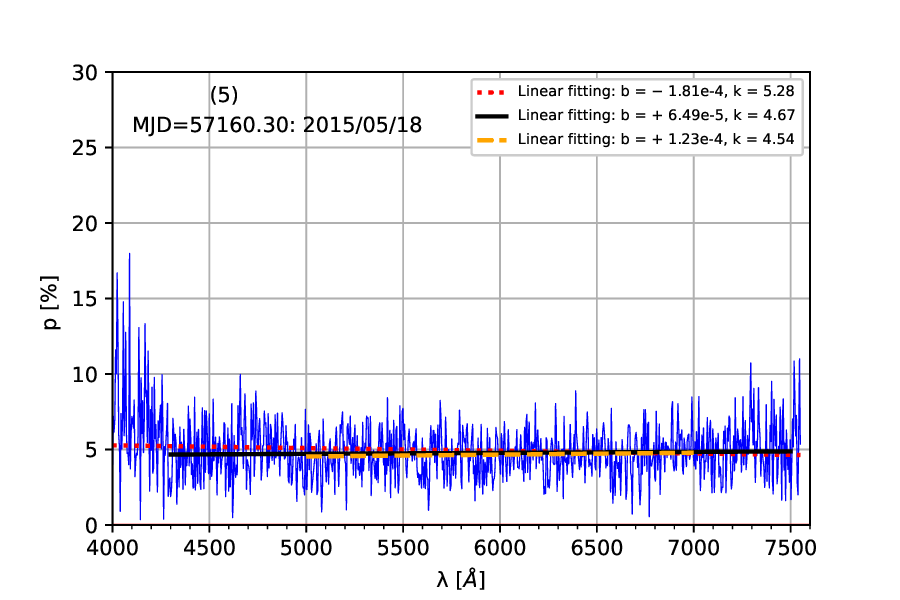}
}
\hbox{
\includegraphics[width=0.5\textwidth]{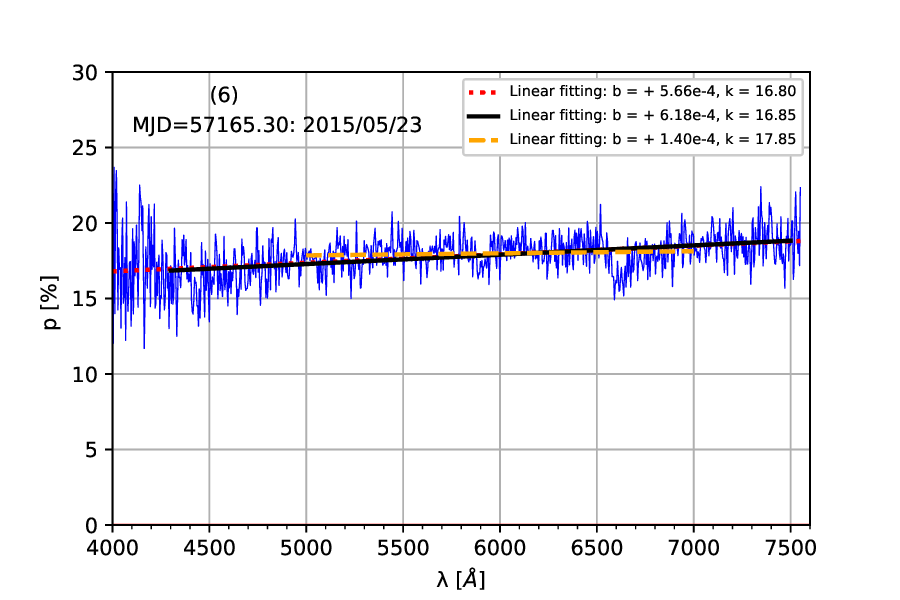}
\includegraphics[width=0.5\textwidth]{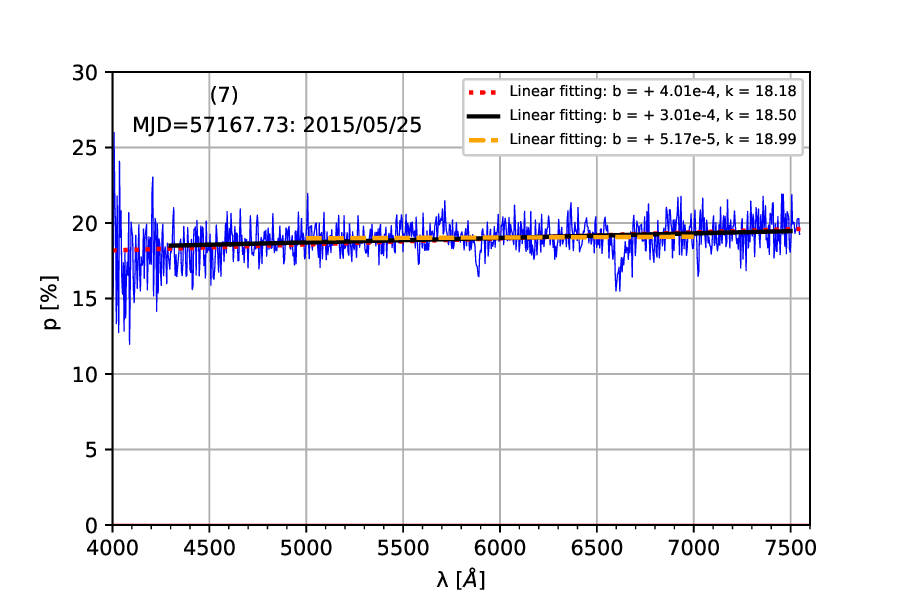}
}
\hbox{
\includegraphics[width=0.5\textwidth]{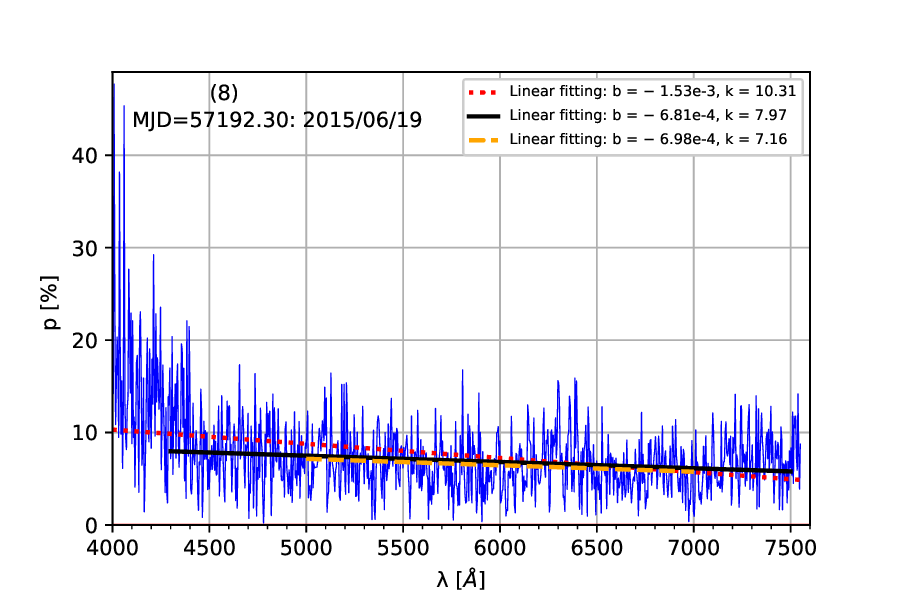}
\includegraphics[width=0.5\textwidth]{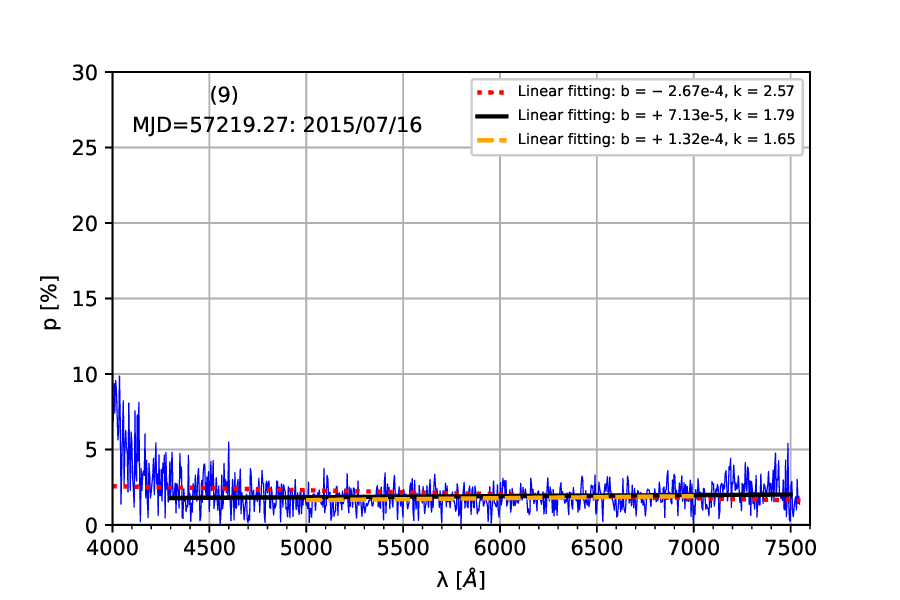}
}

\caption{Dependence of polarization degree on the wavelengths covered by optical spectropolarimetric observations. A linear fitting of form $p(\lambda)=b\lambda+k$ is applied to three wavelength ranges, [4000$-$7550]~\AA\,, [4300$-$7500]~\AA\, and [5000$-$7000]~\AA\,, shown as the red dotted, black solid as well as the orange dashed lines, respectively.}
\label{fig:p-lambda-2}
\end{figure*} 

The spectropolarimetric flux is a composite of the non-thermal emission from synchrotron, thermal emission from the AD, and various emission lines from the BLR. These components generally have distinct flux level in different flaring states. In the optical spectra shown here, emission lines are prominent, as illustrated in Figures \ref{fig:flux-nu-1} and \ref{fig:flux-nu-2}. Two notable spikes are observed at rest-frame wavelengths of 4861.3~\AA\, and 4340.48~\AA\,, corresponding to $H_{\beta}$ and $H_{\gamma}$ lines, respectively, both are marked by vertical red-dashed lines. These findings are consistent with the results presented in Figure \ref{fig:SED-3} by \cite{2023ApJ...952L..38A}, where additional emission lines, such as Fe~{\sc ii}, H$_{\delta}$, Ne~{\sc iii}, and others, may contribute. However, the most discernible features remain $H_{\beta}$ and $H_{\gamma}$. Similar observational results are also presented by \cite{2024ApJ...968..130P}, using the data sets gathered by the Robert Stobie Spectrograph on the Southern African Large Telescope (SALT).   
 
As indicated by our SED modelings, the polarization degree would decrease with increasing frequency in optical band due to the dilution from the unpolarized thermal emission components. To further verify this influence from the observations, we retrieved the Stokes q and u spectra from the Steward Observatory archive database. These spectra are wavelength-calibrated and normalized linear polariztion parameters, related to the linear polarization parameters Q and U in spectral form as $q(\lambda)\equiv Q(\lambda)/I(\lambda)$, $u(\lambda)\equiv U(\lambda)/I(\lambda)$, where $I(\lambda)$ is the total intensity. Such that the polarization degree is determined by $p(\lambda)=100\sqrt{q(\lambda)^{2}+u(\lambda)^{2}}$, and the resulting distributions are presented in Figures \ref{fig:p-lambda-1} and \ref{fig:p-lambda-2}. Since the spectral polarization distribution shows significant variation across the entire wavelength range, we performed a linear fitting to determine trend of the polarization degree with the wavelength, using a linear function $p(\lambda)=b\lambda+k$. 

The spectral fittings were applied to three wavelength ranges. The first one covers 4000 $-$ 7550~\AA\, and shows that only 4 out of 9 spectra possess a positive spectral slope (though small), that is, the polarization degree gradually decreases toward shorter wavelengths. This result indicates the presence of a weak AD contribution, consistent with our SED modelings. At the same time, we note that for the data sets in panels (3), (4), (5) and (8), they show strong fluctuations in the 4000 $-$ 4300~\AA\, range. Most likely, these fluctuations are caused by the edge noise \citep{2018MNRAS.479.2037P,2023MNRAS.518.5788O}. Thus, the second fitting, which removes this part and focuses on a short wavelength range from 4300~\AA\, to 7500~\AA\,, the resulting fittings are shown in Figures \ref{fig:p-lambda-1} and \ref{fig:p-lambda-2}. In such case, three additional spectra, i.e., in panels (3), (5) and (9), also show a positive slope, whereas panels (4) and (8) present an opposite situation, where the polarization degree decreases with increasing wavelength. This contradicts with the SED modelings of Pr-19(A) and Pr-19(Q2), where the synchrotron emission overlaps with the rising part of the BBB, producing the observed optical spectrum; therefore, a drop in polarization degree would be expected. For the third case, the fitted wavelength range is more shorter, spanning 5000 $-$ 7000~\AA\,, it aligns with that used in calculating the optical polarization degree Q in the Steward Observatory archive. In contrast with the second fitting, the third one gives an opposite slope for panels (1), (3) and (5); specifically, the slopes in the first two panels are negative, whereas the slope in the last panel is positive. Overall, the spectral polarization distributions are statistically flat across both the 4300 $-$ 7500~\AA\, and 5000 $-$ 7000~\AA\, ranges, resulting in a rather mild slope in the linear fittings, this indicates that the optical polarization degree has only weak wavelength-dependent variation. Here, we further calculate the corresponding average polarization degree as $P=100\sqrt{<q>^{2}+<u>^{2}}$, where $<q>$ and $<u>$ are the average values of the normalized Stokes parameters q and u distributions. For the 4300 $-$ 7500~\AA\, range, three higher $P$ values of approximately 12.28\%, 17.80\% and 18.96\% are obtained from panals (1), (6) and (7), respectively. Four intermediate $P$ values of approximately 7.68\%, 6.36\%, 5.27\% and 4.94\% are derived from panels (2), (3), (4) and (8), respectively. In contrast, panels (5) and (9) exhibit lower $P$ values of approximately 4.28\% and 1.54\%. As for the 5000 $-$ 7000~\AA\, range, the resulting $P$ values are approximately 12.43\%, 7.59\%, 6.51\%, 5.20\%, 4.40\%, 17.97\%, 19.02\%, 5.07\% and 1.53\%, corresponding to panels (1) through (9), respectively. 

\section{Discussion} \label{sec:part-6}
This paper studied the origin of multiwavelength emissions of PKS\,1510$-$089 by fitting 12 SEDs from four periods spanning from 2008 to 2015. In the following subsections, we discuss the main results and their implications.

\subsection{The BH mass and inner radius of the AD}
For the 12 SEDs studied in this paper, both the spectral shape and notable flux excess, shown by the data sets, indicate significant contributions from the AD at optical/UV frequencies. This is supported by several studies with similar spectral morphologies \citep{2018A&A...619A.159M,2021MNRAS.504.1103R,2023ApJ...952L..38A}, which provide meaningful references. Fitting optical/UV data sets from different flaring episodes will impose robust constraint on $L_{\rm d}$ and $M_{\rm BH}$. However, as shown by these 12 SEDs, the peak location of the BBB is variable. Consequently, if $M_{\rm BH}$ is treated as a free parameter and the BBB component is modeled using a multichromatic blackbody spectrum, a range of the BH mass will be required. Actually, the fitting to Ab-10(A), Ab-10(B) and Ab-10(C) yields the same BH masses of $8.0\times 10^{8}$\,$M_{\odot}$, whereas the modelings of Na-12(A) and Na-12(B) require a BH masses of $5.4\times 10^{8}$\,$M_{\odot}$. The BH masses obtained for Ah-17(A) and Ah-17(B) are $1.6\times 10^{9}$\,$M_{\odot}$ and $2.7\times 10^{9}$\,$M_{\odot}$, respectively. In contrast, the resulting BH masses obtained from five SEDs presented in \cite{2019ApJ...883..137P} are more dispersed, with values of $2.2\times 10^{9}$\,$M_{\odot}$, $3.3\times 10^{9}$\,$M_{\odot}$, $2.2\times 10^{9}$\,$M_{\odot}$, $3.3\times 10^{9}$\,$M_{\odot}$ and $2.7\times 10^{9}$\,$M_{\odot}$, corresponding to Pr-19(A), Pr-19(B), Pr-19(Q2), Pr-19(C) and Pr-19(D), respectively. Notably, the obtained BH mass varies across different SED observations. Furthermore, even within the same observational period, the BH mass may differ. This change poses a significant challenge to current BH physics and is difficult to reconcile with standard accretion or merging processes.

Moreover, it is worth noting that the central BH mass of PKS\,1510$-$089 remains under the dabate. Several methods have been employed to estimate the BH mass, including RM technique, spectroscopic monitoring, SED modeling and the quasi-periodic oscillation (QPOs) analysis. These methods provide a range of BH mass spanning nearly three orders of magnitude, from $4.19\times 10^{7}$\,$M_{\odot}$ to $8.2\times 10^{10}$\,$M_{\odot}$ \citep{2020A&A...642A..59R,2022MNRAS.510.3641R}. To adopt a more appropriate value for the BH mass, two aspects have been considered: first, we note that if the central BH has mass $\lesssim 1.0\times 10^{8}$\,$M_{\odot}$, according to Equations (\ref{eq:BZ}) and (\ref{eq:BP}), the maximum jet power produced through BZ or BP mechanism would be less that derived from the SED modelings. Instead, if a larger BH mass were used ($\gtrsim 1.0\times 10^{9}$\,$M_{\odot}$), the resulting peak of the BBB would be located at a lower frequency, i.e., $\nu_{\rm p,AD}^{\rm obs}\simeq 77.56/(1+z)~r_{\rm in}^{-3/4}(\dot{m}_{\rm acc}/0.01)^{1/4}M_{\rm BH,8}^{-1/4}(12\eta_{\rm f})^{-1/4}$\,eV, where $r_{\rm in}=R_{\rm in}/R_{\rm g}$. Under such case, it is difficult to well reproduce the optical/UV spectrum compatile with a concave profile, as clearly shown in Na-12(B). Taking both aspects into account, we fix $M_{\rm BH}$ at $5.4\times 10^{8}$\,$M_{\odot}$, which is consistent with the value obtained by \cite{2010{\natexlab{b}}ApJ...721.1425A} and close to $3.86\times 10^{8}$\,$M_{\odot}$ given by \cite{2002ApJ...576...81O}. Subsequently, we fine-tune the parameter $f_{\rm dic}$ to satisfactorily fit the optical/UV data sets. The resulting $f_{\rm dic}$ values, ranging between $1.0$ and $6.0$, are listed in Table \ref{tab:mod-parameters}. Therefore, the $R_{\rm ISO}$ of the AD of PKS\,1510$-$089 fluctuates within $3R_{\rm S}$ and $18R_{\rm S}$, rather than being fixed at $3R_{\rm S}$. This fluctuation could be responsible for the dramatical variation in the maximum temperature of the AD across different flaring states. We also note from Tables \ref{tab:mod-parameters} and \ref{tab:Der-quan-sed} that a larger value of $f_{\rm dic}$ roughly corresponds to a relatively higher value of $\dot{M}_{\rm acc}$ and $P_{\rm B}$. Since the magnetic fields play a vital role in production, acceleration as well as the collimation of the relativistic jet, this implies that the state of the jet is closely related to the activity of the central engine; specifically, the energy extraction through the jet may influence the size of the innermost radius of the AD. 

Physically, if a rotating BH has high retrograde spin, the maximal size of $R_{\rm ISO}$ can reach up to $9R_{\rm g}$ \citep{1972ApJ...178..347B}. Besides, the modeling of the \nustar\ data for radio quasar 4C+74.26 also requires a truncated disk with inner radius at $R_{\rm in}/R_{\rm ISO}=35_{\rm -16}^{+40}$ \citep{2018ApJ...866..132B}. Similar to other FSRQs, such as Quasars 3C 273 and 3C 279, PKS 1510$-$089 is well known for commonly showing the QPOs, and a quasi-period of $\sim1330$\,d has been observed in the radio light curve \citep{2023MNRAS.519.4893L}, whereas in $\gamma$-rays, a QPO of 3.6\,d was observed during the 2009 outburst, and a periodicity of $92$\,d was detected between 2018 and 2020 \citep{2022MNRAS.510.3641R}. These QPOs may be related to the perturbations in $R_{\rm ISO}$. If this is the case, the activity of the AD could also leave imprints on the relativistic jets. The timescales of the QPOs induced by disk motion can be approximated as $\tau_{\rm QPO}\simeq 89.6(f_{\rm dic}/5.0)(M_{\rm BH,8}/5.4)(v_{\rm d,m}/0.07c)^{-1}$\,d, where $v_{\rm d,m}$ is the motion speed of disk matter at the innermost radius. A shorter timescale on the order of days can be achieved by decreasing $f_{\rm dic}$, but increasing $v_{\rm d,m}$, whereas the shortest timescale occurs when the $R_{\rm ISO}$ is approximately $3R_{\rm S}$.

In practice, it is difficult to quantify the truncation of the innermost radius of the AD to PKS\,1510-089, but a spectral feature may be related to it, i.e., the peak frequency of the BBB notably changes in different observational periods. For a canonical AD, with $R_{\rm ISO}=3R_{\rm S}$ and $\eta_{\rm f}=1/12$, which emits radiation in a blackbody spectrum, the peak frequency is given by
\begin{eqnarray}\label{eq:pf_AD}
\nu_{\rm p,AD}^{\rm obs} &\simeq& \frac{7.2\times 10^{15}}{1+z}\biggr(\frac{M_{\rm BH}}{10^{9}~M_{\odot}}\biggr)^{-1/2}
\nonumber\\
&\times& \biggr(\frac{L_{\rm d}}{5\times 10^{46}\,\rm erg~s^{-1}}\biggr)^{1/4} \rm Hz.
\end{eqnarray}
Clearly, $M_{\rm BH}$ can be determined using $\nu_{\rm p,AD}^{\rm obs}$ and $L_{\rm d}$, where $L_{\rm d}$ depends on the amplitude of the BBB and ratio of thermal to non-thermal contributions in it. This ratio can be disentangled by combining the spectropolarimetric observations. However, if there are multiple SED observations spanning several years and their optical/UV data sets clearly demonstrate that $\nu_{\rm p,AD}^{\rm obs}$ is not constant. According to Equation (\ref{eq:pf_AD}), this would lead to a series of distinct BH masses. Given that a BH cannot change its mass within such a short period, then the peak shift of the BBB likely stems from the variation of $R_{\rm ISO}$, instead of the BH itself. Therefore, the well sampling SED data sets and the simultaneous spectropolarimetric observations, both have long-term time span, are very important to check this inference in the future.

\subsection{The origin of broadband emissions and their intra-band correlations}
The radiation contribution from the corona is also considered in Figures \ref{fig:SED-1} $-$ \ref{fig:SED-4}, where $\xi_{\rm Corona}=0.06$ provides a good fit to the X-ray data sets. Because the lower limit of the emitted photon frequencies is set at $5.0\times 10^{16}$\,Hz, so an abrupt spectral cutoff appears at the low-energy end. At these frequencies, the corona's contribution is comparable to that of the ERC-DT and SSC emissions. Our model has also considered the Compton scattering of soft photons from the AD (ERC-AD), but since the relatively large distance of $\gamma$-ray emitting region, this process makes an almost negligible contribution to the observed $\gamma$-ray flux.   

Figures \ref{fig:SED-1} $-$ \ref{fig:SED-4} have clearly demonstrated that the low-energy hump of the SED primarily arises from synchrotron emissions. Only a small portion, mainly at infrared and optical/UV frequencies, is mixed with some contributions from the DT and AD. Among both components, the AD emissions play a main role in shaping the BBB; meanwhile, it will also dilute the non-thermal emissions and reduce the polarization degree. This point has been revealed in Section \ref{sec:part-5}, where part of the spectral polarization distributions show a slight decreasing trend toward the shorter wavelengths. Additionally, the AD emissions will smooth the variability, this is consistent with the observations that the optical/UV flux does not shows notablely coordinated variability with the X- and $\gamma$-rays \citep{2017A&A...603A..29A,2019ApJ...883..137P}. At the low-energy end of the X-ray spectrum, where the SED spectrum is composed of multiple components, including the SSC, ERC-DT, and the corona emissions. In the high-energy band, the spectrum is well dominated by ERC-BLR process, although the contribution from the ERC-DT process is also significant, especially for Na-12(B), where the ERC-DT fit exceeds the ERC-BLR at X-ray energies. 

From the perspective of variability, since the synchrotron, ERC-DT and ERC-BLR processes have a linear dependence on electron distribution ($F_{\nu}^{\rm syn}\propto n_{\rm e}^{'}(\gamma)$, $F_{\nu}^{\rm BLR}\propto n_{\rm e}^{'}(\gamma)$, $F_{\nu}^{\rm DT}\propto n_{\rm e}^{'}(\gamma)$), thus the correlated variability will be expected between the synchrotron emission and $\gamma$-rays. A similar variability pattern may also appear between hard X- and $\gamma$-rays, given that the former originates from the ERC-BLR or/and ERC-DT, whereas the latter stems from the ERC-BLR. At the low-energy X-ray band, the spectrum results from multiple components, including SSC emission, which depends quadratically on the electron distribution, $F_{\nu}^{\rm SSC}\propto n_{\rm e}^{'2}(\gamma)$. This superposition leads to more complex variability behavior. As for radio wavelength, because of the synchrotron opacity in the nuclear region, these emissions likely originate some location distinct from the one of the $\gamma$-ray emissions. Alternatively, they may arise from the same region with the $\gamma$-ray emissions but radiate them at different sites. In fact, both scenarios have been supported by previous correlation studies between $\gamma$-ray and radio bands, which suggested that the radio variability usually shows time lags of tens of days or several months relative to the corresponding $\gamma$-ray flare \citep{2010ApJ...722L...7P,2014MNRAS.445..428M,2014MNRAS.441.1899F,2015MNRAS.452.1280R,2016MNRAS.462.2747K,2016A&A...591A..83S,2017A&A...606A..87B,2018MNRAS.480.5517L,2019ApJ...877..106P,2020MNRAS.493.3757J}. An exception is that when the emission region is far from the nuclear region, where all of emissions can be simultaneously observed \citep{2011MNRAS.417..359O,2014{\natexlab{a}}A&A...569A..46A}. However, for the 12 SEDs studied here, the $\gamma$-rays are arised near the BLR, so the correlations between radio and other bands are not taken into account.

\begin{figure*}
\hbox{
\includegraphics[width=0.5\textwidth]{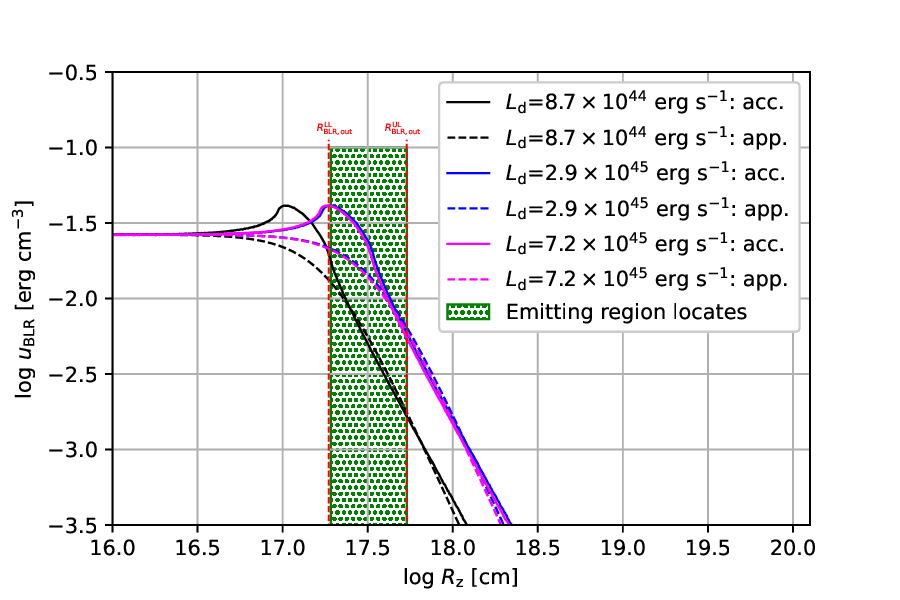}\\
\includegraphics[width=0.5\textwidth]{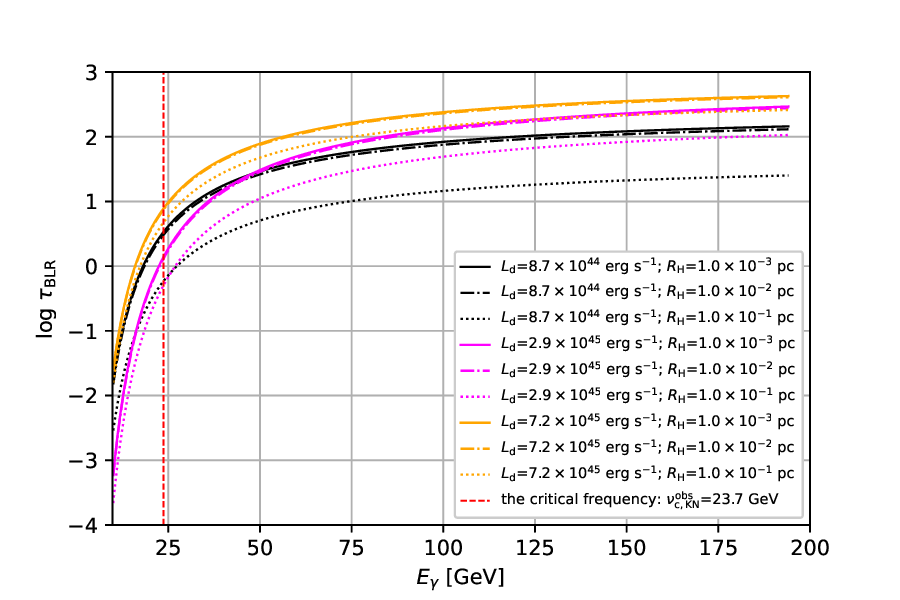}\\
}
\caption{Left panel: Dependence of $u_{\rm BLR}$ on $R_{\rm z}$, The solid lines of different colors represent accurate calculations (acc.) for three cases: Ab-10(A), Pr-19(B) and Pr-19(C). The dashed lines are approximate representations (app.) using Equation (\ref{eq:u_BLR_disk_RF}). The shaded region indicates the span of the location of the $\gamma$-ray emitting regions, where $R_{\rm BLR,out}^{\rm LL}$ and $R_{\rm BLR,out}^{\rm UL}$ are the corresponding lower and upper limits of the outer radius of the BLR. Right panel: Dependence of $\log\tau_{\rm BLR}$ on $E_{\gamma}$. The $\gamma$-ray emitting region is assumed to be located at $1.0\times 10^{-3}$\,pc, $1.0\times 10^{-2}$\,pc and $1.0\times 10^{-1}$\,pc, respectively. The vertical red-dashed line marks the critical frequency, $\nu_{\rm c,KN}^{\rm obs}=23.7$\,GeV.}
\label{fig:u_BLR_disk_RF}
\end{figure*}

To gain insight into the origin of broadband emissions, each SED has been split up using five quantities, $\nu_{\rm br}^{\rm syn}$, $\nu_{\rm max}^{\rm syn}$, $\nu_{\rm BLR,min}^{\rm ERC}$, $\nu_{\rm BLR,br}^{\rm ERC}$ and $\nu_{\rm BLR,max}^{\rm ERC}$, as indicated in Figures \ref{fig:SED-1} to \ref{fig:SED-55} in vertically sandybrown-dashed lines. The last three quantities are used because the high-energy hump is primarily dominated by the ERC-BLR. These quantities are related to the characteristic electron Lorentz factors as follows,
\begin{eqnarray}
\nu_{\rm br}^{\rm syn} &=& \frac{\delta_{\rm D}}{1+z}~\frac{4}{3}~\gamma_{\rm br}^{2}~\nu_{\rm L},
\nonumber\\
\nu_{\rm max}^{\rm syn} &=& \frac{\delta_{\rm D}}{1+z}~\frac{4}{3}~\gamma_{\rm max}^{2}~\nu_{\rm L},
\nonumber\\
\nu_{\rm BLR,min}^{\rm ERC} &=& \frac{\delta_{\rm D}}{1+z}~\frac{4}{3}~\gamma_{\rm min}^{2}~(\Gamma_{\rm j}\nu_{\rm o}^{*}),
\nonumber\\
\nu_{\rm BLR,br}^{\rm ERC} &=& \frac{\delta_{\rm D}}{1+z}~\frac{4}{3}~\gamma_{\rm br}^{2}~(\Gamma_{\rm j}\nu_{\rm o}^{*}),
\nonumber\\
\nu_{\rm BLR,max}^{\rm ERC} &=& \frac{\delta_{\rm D}}{1+z}~\frac{m_{\rm e}c^{2}}{h}~\gamma_{\rm max}~f(\alpha_{\rm 1},\alpha_{\rm 2}),
\end{eqnarray}
where $\nu_{\rm L}=eB^{'}/(2\pi m_{\rm e}c)$ is the Larmor frequency. For $\nu_{\rm BLR,max}^{\rm ERC}$, since it is well in the Klein-Nishina (KN) regime, the KN correction has been applied \citep{1998ApJ...509..608T}. As a result, each section of the SED can be roughly related to the corresponding segment of electron distribution. For the synchrotron spectrum with $\nu<\nu_{\rm br}^{\rm syn}$, the emission is contributed by the electrons with $\gamma$ ranging from $\gamma_{\rm min}$ to $\gamma_{\rm br}$, whereas the synchrotron spectrum \big[$\nu_{\rm br}^{\rm syn}, \nu_{\rm max}^{\rm syn}$\big] is contributed by the electrons with $\gamma$ from $\gamma_{\rm br}$ to $\gamma_{\rm max}$. The spectrum in the range \big[$\nu_{\rm BLR,min}^{\rm ERC}, \nu_{\rm BLR,br}^{\rm ERC}$\big] originates from the ERC-BLR of the electrons with $\gamma$ from $\gamma_{\rm min}$ to $\gamma_{\rm br}$, and the spectrum \big[$\nu_{\rm BLR,br}^{\rm ERC}, \nu_{\rm BLR,max}^{\rm ERC}$\big] originates from the ERC-BLR of the electrons with $\gamma$ from $\gamma_{\rm br}$ to $\gamma_{\rm max}$. To further explore whether the interactions of the electrons and photons occur in the Thomson or KN regimes for the ERC-BLR and ERC-DT processes, it is necessary to compare the spectral shape between the high-energy and the synchrotron bumps. To do this, we have derived the spectral slopes using the $\delta$-approximation, and given in Appendix \ref{sec:Flux-syn} and \ref{sec: Flux-ERC}. It is clear that if the interactions take place in the Thomson regime, the observed flux $\nu F_{\nu}$  will have the same slope $-(s_{\rm 1}-3)/2$ for the spectrum below $\nu_{\rm br}^{\rm syn}$ and for the range \big[$\nu_{\rm BLR,min}^{\rm ERC}$, $\nu_{\rm BLR,br}^{\rm ERC}$\big], and will have slope $-(s_{\rm 2}-3)/2$ for the spectrum in the range \big[$\nu_{\rm br}^{\rm syn}$, $\nu_{\rm max}^{\rm syn}$\big] as well as for \big[$\nu_{\rm BLR,br}^{\rm ERC}$, $\nu_{\rm BLR,max}^{\rm ERC}$\big]. In contrast, \cite{2005MNRAS.363..954M} has showed that if the Compton scattering occurs in the KN regime, there will be different spectral slopes for the spectra above the break frequencies $\nu_{\rm br}^{\rm syn}$ and $\nu_{\rm BLR,br}^{\rm ERC}$ in terms of the synchrotron and Compton emissions. Therefore, we first determine the critical frequency, $\nu_{\rm c,KN}^{\rm obs}$, where the transition takes place from the Thomson into the KN regimes. This frequency in the observer's frame has been given in Appendix \ref{sec: nu-cri-fre} as  
\begin{eqnarray}
\nu_{\rm c,KN}^{\rm obs} &\simeq& \frac{23.7}{\nu_{\rm o,15}^{*}}\frac{\delta_{\rm D}}{\Gamma_{\rm j}(1+z)}~\,\rm GeV,
\end{eqnarray}
where $\nu_{\rm o,15}^{*}$ is the characteristic frequency of the target photon field, in units of $10^{15}$\,Hz. Due to the high-energy hump is dominated by the ERC-BLR, we have taken the peak frequency of diffuse photon distribution of the BLR as $\nu_{\rm o}^{*}$. In Figures \ref{fig:SED-1} $-$ \ref{fig:SED-4}, the $\nu_{\rm c,KN}^{\rm obs}$ is marked by a vertical red-dashed line, where the value of $\nu_{\rm o}^{*}$ is obtained from the SED modeling, and is presented in the last third column of Table \ref{table_linfit_SED} in Appendix.   

We have adopted a linear function $\xi(x)=\kappa x+\rho$ and performed the fitting to the spectra in the frequency intervals \big[$\nu_{\rm br}^{\rm syn}$, $\nu_{\rm max}^{\rm syn}$\big] and \big[$\nu_{\rm BLR,br}^{\rm ERC}$, $\nu_{\rm c,KN}^{\rm ERC}$\big]. The corresponding spectral slopes are $\kappa_{\rm 1}$ and $\kappa_{\rm 2}$, and the intercepts are designated by $\rho_{\rm 1}$ and $\rho_{\rm 2}$, respectively. Here, we use $\nu_{\rm c,KN}^{\rm obs}$ instead of $\nu_{\rm BLR,max}^{\rm ERC}$, because the latter is well in the KN regime. Furthermore, the linear fitting is also applied to the straight-line part of the ERC-DT spectrum, and the obtained spectral slope and intercept are represented as $\kappa_{\rm 3}$ and $\rho_{\rm 3}$, respectively. The fitting curves are plotted as the dotted, dashed and dot-dashed lines in Figures \ref{fig:SED-1} $-$ \ref{fig:SED-4}, the corresponding frequency intervals, along with the fitting results are presented in Table \ref{table_linfit_SED} in Appendix. We notice that $\kappa_{\rm 3}$ and $\kappa_{\rm 1}$ have very close values, but $\kappa_{\rm 2}$ shows a relatively significant deviation. This suggests that the ERC-DT process takes place well within the Thomson regime, while the scatterings of electron population in the range \big[$\gamma_{\rm br}$, $\gamma_{\rm max}$\big] off the BLR photons have already entered the KN regime. For the spectrum, with frequencies less than $\nu_{\rm br}^{\rm syn}$, in view of the synchrotron self-absorption, the linear fitting is not performed, and we merely give the spectral slope $-(s_{\rm 1}-3.0)/2$ from the $\delta$-approximation, which determines the rising trend of the spectrum within the frequency interval \big[$\nu_{\rm BLR,min}^{\rm ERC}$, $\nu_{\rm BLR,br}^{\rm ERC}$\big]. 

\subsection{The location of $\gamma$-ray emitting region}
The inner radius $R_{\rm BLR,in}$ of the BLR is determined according to Equation (\ref{eq:R-BLR-in}), where $\tau_{\rm BLR}$ and $R_{\rm BLR,out}$ have been fixed based on the aforementioned considerations. Consequently, its energy density $u_{\rm BLR}$ merely relies on the disk luminosity. Once $L_{\rm d}$ is obtained, $u_{\rm BLR}$ can be roughly determined. In Figure \ref{fig:u_BLR_disk_RF} left panel, we present $u_{\rm BLR}$ for the cases of Ab-10(A), Pr-19(B) and Pr-19(C), where the disk luminosities, $L_{\rm d}=8.7\times 10^{44}$\,erg~s$^{-1}$, $2.9\times 10^{45}$\,erg~s$^{-1}$, $7.2\times 10^{45}$\,erg~s$^{-1}$, represent the lower, intermediate, as well as the upper limits, respectively. Meanwhile, we also show the range of variation of $R_{\rm BLR,out}$ throughout $R_{\rm BLR,out}^{\rm LL}=1.87\times 10^{17}$\,cm and $R_{\rm BLR,out}^{\rm UL}=5.36\times 10^{17}$\,cm. It is clear that $u_{\rm BLR}$ remains approximately at $2.69\times 10^{-2}$\,erg $\rm cm^{-3}$ within the BLR and then decreases roughly in power-law form beyond $R_{\rm BLR,out}$. From Table \ref{tab:Der-quan-sed} and Figure \ref{fig:u_BLR_disk_RF} (left panel), we can see that $R_{\rm H}$ is located at right hand of $R_{\rm BLR,out}$ for all $\gamma$-ray emitting regions. In Figure \ref{fig:u_BLR_disk_RF}, the dashed lines represent the approximation of the $u_{\rm BLR}$ distribution by using the following function
\begin{equation} \label{eq:u_BLR_disk_RF}
u_{\rm BLR}(R_{\rm z})\simeq \frac{\xi_{\rm BLR}L_{\rm d}}{4\pi c R_{\rm BLR,out}^{2}}\frac{1}{1+(R_{\rm z}/R_{\rm BLR,out})^{\beta_{\rm ext}}}.
\end{equation}
We found only when $\beta_{\rm ext}$ takes 2.5, Equation (\ref{eq:u_BLR_disk_RF}) can provide a good approximation to the accurate numerical result (i.e., solid lines). We also note that numerical results significantly deviate from the approximation in the interval from $R_{\rm BLR,in}$ to $R_{\rm BLR,out}$, where the former shows a notable bulge, while the latter gives a rather flat transition. Such that if the $\gamma$-ray emitting region is located well beyond the BLR, Equation (\ref{eq:u_BLR_disk_RF}) can be used to characterize the energy density of the BLR in the stationary frame. Here we emphasize that as the thickness of the BLR increases, the deviation of this approximation from the accurate numerical result becomes more significant. Thus, Equation (\ref{eq:u_BLR_disk_RF}) is effective only for a relatively thin BLR structure.

In FSRQs, due to the occurring of photon-annihilation processes, the observed high-energy $\gamma$-rays are closely related to the BLR radiation strength and its structure \citep[][]{2006ApJ...653.1089L,2014PASJ...66....7L}. For a spherical BLR structure, we can roughly estimate the energy of escaped $\gamma$-rays by assuming $\gamma$-ray emitting region is located inside the BLR and the diffuse photons of the BLR peak at $E_{\rm 0}$. A $\gamma$-ray photon with the energy $E_{\gamma}$, passing through the BLR radiation field with the photon column density $N_{\rm e}=u_{\rm BLR}/E_{\rm 0}$, must satisfy the condition $\tau(E_{\gamma},E_{\rm 0})=\tau_{\rm T}\sigma_{\gamma\gamma}(s)/\sigma_{\rm T}\lesssim 1$ to escape. Here, $\tau_{\rm T}=110(L_{\rm d,45}/R_{\rm BLR,in,17})(10\,\rm eV/E_{\rm 0})$ \citep{2010ApJ...717L.118P}, and $R_{\rm BLR,in,17}$ is the inner radius of the BLR in units of $10^{17}$\,cm. This constraint implies that photons with energies above tens of GeV cannot escape the BLR, with a threshold energy $E_{\rm th}\simeq 26(10\,\rm eV/E_{\rm 0})$\,GeV. The actual optical depth $\tau_{\rm BLR}$ has been calculated numerically and is presented in Figure \ref{fig:u_BLR_disk_RF} right panel, where three cases, $L_{\rm d}=8.7\times 10^{44}$\,erg~s$^{-1}$, $2.9\times 10^{45}$\,erg~s$^{-1}$, $7.2\times 10^{45}$\,erg~s$^{-1}$, are plotted in different colors, whereas each $L_{\rm d}$ corresponds to three locations again, i.e., $R_{\rm H}=1.0\times 10^{-3}$\,pc, $1.0\times 10^{-2}$\,pc and $1.0\times 10^{-1}$\,pc. In the panel, the critical frequency, $\nu_{\rm c,KN}^{\rm obs}=23.7$\,GeV, is also plotted in vertical red-dashed line. Since the value of $\nu_{\rm o,15}^{*}$ falls within the range [0.97, 2.06], thus the $\gamma$-ray photons with energies from $\sim 11.5$\,GeV to $\sim 24.4$\,GeV can be observed in different flaring states. From $\log \tau_{\rm BLR}$ versus $E_{\gamma}$ distribution, it is evident that the optical depth $\tau_{\rm BLR}$ critically depends on $L_{\rm d}$ and $R_{\rm H}$. As $L_{\rm d}$ increasing, the inner radius $R_{\rm BLR,in}$ of the BLR enlarges, i.e., $R_{\rm BLR,in}\varpropto L_{\rm d}^{1/2}$, this leads the high-energy photons generated at larger distance to be affected by $\gamma$-ray absorption. In addition, for a fixed $L_{\rm d}$, a decrease in $R_{\rm H}$ will cause high-energy photons to suffer from more serious absorption. Utimately, the changes in both $L_{\rm d}$ and $R_{\rm H}$ determine the systematic distribution of these curves, as shown in the panel. Remarkably, if $\gamma$-ray emitting region is located inside the BLR, only the $\gamma$-ray photons with the lower energy can escape the BLR and be observed. However, seeing from Figures \ref{fig:SED-1} $-$ \ref{fig:SED-4}, the higher energy of $\gamma$-ray photons have been detected, implying that the $\gamma$-ray emitting regions are located at greater distance. This agrees well with our SED modelings, by which the best fitting is obtained and some derived quantities are given in Table \ref{tab:Der-quan-sed}, The table suggests that 6 out of 12 SEDs are associated with the $\gamma$-ray emitting region near the outer boundary of the BLR, while the remaining 6 SEDs have $\gamma$-ray emitting region well beyond the BLR, i.e., $R_{\rm H}/R_{\rm BLR,out}\geq1$. However, all $\gamma$-ray emitting regions are located well within the DT. 

\begin{figure*}
\hbox{
\includegraphics[width=0.5\textwidth]{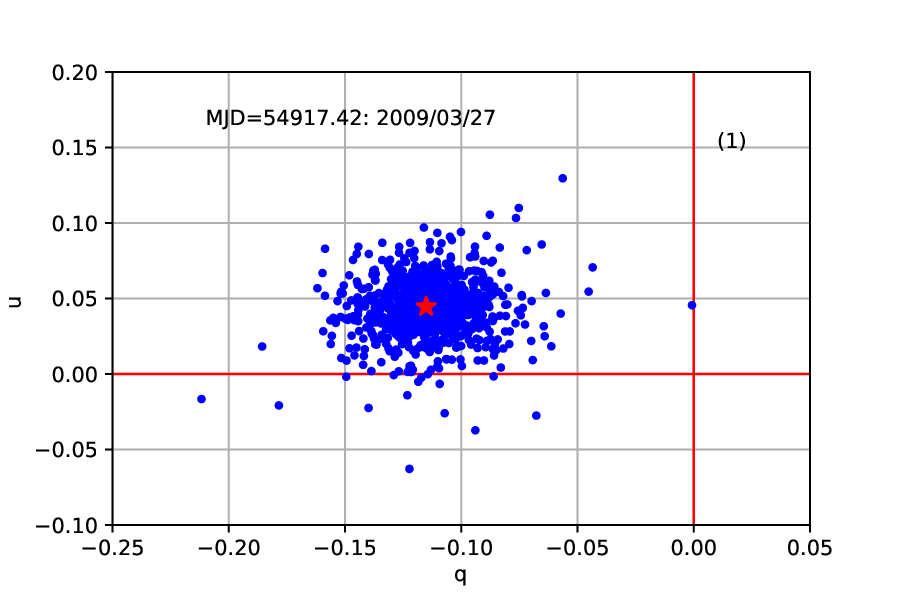}
\includegraphics[width=0.5\textwidth]{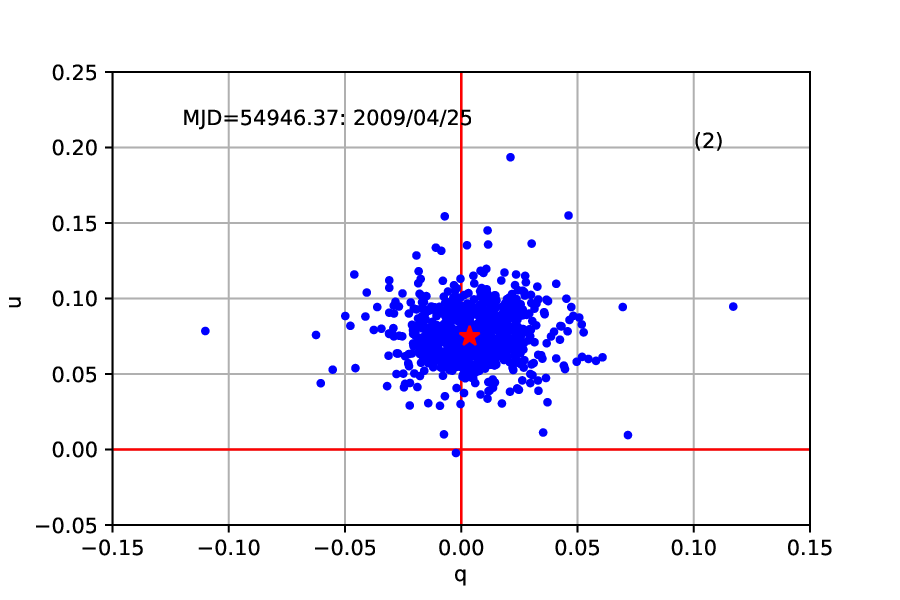}
}
\hbox{
\includegraphics[width=0.5\textwidth]{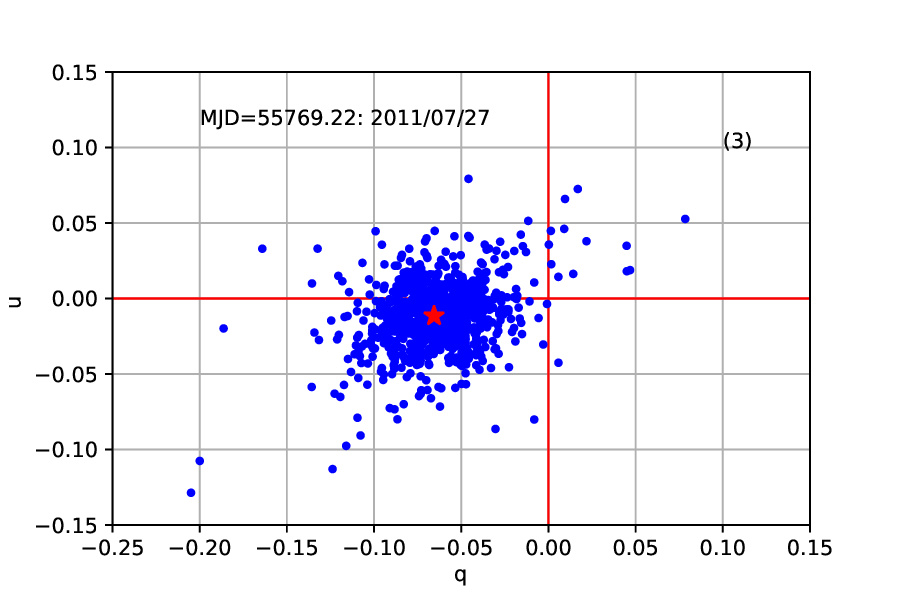}
\includegraphics[width=0.5\textwidth]{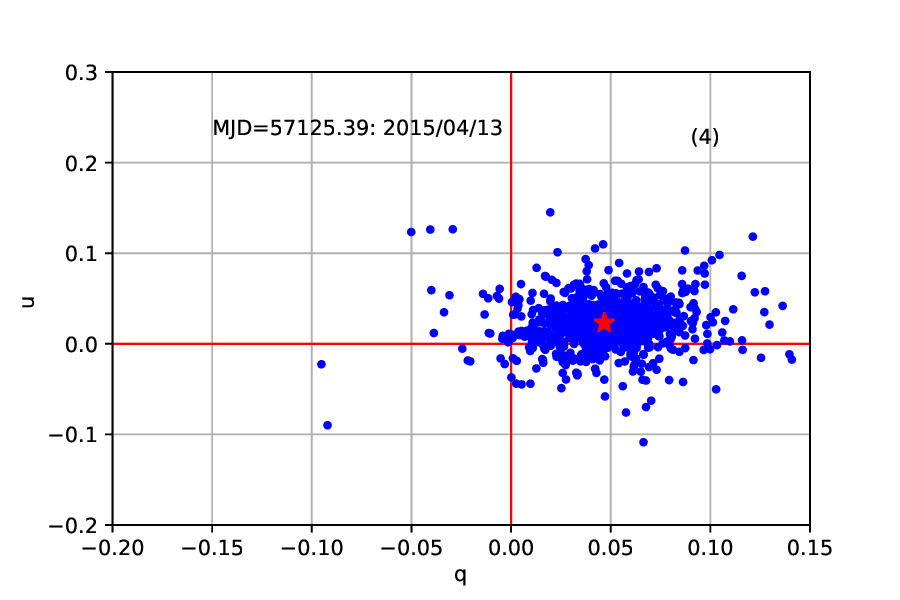}
}
\hbox{
\includegraphics[width=0.5\textwidth]{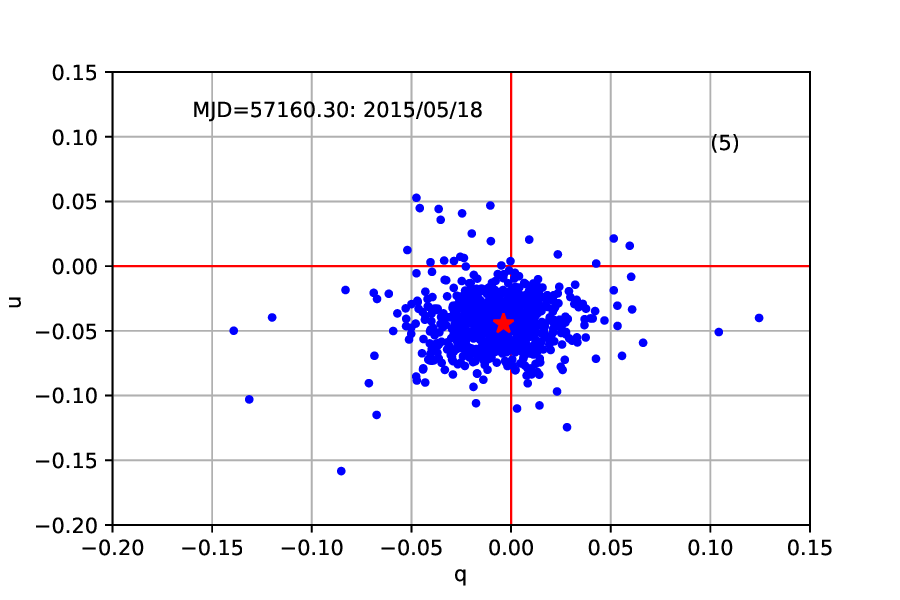}
\includegraphics[width=0.5\textwidth]{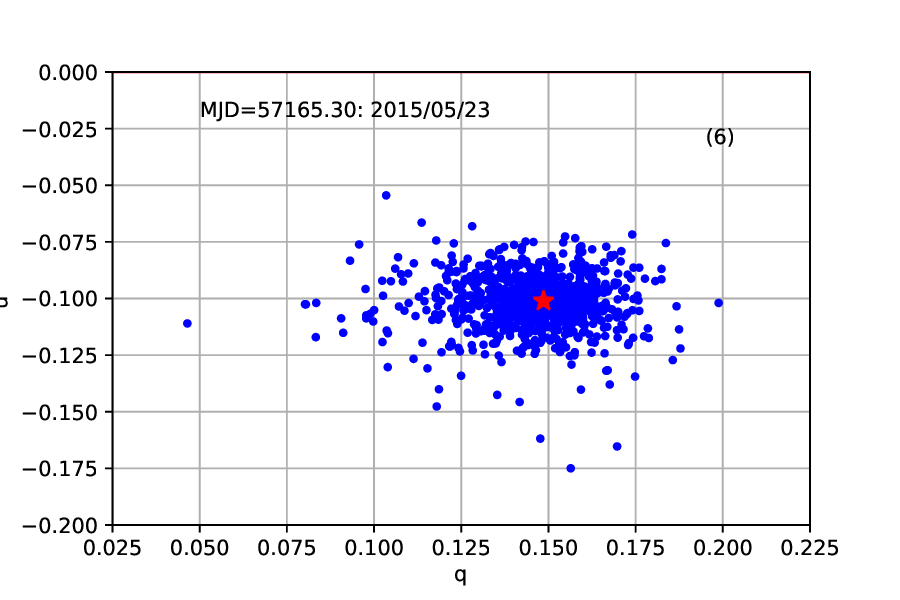}
}
\caption{Distribution plane of normalized Stokes parameters u and q, in which the red star represents the median point of u and q distributions, and the red cross corresponds to u=0 and q =0.}
\label{fig:uq_dis_1}
\end{figure*}

\begin{figure}
\hbox{
\includegraphics[width=0.5\textwidth]{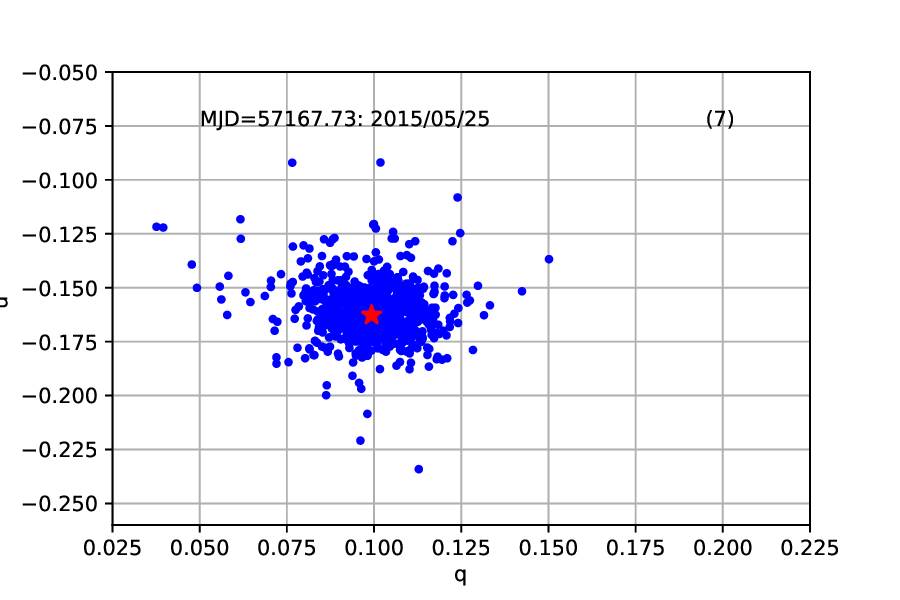}
}
\hbox{
\includegraphics[width=0.5\textwidth]{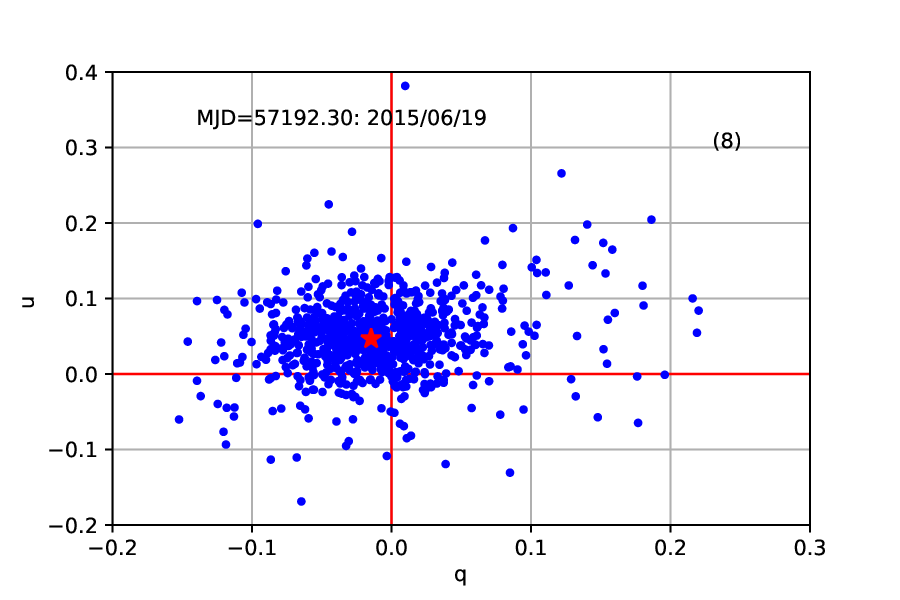}
}
\hbox{
\includegraphics[width=0.5\textwidth]{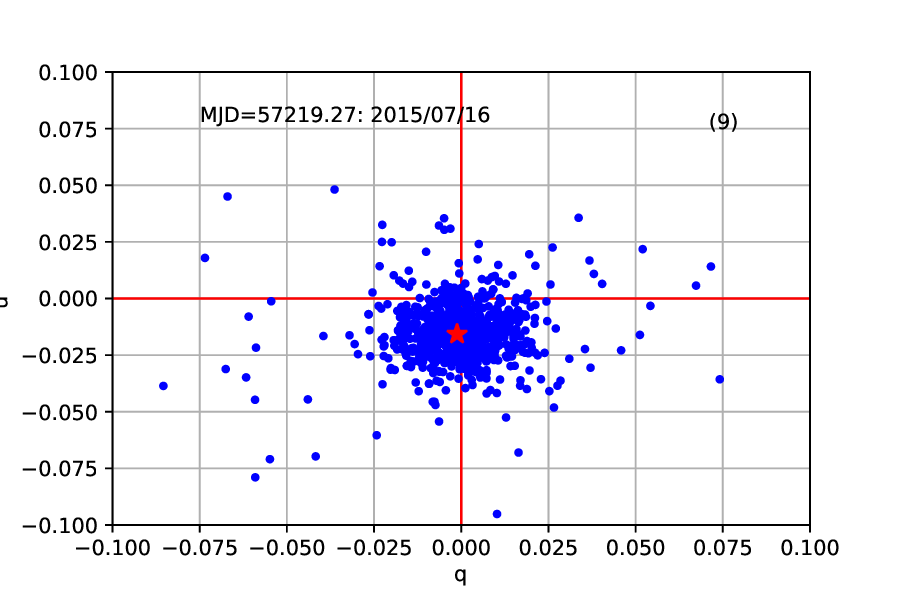}
}

\caption{Distribution plane of normalized Stokes parameters u and q, in which the red star represents the median point of u and q distributions, and the red cross corresponds to u=0 and q =0.}
\label{fig:uq_dis_2}
\end{figure}

\subsection{Implications from polarization observations} \label{sec:}
Seeing from the correspondence given in Table \ref{tab:table-uqQ}, the higher optical polarization degrees are associated with three SEDs, that is, Ab-10(B), Ah-17(B) and Pr-19(B), with corresponding Q values are 12.34\,\%, 17.98\,\% and 19.06\,\%, respectively. As illustrated in the top and bottom panels of Figure \ref{fig:Q-Time}, the source was in a high polarization state during these SED observations. This elevated polarization could be influenced by the underlying acceleration processes responsible for the high polarization states. Although there are some discrepancies in both spectropolarimetric and SED fluxes, our SED modeling clearly indicates that the non-thermal emissions play a dominant role at optical band, except for Na-12(A) and Na-12(B), where the AD contributions are nearly equal and slightly higher, respectively. Additionally, a minimum polarization degree of 1.57\,\% was observed for Pr-19(C) on 2015 July 16, when spectropolarimetric and SED observations reached nearly the same flux level. This suggests that during this episode, the polarization degree may well reflect the contribution of the synchrotron relative to thermal emissions. This is consistent with the SED modeling shown in Figure \ref{fig:SED-4}, where the synchrotron flux is slightly higher than the AD emission. Combining the Q values and the SED modeling results, it is evident that higher polarization degree correspond to higher contribution from synchrotron emission. In scenario where model parameters are degenerate, the spectropolarimetric observations can serve as an efficient auxiliary to unveil the origin of broadband emissions. As previously shown, although Figures \ref{fig:SED-2} and \ref{fig:SED-55} provide satisfactory fits to the SED data set of Na-12(A), but they have significantly different polarization properties at optical frequencies. Specifically, Figure \ref{fig:SED-2} predicts a higher polarization degree due to an extended upward synchrotron spectrum, indicating a higher ratio of non-thermal to thermal contributions, and is consistent with the spectropolarimetric observation with a polarization degree approximately $6.66$\,\%. Conversely, Figure \ref{fig:SED-55} indicates a lower polarization degree due to the lower cutoff of the electron spectrum. It is evident that the model parameters used in Figure \ref{fig:SED-2} are more favorable from the point of polarization. A similar scenario may apply to Na-12(A), given the very similar spectral shape in this energy band. 

For the 12 SEDs studied in this paper, nine of them have corresponding spectropolarimetric observations. The associated optical spectra generally display a flat variation trend with wavelength, except at spectral boundaries, where the edge noise likely becomes more pronounced. However, when we focus only on the intermediate wavelength range, i.e., 4300 $-$ 7500~\AA\,, a slight increase in polarization degree toward longer wavelengths is observed. This trend aligns with the presence of the AD emissions, which dilute the non-thermal jet emssions and reduce the polarization levels. A comprehensive investigation into the composition of optical spectrum requires more sophisticated model setups \citep{2022ApJ...925..139S,2023ApJ...952L..38A}, capable of disentangling the composite spectra and quantifying individual contributions through detalied analysis of polarization data, this will be addressed in forthcoming work. 

In Section \ref{sec:part-5}, we have calculated the average polarization degree $P$ for both 4300 $-$ 7500~\AA\ and 5000 $-$ 7000~\AA\, ranges. The results revealed two notably different $P$ values for each spectral polarization distribution. However, given the strong dependence of these polarization degrees on the selected wavelength range, as well as the relatively stable spectral polarization distribution within 5000 $-$ 7000~\AA\,, it is appropriate to adopt the polarization degree Q, derived from the medians of the normalized Stokes parameters u and q (provided in Table \ref{tab:table-uqQ}), as the representative optical polarization degree. This approach minimizes sensitivity to outliers caused by edge noise and better reflects the typical polarization behavior. Determining the polarization degree will help elucidate the magnetic properties of the $\gamma$-ray emitting region. For illustration, we use Pr-19(B) and Pr-19(C) as examples, as they exhibit the highest and lowest optical polarization degrees, respectively. During Pr-19(B) observation, the source was in a flaring state with relatively high optical flux, and the corresponding polarization degree was $Q\simeq 19$\,\%. Such a high value suggests that the emission region was dominated by a more orderly magnetic field. In contrast, during Pr-19(C) observation, the optical flux declined, and polarization degree Q dropped to $\sim 1.57$\,\%. This decrease is likely attributed to a change in magnetic configuration caused by underlying acceleration processes, resulting in a more tangled magnetic field permeating the emission region. 

On the other hand, by applying the normalized Stokes parameters u and q to construct the u$-$q plane, as shown in Figures \ref{fig:uq_dis_1} and \ref{fig:uq_dis_2}, the red star represents the position of the median values ($u_{\rm c}$ and $q_{\rm c}$). Discarding some points with large offset from the red star, a clear clustering of distribution is observed. The distribution of the data points significally deviates from the central point ($q=0$ and $u=0$) in the same direction. Specifically, in panels (2) and (8), the data cluster is overall distributed above the $u=0$ axis, and the red star is very close to the $q=0$ axis. In contrast, panels (3), (5) and (9) show data clusters roughly located either in the negative and positive direction of the u=0 axis or below it. In panel (1), the data cluster are overall distributed in second quadrant, while in panel (6) and (7), the data sets concentrate in the fourth quadrant. These asymmetrical distributions of u and q around the central point could be associated with variable components related to the propagation of shocks within the jet \citep{1984MNRAS.211..497H,2008ApJ...672...40H,2019ApJ...881..125D}.

\begin{deluxetable*}{cccccccccccccccc}
\tablenum{4}
\tablecaption{Derived quantities obtained through SED modeling \label{tab:Der-quan-sed}}
\tablewidth{0pt}
\tabletypesize{\small}
\tablehead{
 \colhead{Parameters} &  \colhead{Ab-10} &  \colhead{Ab-10} &  \colhead{Ab-10} &  \colhead{Na-12} &  \colhead{Na-12} & \colhead{Ah-17}  & \colhead{Ah-17} & \colhead{Pr-19} & \colhead{Pr-19} & \colhead{Pr-19} & \colhead{Pr-19} & \colhead{Pr-19}\\
\colhead{}  & \colhead{(A)}  & \colhead{(B)} &   \colhead{(C)} &    \colhead{(A)} & \colhead{(B)} & \colhead{(A)} &  \colhead{(B)} & \colhead{(A)} &   \colhead{(B)} &   \colhead{(Q2)} &   \colhead{(C)}&   \colhead{(D)}
}
\startdata
$U_{\rm e}^{'}$($\rm erg~s^{-1}$)  & 1.05e-2  & 9.48e-3  & 9.62e-3 & 2.71e-2  &  2.64e-2  & 2.37e-2  & 2.90e-2  & 3.68e-2   & 3.98e-2  & 4.15e-2  &  3.34e-2  &7.38e-2\\
$U_{\rm B}^{'}$($\rm erg~s^{-1}$)  & 2.55e-2  & 4.81e-2  & 1.02e-1 & 5.45e-3  &  5.16e-3  & 1.29e-2  & 2.48e-2  & 4.81e-2   & 6.72e-2  & 3.98e-2  & 1.93e-1  &6.37e-3\\
$\eta_{\rm eB}(U_{\rm e}/U_{\rm B})$  & 0.41  & 0.20  & 0.09 & 4.98  &  5.12  & 1.84  &  1.17 &  0.76  & 0.59 & 1.04   &  0.17  &11.60\\
$P_{\rm rad}(\rm erg~s^{-1})$  & 6.31e44  & 9.32e44  & 7.46e44 & 9.39e44  & 3.76e44   & 1.73e45  & 9.77e44  &   8.21e44 & 1.08e45  &  4.50e44 & 7.77e44   &7.46e44\\
$P_{\rm kin}(\rm erg~s^{-1})$  & 1.19e46  & 1.17e46  & 1.24e46 & 3.02e46  & 3.48e46   & 2.67e46  & 2.23e46  &   1.83e46 & 1.94e46  & 2.23e46  &  1.68e46  &3.47e46\\
$P_{\rm e}(\rm erg~s^{-1})$  & 4.47e44  & 4.42e44  & 4.49e44 & 1.22e45  &  1.40e45  & 9.26e44  & 1.13e45  &   6.73e44 & 7.28e44  & 8.33e44  &  6.11e44  &1.48e45\\
$P_{\rm p}(\rm erg~s^{-1})$  & 1.15e46  & 1.12e46  & 1.19e46 & 2.90e46  & 3.34e46   & 2.58e46  & 2.11e46  &   1.76e46 & 1.87e46  & 2.14e46  & 1.62e46   &3.32e46\\
$P_{\rm B}(\rm erg~s^{-1})$  & 1.08e45  & 2.24e45  & 4.75e45 & 2.44e44  & 2.73e44   & 5.04e44  & 9.69e44  &   8.81e44 & 1.23e45  & 7.99e44  &  3.52e45  &1.28e44\\
$P_{\rm jet}(\rm erg~s^{-1})$  & 1.37e46  & 1.48e46  & 1.79e46 & 3.14e46  & 3.55e46   & 2.90e46  & 2.42e46  &   2.0e46 & 2.17e46  & 2.35e46  & 2.11e46  &3.56e46\\
$\ell_{\rm Edd}(L_{\rm d}/L_{\rm Edd})$ & 1.28e-2  & 1.77e-2  & 3.24e-2 & 3.98e-2  & 3.68e-2   & 3.98e-2  & 3.98e-2  & 2.06e-2  & 4.27e-2  & 3.98e-2  & 1.06e-1   &6.19e-2\\
$\zeta_{\rm jet}(P_{\rm jet}/L_{\rm Edd})$  & 2.01e-1  & 2.19e-1  & 2.63e-1 & 4.62e-1  & 5.22e-1   & 4.27e-1  &  3.56e-1  &  2.94e-1  & 3.20e-1  & 3.46e-1  &  3.11e-1  &5.24e-1\\
$P_{\rm BP}(\rm erg~s^{-1})$ & 4.24e43  & 8.06e43  & 2.71e44 &4.08e44   &  3.50e44  & 4.08e44  &  4.08e44 &   1.10e44 & 4.71e44  & 4.08e44 & 2.90e45   &9.88e44\\
$P_{\rm BZ}(\rm erg~s^{-1})$ & 1.40e46  & 1.93e46  & 3.54e46 & 4.34e46  &  4.02e46  & 4.34e46  & 4.34e46  &   2.25e46 & 4.66e46 & 4.34e46 &  1.16e47  &6.75e46\\
$\dot{M}_{\rm acc}(\rm M_{\odot}~yr^{-1})$ & 0.18  & 0.25  & 0.47 & 0.57  & 0.53   & 0.57  &  0.57 &  0.30  &  0.61 & 0.57 &  1.52  &0.89\\
$R_{\rm H}/R_{\rm BLR,out}$ & 1.03  & 1.00  & 1.01 & 1.13  & 1.36   & 1.04  & 1.16  &  1.00 & 1.00  & 1.10  &   1.00 &1.20\\
$R_{\rm BLR,in}(\rm cm)$  & 9.33e16  & 1.10e17  & 1.48e17 & 1.64e17  & 1.58e17   & 1.64e17  & 1.64e17  & 1.18e17   & 1.70e17  & 1.64e17  &  2.68e17  &2.05e17\\
\enddata
\end{deluxetable*}

\subsection{Jet Power} \label{sec:jet-power}
We estimated the power carried by the jet in the form of relativistic electrons ($P_{\rm e}$), cold protons ($P_{\rm p}$) and magnetic field ($P_{\rm B}$), according to \citep{2008MNRAS.385..283C}
\begin{eqnarray}
P_{\rm i} &\simeq& \pi R_{\rm b}^{'2}\Gamma_{\rm j}^{2}\beta_{\rm j}c U_{\rm i}^{'},
\end{eqnarray}
where $U_{\rm i}^{'}$ is energy density of the i'th component in the comoving frame. For the jet power contributed by relativistic electrons, it can be calculated by
\begin{eqnarray}\label{eq:Pe}
P_{\rm e} &\simeq& \pi R_{\rm b}^{'2}\Gamma_{\rm j}^{2}\beta_{\rm j}cU_{\rm e}^{'},
\end{eqnarray}
where $U_{\rm e}^{'}\simeq<\gamma>m_{\rm e}c^{2}N_{\rm e}$ is the electron energy density, $N_{\rm e}$ is the total electron number per unit volume. For the electron distribution described by Equation (\ref{eq:1}), as in general case, $s_{\rm 1}$ and $s_{\rm 2}$ are larger than 1. If $\gamma_{\rm max}\gg \gamma_{\rm br}\gg \gamma_{\rm min}$, then $N_{\rm e}$ can be approximated as
\begin{eqnarray}\label{eq:Ne}
N_{\rm e} &=& \int_{\gamma_{\rm min}}^{\gamma_{\rm max}}n_{\rm e}^{'}(\gamma)d\gamma
\nonumber\\
&\simeq& n_{\rm 0}\Big[\frac{1}{s_{\rm 1}-1}\gamma_{\rm min}^{1-s_{\rm 1}}+\frac{1}{s_{\rm 2}-1}\gamma_{\rm br}^{1-s_{\rm 1}}\Big]
\nonumber\\
&\approx& \frac{n_{\rm 0}}{s_{\rm 1}-1}\gamma_{\rm min}^{1-s_{\rm 1}}.
\end{eqnarray}
Apparently, the lower energy of the electrons will dominate the kinematic energy of the jet. Similarly, the jet power contributed by cold protrons is given by
\begin{eqnarray}
P_{\rm p} &\simeq& \pi R_{\rm b}^{'2}\Gamma_{\rm j}^{2}\beta_{\rm j}cU_{\rm p}^{'},
\end{eqnarray}
where $U_{\rm p}^{'}\simeq <N_{\rm p}>m_{\rm p}c^{2}$ is the cold proton energy density, $<N_{\rm p}>$ is the total proton number per unit volume, $m_{\rm p}$ is the rest mass of proton. In calculation to $P_{\rm p}$, one protron per ten electrons has been assumed \citep{2014Natur.515..376G}. The jet power carried by Poynting flux is given as
\begin{eqnarray}
P_{\rm B} &\simeq& \pi R_{\rm b}^{'2}\Gamma_{\rm j}^{2}\beta_{\rm j}cU_{\rm B}^{'},
\end{eqnarray}
where $U_{\rm B}^{'}=B^{'2}/8\pi$ is the magnetic energy density. 

In the model, both the radiated power ($P_{\rm rad}$) and kinetic power ($P_{\rm kin}$) of the jet are also taken into account, in which $P_{\rm rad}$ can be approximated as \citep{2010MNRAS.402..497G}
\begin{eqnarray}
P_{\rm rad} &\simeq& L^{'}\frac{\Gamma_{\rm j}^{2}}{4}=L\frac{\Gamma_{\rm j}^{2}}{4\delta_{\rm D}^{4}},
\end{eqnarray}
here the total non-thermal radiation luminosity in the comoving frame is related to that in the observer's frame via $L=\delta_{\rm D}^{4}L^{'}$, where $L^{'}\simeq 4\pi r^{2}c U_{\rm rad}^{'}$, $U_{\rm rad}^{'}$ is the enegy density of the total non-thermal radiation and $r$ is the spread distance of the radiation from the emitting region. For the kinematic power, governed by both the relativistic electrons and the cold protons, it can be determined by $P_{\rm kin}\simeq P_{\rm e}+P_{\rm p}$. At this point, it can provide an estimation of the total jet power, $P_{\rm jet}$, through summing up $P_{\rm e}$, $P_{\rm p}$, $P_{\rm B}$ and $P_{\rm rad}$.

In addition, we calculate the equipartition parameter as the ratio of the electron's and magnetic energy densities, $\eta_{\rm eB}\equiv U_{\rm e}^{'}/U_{\rm B}^{'}$. Generally, the equipartition parameter can be regarded as an indicator to the particle acceleration mechanism. Specifically, the relativistic electrons may originate from the magnetic reconnection if $\eta_{\rm eB}\ll 1$ or from the Fermi-type acceleration processes if $\eta_{\rm eB}>1$. Using the best-fitting model parameters, $\eta_{\rm eB}$ has been calculated and is presented in Table \ref{tab:Der-quan-sed}. We note that 6 out of 12 SED modelings have $\eta_{\rm eB}\ge 1$, with a large deviation observed in Pr-19(D), where $\eta_{\rm eB} \simeq 11.6$. Overall, the $\gamma$-ray emitting regions are matter-dominated (electrons and cold protons). Given the large $\eta_{\rm eB}$ and $R_{\rm H}$, along with the power-law electron distributions, this could imply that the diffusive shock accelerations are responsible for the flaring states of PKS\,1510$-$089. These flaring states likely originate from the interactions either between the intermittent jetted matter or between the jet and ambient matter. Based on the internal shock scenario, \cite{2001MNRAS.325.1559S} gave a lower limit to the distance at which the collisions take place, i.e.,
\begin{eqnarray}
R_{\rm min} &\simeq& \frac{2\alpha_{\Gamma}^{2}}{\alpha_{\Gamma}^{2}-1}\Gamma_{\rm m}^{2}ct_{\rm v,min},
\end{eqnarray}
where $\alpha_{\Gamma}=\Gamma_{\rm M}/\Gamma_{\rm m}$, $\Gamma_{\rm M}$ and $\Gamma_{\rm m}$ are the maximum and minimum bulk Lorentz factors of the colliding shells, respectively. Assuming $\Gamma_{\rm M}=25$ and $\Gamma_{\rm m}=10$, for the parameters in Table \ref{tab:mod-parameters}, the value of $R_{\rm min}$ is on the order of $\sim 10^{17}$\,cm, which is very close to $R_{\rm H}$ obtained through our SED modeling.  

Table \ref{tab:Der-quan-sed} presents the total jet power, $P_{\rm jet}$, which constitutes a significant fraction of the Eddington luminosity, $L_{\rm Edd}=4\pi G M_{\rm BH}m_{\rm p}c/\sigma_{\rm T}\simeq 6.79\times 10^{46}$\,erg~s$^{-1}$ for BH mass of $5.4\times 10^{8}$\,$M_{\odot}$. Our model suggests that $\zeta_{\rm jet}\equiv P_{\rm jet}/L_{\rm Edd}\simeq 0.2\sim 0.52$. Using $L_{\rm d}$ and $L_{\rm Edd}$, the accretion rate of the BH can be estimated as $\dot{m}_{\rm acc}\simeq 1.3\times 10^{-2}\sim 0.11$, which are typical values for FSRQs. Moreover, the model calculates the total radiative power $P_{\rm rad}$, ranging from $\sim 3.8\times 10^{44}$ to $\sim 1.7\times 10^{45}$\,erg s$^{-1}$. From our calculations, it is clear that the radiative power is only a minor fraction of the total jet power, with the ratio ranging from $\sim1.1$\% to $\sim6.3$\%. At the same time, we also note that $P_{\rm jet}$ is larger than $L_{\rm d}$ by a factor of $\sim 2.9$ to $\sim 15.8$. This is possible if there exists an efficient mechanism to extract energy from the central spinning BH \citep{2010MNRAS.402..497G}.

To explore the production mechanism of the relativistic jets, the maximal jet powers that can be provided through BZ and BP mechanisms are also given in Table \ref{tab:Der-quan-sed}. In comparision of $P_{\rm jet}$ with $P_{\rm BZ}$ and $P_{\rm BP}$, a clear conclusion can be drawn that the BZ mechanism is a promising mechanism at work in the central engine of PKS\,1510$-$089, whereas the energy extracted by the BP mechanism is insufficient to power the jet activity. Here, we note that $P_{\rm jet}$ constitutes a substantial part of $P_{\rm BZ}$, suggesting a higher conversion efficiency from accretion of the matter to the jet power. This is consistent with the investigation of radio-loud AGNs, where the BZ mechanism may dominate over the BP mechanism for jet acceleration \citep{2012ApJ...759..114C}. Moreover, \cite{2012ApJ...759..114C} showed that the faster-moving jets are magnetically accelerated by the magnetic fields threading the horizon of more rapidly rotating black holes, which is consistent with the observations that the apparent speed of the jet of PKS\,1510$-$089 is 20\,c, and can even reach up to $\sim 46$\,c \citep{2002ApJ...580..742H,2005AJ....130.1418J}.

\section{Conclusions} \label{sec:part-7}
PKS\,1510$-$089 is a bright FSRQ-type $\gamma$-ray source and shows the extreme activity of nearly acrossing entire electromagnetic spectrum. Thus, this source has been the target of extensive multiwavelength observations, making it one of the best-studied sources. To date, multiple SED observations have been presented. Among FSRQs, PKS\,1510$-$089 is one of the most unique sources, exhibiting a notable BBB hump. This provides a excellent opportunity to explore the coupling between the activity of the relativistic jet and the central engine, offering further insight into the origin of multiwavelength emissions.

In this paper, we collected 12 SEDs spanning four flaring periods from 2008 to 2015, and reproduced them using one-zone homogeneous leptonic model, which appropriately accounts for the external radiation fields from the BLR and DT. By combining the SED modeling and spectropolarimetric observations, the following conclusions can be drawn:
\begin{enumerate}
\item The distribution of the SED data sets indicates that the peak location of the BBB changes in different flaring states. However, we do not expect the BH  to change measurably within the time span of these SED observations ($\sim 7$ years). Therefore, the BH mass is considered constant and fixed at $5.4\times 10^{8}$\,$M_{\odot}$ during all the observational epoches, but allowing $R_{\rm ISO}$ to vary. Then the fitting to optical/UV data sets requires that $R_{\rm ISO}$ varies between $3R_{\rm S}$ to $18R_{\rm S}$. If this is true, it implies that the energy extraction of the relativistic jet could influence $R_{\rm ISO}$, causing it to increase in size. Here we emphasize that the AD with a truncated innermost stable orbit is likely to occur under certain conditions, e.g., when it orbits around a highly spinning BH. On the other hand, if we apply a canonical AD with $R_{\rm ISO}=6R_{\rm g}$ and $\eta_{\rm f}=1/12$ to fit the optical/UV data of 12 SEDs, the obtained BH mass spans a large range, from $5.4\times 10^{8}M_{\odot}$ to $3.3\times 10^{9}M_{\odot}$. Given that the value of $\ell_{\rm Edd}$ ranges from $\sim 1.3\times 10^{-2}$ to $\sim 1.1\times 10^{-1}$ from the SED modelings, such that the peak of the BBB will span a range from $\sim 9.0\times 10^{14}$\,Hz to $\sim 2.6\times 10^{15}$\,Hz. Therefore, to verify the change in the peak of the BBB in different observational epoches, the future observations focusing on the optical/UV band will be critical.

\item In the optical band, the polarization degree as a whole exhibits a flat variation with wavelength, part of them has a slight decreasing trend at shorter wavelengths, this implies that there is likely a contributions from the AD, which is consistent with our SED modelings. On the other hand, the high-energy hump of the SEDs is overall dominated by ERC-BLR process, the ERC-DT process has also contribution, albeit not very significantly, except for Na-12(B), where the ERC-DT fit is above the ERC-BLR at X-ray energies. As for the low-energy end of the X-rays, it may comprise multiple components, including SSC, ERC-DT and corona emissions. Linear fittings suggest that the ERC-BLR spectrum has a steeper slope compared to the synchrotron and ERC-DT ones, the last two components have the nearly identical spectral slopes. This indicates that the ERC-BLR process occurs in the KN regime, while the ERC-DT process takes place in the Thomson regime. Our SED modeling also considers the ERC-AD process, but due to the large distance of $\gamma$-ray emitting region, it contributes almost negligible to the observed high-energy emission.

\item The SED modelings of seven multiwavelength observations, i.e., Ab-10(A), Ab-10(B), Ab-10(C), Ah-17(A), Pr-19(A), Pr-19(B) and Pr-19(C), require $\gamma$-ray emitting regions are located in the immediate vicinity of the outer boundary of the BLR, with $R_{\rm H}/R_{\rm BLR,out}\simeq 1$. However, for other four SEDs, i.e., Na-12(A), Na-12(B), Ah-17(B) as well as Pr-19(D), the modeling constrains $\gamma$-ray emitting regions to be well beyond the BLR, with $R_{\rm H}/R_{\rm BLR,out}> 1$. Nevertheless, all $\gamma$-ray emitting regions are located inside the DT. Specifically, the locations of the $\gamma$-ray flare associated with the 12 SEDs are limited to distance of $\sim 6.2\times 10^{-2}$\,pc to $\sim1.7$\,pc from the central engine. At such distances, diffusive shock acceleration is likely the dominant acceleration mechanism, producing high-energy electrons with power-law distribution, as we have been adopted in SED modelings. This scenario is supported by following three aspects: the first is the low magnetic field strength in the $\gamma$-ray emitting region, ranging from 0.36\,G to 1.6\,G, corresponds to a low magnetization parameter $\sigma$, with the minimum $\sigma_{\rm min}\simeq4.7\times 10^{-5}$ and maximum $\sigma_{\rm max}\simeq 6.1\times 10^{-3}$, while an efficient magnetic reconnection process typically occurs in strongly magnetized plasma ($\sigma\gtrsim$ 1). The second is that the $\gamma$-ray emitting regions are very close to the outer boundary of the BLR, likely arising from the internal shock processes. In principle, the dynamics of the relativistic jet are influenced by the inhomogeneous distributions of the jet matter (with significant velocity gradients radially and laterally), the clouds of the BLR, and the radiation pressure from the BLR and DT. Under these conditions, internal shocks occur inside the jet, accelerating the particles to higher energy. Another one comes from the spectropolarimetric observations, which show that the polarization degree at optical band significantly increases over time during the flaring states from 2008 to 2015. This could be attributed to a series of shocks propagating down the relativistic jet, compressing the turbulent magnetic field and resulting in a higher polarization degree.

\item For the studied 12 SEDs, 6 out of them are associated with a $\gamma$-ray emitting region that is matter-dominated (electrons and cold protons), with the equipartition parameter $\eta_{\rm eB}$ varies from 1.04 to 11.6. For the other 6 SEDs, the $\gamma$-ray emitting region has slightly higher magnetic energy density, with $\eta_{\rm eB}$ varying within the range of 0.09 to 0.76. Based on our SED modeling, the X-ray flux is primarily a superposition of SSC, corona, ERC-BLR and ERC-DT emissions, in which a large $B^{'}$ would induce a high SSC flux, thereby elevating the resulting X-ray flux, particularly at low-energy end. Thus, a larger $B^{'}$ would result in a poor fit to the X-ray data. Besides, all the SEDs exhibit a large Compton dominance, implying that a higher electron number density is required. Therefore, a relatively higher matter energy density provides a better fit in the SED modeling. 

\item Our SED modelings show that the energy provided by BP mechanism is on the order of magnitude of $10^{44}$\,erg s$^{-1}$, with the highest value required for Pr-19(C) is $2.9\times 10^{45}$\, erg~s$^{-1}$. In comparision, the energy provided by BZ mechanism is significantly higher by about one to two orders of magnitude, it slightly exceeds the corresponding jet power, which is on the order of magnitude of $10^{46}$\,erg s$^{-1}$. Therefore, the relativistic jet indicates a very high conversion efficiency of the BH spinning energy into the energy that governs the jet dynamics, with the highest value is $\sim 97.9$\% for Ab-10(A) and the lowest value of $\sim 18.2$\% for Pr-19(C). For PKS\,1510$-$089, we conclude that the maximum energy extracted by the BP mechanism is insufficient to sustain the jet activity, and the relativistic jet can be only powered through the BZ mechanism. Moreover, since a high spin $a=0.95$ is required to power the jet, this implies that such high conversion efficiency is likely related to a highly spinning BH. 
\end{enumerate}

\begin{acknowledgments}

We thank the journal referee for insightful and constructive comments, which helped to improve the manuscript significantly. We also sincerely thank Andrea Tramacere, Julian Sitarek, Krzysztof Nalewajko, Raj Prince for sending us the multiwavelength data. This work is supported by the National Natural Science
Foundation of China (NSFC-12463004), The Guangdong Major Project of Basic and Applied
Basic Research (grant No. 2024A1515013169), the Growth project of young scientific and technological talents in Colleges and universities in Guizhou Province (Qianjiaohe$-$KY$-$Zi[2020]221), as well as the Scientiﬁc Research Foundation for Doctoral Program of Minzu Normal University of Xingyi (20XYBS16).

\end{acknowledgments}

\appendix

\section{The $\nu F_{\nu}$ flux of the synchrotron emission in $\delta$ approximation}
\label{sec:Flux-syn}
The synchrotron emissivity $\epsilon_{\rm syn}^{'}(\nu^{'})$ (erg~cm$^{-3}$~s$^{-1}$~sr$^{-1}$~Hz$^{-1}$), for a relativistic electron population $N^{'}(\gamma)$, can be approximately given by using a $\delta$-function $\delta(\nu^{'}-\gamma^{2}\nu_{\rm L})$ as \citep{2012MNRAS.419.1660S}
 \begin{eqnarray}
\epsilon_{\rm syn}^{'}(\nu^{'}) &\simeq&  \frac{\sigma_{\rm T}cB^{'2}}{48\pi^{2}}\nu_{\rm L}^{-3/2}~ N^{'}\bigg(\sqrt{\frac{\nu^{'}}{\nu_{\rm L}}}\bigg)~\nu^{'1/2},
\end{eqnarray}
where $\nu_{\rm L}=eB^{'}/(2\pi m_{\rm e}c)$ is the Larmor frequency. Then the observed synchrotron flux can be determined as follows,
\begin{eqnarray}
F_{\rm syn}^{\rm obs}(\nu) &=& \frac{\delta_{\rm D}^{3}(1+z)}{d_{\rm L}^{2}}~V^{'}~\epsilon_{\rm syn}^{'}\bigg[\frac{1+z}{\delta_{\rm D}}\nu\bigg],
\nonumber\\
&=& (1+z) \frac{\sigma_{\rm T}c}{48\pi^{2}d_{\rm L}^{2}}~\big(\delta_{\rm D}^{3}B^{'2}\big)~\nu_{\rm L}^{-3/2}~ V^{'} ~ N^{'}\bigg(\sqrt{\frac{\nu^{'}}{\nu_{\rm L}}}\bigg)~\nu^{'1/2},
\end{eqnarray}
where $V^{'}$ is the volume of $\gamma$-ray emitting region. In present paper, a spherical emitting region is adopted, $V^{'}=(4\pi/3) R_{\rm b}^{'3}$. Subsequently, the $\nu F_{\nu}$ flux can be obtained by 
\begin{eqnarray}
\nu F_{\nu}^{\rm syn} &\equiv& \nu F_{\rm syn}^{\rm obs}(\nu)= \big[(1+z)\nu\big] \frac{\sigma_{\rm T}c}{48\pi^{2}d_{\rm L}^{2}}~\big(B^{'2}\delta_{\rm D}^{3}\big)~\nu_{\rm L}^{-3/2}~ V^{'} ~ N^{'}\bigg(\sqrt{\frac{\nu^{'}}{\nu_{\rm L}}}\bigg)~\nu^{'1/2},
\nonumber\\
&=& \frac{\sigma_{\rm T}c}{48\pi^{2}d_{\rm L}^{2}}~\big(B^{'2}\delta_{\rm D}^{4}\big)~\nu_{\rm L}^{-3/2}~ V^{'} ~ N^{'}\bigg(\sqrt{\frac{\nu^{'}}{\nu_{\rm L}}}\bigg)~\nu^{'3/2}.
\end{eqnarray}
For the electron distribution described by Equation (\ref{eq:1}), this expression is sub-divided into two segments:\\
\begin{equation} \label{eq:B6}
\nu F_{\nu}^{\rm syn} = \left\{\begin{array}{ll}
\frac{\sigma_{\rm T}c}{48\pi^{2}d_{\rm L}^{2}}~\big(n_{\rm 0}B^{'2}\delta_{\rm D}^{4}\big)~\nu_{\rm L}^{-(3-s_{\rm 1})/2}~ V^{'} ~\nu^{'-(s_{\rm 1}-3)/2} & \hspace{5mm}  \gamma_{\rm min}\leq \gamma \leq \gamma_{\rm br}, \\
\frac{\sigma_{\rm T}c}{48\pi^{2}d_{\rm L}^{2}}~\big(n_{\rm 0}B^{'2}\delta_{\rm D}^{4}\big)~\nu_{\rm L}^{-(3-s_{\rm 2})/2}~\gamma_{\rm br}^{(s_{\rm 2}-s_{\rm 1})}~ V^{'} ~\nu^{'-(s_{\rm 2}-3)/2}  & \hspace{5mm} \gamma_{\rm br}<\gamma \leq \gamma_{\rm max}. \\
\end{array} \right.
\end{equation}

\section{The $\nu F_{\nu}$ flux of the ERC emission in $\delta$ approximation}
\label{sec: Flux-ERC}
The ERC emissivity $\epsilon_{\rm ERC}^{'}(\nu^{'})$ (erg~cm$^{-3}$~s$^{-1}$~sr$^{-1}$~Hz$^{-1}$), for scattering an externally isotropic monochromatic radiation field with characteristic frequency $\nu_{\rm o}^{*}$ and energy density $u_{\rm o}^{*}$, can be approximately given by using a $\delta$-function $\delta(\nu^{'}-\gamma^{2}\Gamma_{\rm j}\nu_{\rm o}^{*})$ as \citep{2012MNRAS.419.1660S}
\begin{eqnarray}
\epsilon_{\rm ERC}^{'}(\nu^{'}) &=& \frac{c\sigma_{\rm T}u_{\rm o}^{*}}{8\pi \nu_{\rm o}^{*}}~\bigg(\frac{\Gamma_{\rm j}\nu^{'}}{\nu_{\rm o}^{*}}\bigg)^{1/2}~N^{'}\bigg[\big(\frac{\nu^{'}}{\Gamma_{\rm j}\nu_{\rm o}^{*}}\big)^{1/2}\bigg].
\end{eqnarray}
Then the observed ERC flux can be given as
\begin{eqnarray}
F_{\rm ERC}^{\rm obs}(\nu) &=& \frac{\delta_{\rm D}^{3}(1+z)}{d_{\rm L}^{2}}~V^{'}~\epsilon_{\rm ERC}^{'}(\nu^{'}),
\nonumber\\
&=& (1+z)\frac{c\sigma_{\rm T}\delta_{\rm D}^{3}}{8\pi d_{\rm L}^{2}}~\bigg(\frac{u_{\rm o}^{*}}{\nu_{\rm o}^{*}}\bigg)~\bigg(\frac{\Gamma_{\rm j}\nu^{'}}{\nu_{\rm o}^{*}}\bigg)^{1/2}~N^{'}\bigg[\big(\frac{\nu^{'}}{\Gamma_{\rm j}\nu_{\rm o}^{*}}\big)^{1/2}\bigg].
\end{eqnarray}
Subsequently, the $\nu F_{\nu}$ can be obtained by multiplying $\nu$ to both sides, i.e.,
\begin{eqnarray}
\nu F_{\nu}^{\rm ERC} &\equiv& \nu F_{\rm ERC}^{\rm obs}(\nu)
= (1+z)\nu\frac{c\sigma_{\rm T}\delta_{\rm D}^{3}}{8\pi d_{\rm L}^{2}}~\bigg(\frac{u_{\rm o}^{*}}{\nu_{\rm o}^{*}}\bigg)~\bigg(\frac{\Gamma_{\rm j}\nu^{'}}{\nu_{\rm o}^{*}}\bigg)^{1/2}~N^{'}\bigg[\big(\frac{\nu^{'}}{\Gamma_{\rm j}\nu_{\rm o}^{*}}\big)^{1/2}\bigg],
\nonumber\\
&=& \frac{c\sigma_{\rm T}}{8\pi d_{\rm L}^{2}}~(u_{\rm o}^{*}\nu_{\rm o}^{*-3/2}\delta_{\rm D}^{4}\Gamma_{\rm j}^{1/2})~N^{'}\bigg[\big(\frac{\nu^{'}}{\Gamma_{\rm j}\nu_{\rm o}^{*}}\big)^{1/2}\bigg]~\nu^{'3/2}.
\end{eqnarray}
For the electron distribution described by Equation (\ref{eq:1}), the same as the synchrotron flux, this expression will be further divided into two segments, i.e.,
\begin{equation} \label{eq:B6}
\nu F_{\nu}^{\rm ERC} = \left\{\begin{array}{ll}
\frac{c\sigma_{\rm T}}{8\pi d_{\rm L}^{2}}~\big[n_{\rm 0}u_{\rm o}^{*}\delta_{\rm D}^{4}\Gamma_{\rm j}^{(1+s_{\rm 1})/2}\big]~\nu_{\rm o}^{*-(3-s_{\rm 1})/2}~V^{'}~\nu^{'-(s_{\rm 1}-3)/2}  & \hspace{5mm}  \gamma_{\rm min}\leq \gamma \leq \gamma_{\rm br}, \\
\frac{c\sigma_{\rm T}}{8\pi d_{\rm L}^{2}}~\big[n_{\rm 0}u_{\rm o}^{*}\delta_{\rm D}^{4}\Gamma_{\rm j}^{(1+s_{\rm 2})/2}\big]~\nu_{\rm o}^{*-(3-s_{\rm 2})/2}~\gamma_{\rm br}^{(s_{\rm 2}-s_{\rm 1})}~V^{'}~\nu^{'-(s_{\rm 2}-3)/2}    & \hspace{5mm} \gamma_{\rm br}<\gamma \leq \gamma_{\rm max}. \\
\end{array} \right.
\end{equation}

\section{The critical frequency between the Thomson and KN regimes}
\label{sec: nu-cri-fre}
In leptonic scenarios, the $\gamma$-rays above the high-energy peak of the SED are produced through Compton scattering of the energetic electron population off the ambient photons, and governed by the KN supression, which causes the spectrum to steepen. Considering a monochromatic radiation field with a characteristic frequency $\nu_{\rm o}^{*}$ in the lab frame, then the frequency of the photon in the comoving frame will be Doppler-boosted via $\nu_{\rm o}^{'}=\Gamma(1+\beta_{\rm j})\nu_{\rm o}^{*}\simeq 2\Gamma_{\rm j}\nu_{\rm o}^{*}$. An electron with the Lorentz factor $\gamma$ will scatter these photons in the Thomson regime only if $h\nu_{\rm o}^{'}<(3/4)m_{\rm e}c^{2}$. However, once the condition $h\nu_{\rm o}^{'}>(3/4)m_{\rm e}c^{2}$ holds, the scattering will occur in the KN regime \citep{2005MNRAS.363..954M,2008MNRAS.386..945T,2009MNRAS.397..985G}. At the critical point, where $\gamma=(3/4)m_{\rm e}c^{2}/(h\nu_{\rm o}^{'})$, the scattered frequency of the photon can be expressed as
\begin{eqnarray}
\nu_{\rm c,KN}^{'} &\simeq& \frac{3}{4}\gamma^{2}\nu_{\rm o}^{'}=\frac{3}{4}\bigg(\frac{3}{4}\frac{m_{\rm e}c^{2}}{h\nu_{\rm o}^{'}}\bigg)\nu_{\rm o}^{'}=\frac{3}{8}\bigg(\frac{m_{\rm e}c^{2}}{h}\bigg)^{2}\frac{1}{\Gamma_{\rm j}\nu_{\rm o}^{*}},
\nonumber\\
&=& 5.73\times 10^{24}~\frac{1}{\Gamma_{\rm j}\nu_{\rm o,15}^{*}}~~(\rm Hz),
\nonumber\\
&=& 23.7~ \frac{1}{\Gamma_{\rm j}\nu_{\rm o,15}^{*}}~~(\rm GeV),
\end{eqnarray}
where $\nu_{\rm o}^{'}\simeq 2\Gamma_{\rm j}\nu_{\rm o}^{*}$ has been used, and $\nu_{\rm o,15}^{*}$ is in units of $10^{15}$\,Hz. In observer's frame, the corresponding frequency is given by
\begin{eqnarray}
\nu_{\rm c,KN}^{\rm obs} &=& 5.73\times 10^{24}~\nu_{\rm o,15}^{*}\frac{\delta_{\rm D}}{\Gamma_{\rm j}(1+z)}~~(\rm Hz),
\nonumber\\
&=& 23.7~ \nu_{\rm o,15}^{*}\frac{\delta_{\rm D}}{\Gamma_{\rm j}(1+z)}~~(\rm GeV).
\end{eqnarray}

\section{The linear fitting to the synchrotron, ERC-DT and ERC-BLR humps}
\label{sec: lin-fit-resut}

To quantify the spectral shape and the changes between the spectral indices, a linear function, $\chi(x)=\kappa x+\rho$, is used to fit the $\log(\nu F_{\nu}$) versus $\log(\nu$) distribution for the synchrotron, ERC-BLR and ERC-DT processes. These are labeled respectively by sub- and super-scripts `1', `2' and `3', i.e, $\kappa_{\rm 1}$ and $\rho_{\rm 1}$ represents the spectral slope and the intercept obtained from fitting the synchrotron spectrum, with the fitting spanning the frequency range from $\log(\nu_{\rm d}^{1})$ to $\log(\nu_{\rm u}^{1})$. For the high-energy hump, since the ERC-BLR process dominates over the ERC-DT, ERC-AD and the SSC processes, thus the $\nu_{\rm d}^{2}$ is taken as the peak frequency of the ERC-BLR spectrum. Moreover, $\nu_{\rm u}^{2}$ is fixed at the cutoff frequency where the KN effect becomes significant for the peak frequency of the BLR radiation field, which has a maximum frequency of $\nu_{\rm p,15}^{\rm BLR}$ in terms of $10^{15}$\,Hz in the stationary frame. For comparision, we also present the derived slopes using the $\delta$ approximation, $-(s_{\rm 1}-3.0)/2$ and $-(s_{\rm 2}-3.0)/2$, where the spectra are contributed by the electron population with Lorentz factor $\gamma$ within the range between $\gamma_{\rm min}$ and $\gamma_{\rm br}$, and the electron population with Lorentz factor $\gamma$ within the range between $\gamma_{\rm br}$ and $\gamma_{\rm max}$, respectively. The detailed results of the linear fitting are provided in Table \ref{table_linfit_SED}.

\begin{table*}
\tablenum{A1}
    \scriptsize 
    \setlength{\tabcolsep}{3.5pt}
    \caption{Results of the linear fitting to radiation humps of synchrotron, ERC-BLR and ERC-DT}
    \begin{center}
        \begin{tabular}{ccccccccccccccccccc}
        \hline
        \hline        
         &\multicolumn{4}{c}{Fitting to the synchrotron hump} & \multicolumn{4}{c}{Fitting to the ERC-BLR hump} & \multicolumn{4}{c}{Fitting to the ERC-DT hump}\\       
        \cmidrule(r){2-5}\cmidrule(r){6-9}\cmidrule(r){10-13}
        
         SED-ID & $\kappa_{1}$ & $\rho_{1}$ & log($\nu_{\rm d}^{1}$)& log($\nu_{\rm u}^{1}$) & $\kappa_{2}$ & $\rho_{2}$ & log($\nu_{\rm d}^{2}$) & log($\nu_{\rm u}^{2}$) & $\kappa_{3}$ & $\rho_{3}$ & log($\nu_{\rm d}^{3}$) & log($\nu_{\rm u}^{3}$)&$\nu_{\rm p,15}^{\rm BLR}$ & -($s_{\rm 1}$-3.0)/2 & -($s_{\rm 2}$-3.0)/2 \\        
        \hline
		Ab-10(A)& -0.048 & -10.97 & 12.0 & 12.5& -0.269 & -5.58 & 22.19 & 24.51 & -0.053 & -10.65 &  20.0 & 20.35  & 1.30 & 0.55 & -0.05 \\	
		Ab-10(B)& -0.101 & -10.36 & 12.5 & 15.8  & -0.357 & -3.99 & 22.41 & 22.48  & -0.102 & -9.92  & 20.3  & 23.2& 1.41 & 0.55 & -0.10\\	
		Ab-10(C)& -0.095 & -10.14 & 12.2 & 15.7  & -0.357 & -4.16 & 22.3 & 24.41 & -0.104 & -10.00 & 20.0  & 23.5 & 1.64 & 0.55 & -0.10\\
		Na-12(A)& -0.108 & -10.63 & 12.4 & 15.5 & -0.275 & -5.59 & 22.41 & 24.32 & -0.106 & -9.05 & 20.8 & 23.2 & 2.02 & 0.55 & -0.10\\				
		Na-12(B)& -0.056 & -10.87 & 12.1 & 15.8  & -0.291 & -4.90 & 22.63 & 24.31 & -0.056 & -9.91  & 20.4  & 23.8 & 2.06 & 0.55 & -0.05\\	
		Ah-17(A)& 0.090  & -11.43 & 12.1 & 16.2 & -0.164 & -6.53 & 22.96 & 24.55  & 0.090  & -11.93& 20.4 & 23.8& 1.19 & 0.50 & 0.10\\	
		Ah-17(B)& -0.055 & -10.26 & 12.5 & 16.0 & -0.245 & -5.66 & 22.41 & 24.64 & -0.055 & -9.86 & 20.5 & 23.7 & 0.971 & 0.60 & -0.05\\	
		Pr-19(A)& -0.054 & -10.68 & 12.6 & 15.6 & -0.0298 & -4.50 & 22.30 & 24.38 & -0.054 & -10.67 & 20.0 & 23.5 & 1.75 & 0.55 & -0.05\\	
		Pr-19(B)& -0.005 & -10.70 & 12.5 & 16.0 & -0.223 & -6.05 & 22.19 & 24.68 & -0.006 & -11.21 & 20.0 & 23.7 & 0.887 & 0.55 & 0\\	
		Pr-19(Q2)& -0.107 & -10.35 & 12.5 & 15.8 & -0.302 & -5.22 & 22.19 & 24.46 & -0.103 & -9.85 & 20.5 & 22.5 & 1.45 & 0.55 & -0.10\\	
		Pr-19(C)& -0.101  & -9.91  & 12.5 & 15.8 & -0.354 & -4.27 & 22.19 & 24.55 & -0.102 & -10.15 & 20.0 & 23.0 & 1.18 & 0.55 & -0.10\\	
	    Pr-19(D)& 0.097  &  -11.64 & 12.0 & 15.6 & -0.125 & -7.58 & 22.96 & 24.50 & 0.097 & -12.03 & 20.3 & 23.7 & 1.33 & 0.55 & 0.10\\
        \hline
        \end{tabular}
    \end{center}

    \label{table_linfit_SED}
\end{table*}

\end{document}